 \documentclass[12pt,preprint]{aastex}
 
\usepackage{natbib}

\newcommand{\msun}{\mbox{$M_{\odot}$}}

\def\deg      {{\ifmmode^\circ\else$^\circ$\fi}} 

 \shorttitle{COSMOS-LSS}
 \shortauthors{Scoville et al.}
 
\begin{document}
 
 
 \title{Large Structures and Galaxy Evolution in COSMOS at z $< 1.1$}
 

%
%
%
 \author{ N. Scoville\altaffilmark{1,2},
H. Aussel\altaffilmark{4,10},
A. Benson\altaffilmark{1},
A. Blain\altaffilmark{1},
D. Calzetti\altaffilmark{5},
P. Capak\altaffilmark{1},
R. S. Ellis\altaffilmark{1},
A. El-Zant\altaffilmark{3},
A. Finoguenov\altaffilmark{7},
M. Giavalisco\altaffilmark{5},
L. Guzzo\altaffilmark{6},
G. Hasinger\altaffilmark{7},
J. Koda\altaffilmark{1},
O. Lefevre\altaffilmark{8},
R. Massey\altaffilmark{1},
H. J. McCracken\altaffilmark{9,16},
B. Mobasher\altaffilmark{5},
A. Renzini\altaffilmark{11,19},
J. Rhodes\altaffilmark{1,12},
M. Salvato\altaffilmark{1},
D. B. Sanders\altaffilmark{4},
S. S. Sasaki\altaffilmark{14,17},
E. Schinnerer\altaffilmark{13},
K. Sheth\altaffilmark{1},
P. L. Shopbell\altaffilmark{1},
Y. Taniguchi\altaffilmark{17},
J. E. Taylor\altaffilmark{1},
D. J. Thompson\altaffilmark{15,18}}

\altaffiltext{$\star$}{Based on observations with the NASA/ESA {\em
Hubble Space Telescope}, obtained at the Space Telescope Science
Institute, which is operated by AURA Inc, under NASA contract NAS
5-26555; also based on data collected at : the Subaru Telescope, which is operated by
the National Astronomical Observatory of Japan; the XMM-Newton, an ESA science mission with
instruments and contributions directly funded by ESA Member States and NASA; the European Southern Observatory under Large Program 175.A-0839, Chile; Kitt Peak National Observatory, Cerro Tololo Inter-American
Observatory and the National Optical Astronomy Observatory, which are
operated by the Association of Universities for Research in Astronomy, Inc.
(AURA) under cooperative agreement with the National Science Foundation;
the National Radio Astronomy Observatory which is a facility of the National Science
Foundation operated under cooperative agreement by Associated Universities Inc ;
and the Canada-France-Hawaii Telescope with MegaPrime/MegaCam operated as a
joint project by the CFHT Corporation, CEA/DAPNIA, the NRC and CADC of Canada, the CNRS of France, TERAPIX and the Univ. of
Hawaii.}
\altaffiltext{1}{California Institute of Technology, MC 105-24, 1200 East
California Boulevard, Pasadena, CA 91125}
\altaffiltext{2}{Visiting Astronomer, Univ. Hawaii, 2680 Woodlawn Dr., Honolulu, HI, 96822}
\altaffiltext{3}{Canadian Institute for Theoretical Astrophysics, Mclennan Labs, University of Toronto, 60 St. George St, Room 1403, Toronto, ON M5S 3H8, Canada}
\altaffiltext{4}{Institute for Astronomy, 2680 Woodlawn Dr., University of Hawaii, Honolulu, Hawaii, 96822}
\altaffiltext{5}{Space Telescope Science Institute, 3700 San Martin
Drive, Baltimore, MD 21218}
\altaffiltext{6}{INAF-Osservatorio Astronomico di Brera, via Bianchi 46, I-23807 Merate (LC), Italy}
\altaffiltext{7}{Max Planck Institut f\"ur Extraterrestrische Physik,  D-85478 Garching, Germany}
\altaffiltext{8}{Laboratoire d'Astrophysique de Marseille, BP 8, Traverse
du Siphon, 13376 Marseille Cedex 12, France}
\altaffiltext{9}{Institut d'Astrophysique de Paris, UMR7095 CNRS, Universit\`e Pierre et Marie Curie, 98 bis Boulevard Arago, 75014 Paris, France}
\altaffiltext{10}{Service d'Astrophysique, CEA/Saclay, 91191 Gif-sur-Yvette, France}
\altaffiltext{11}{European Southern Observatory,
Karl-Schwarzschild-Str. 2, D-85748 Garching, Germany}
\altaffiltext{12}{Jet Propulsion Laboratory, Pasadena, CA 91109}
\altaffiltext{13}{Max Planck Institut f\"ur Astronomie, K\"onigstuhl 17, Heidelberg, D-69117, Germany}
\altaffiltext{14}{Astronomical Institute, Graduate School of Science,
         Tohoku University, Aramaki, Aoba, Sendai 980-8578, Japan}
\altaffiltext{15}{Caltech Optical Observatories, MS 320-47, California Institute of Technology, Pasadena, CA 91125}
\altaffiltext{16}{Observatoire de Paris, LERMA, 61 Avenue de l'Observatoire, 75014 Paris, France}
\altaffiltext{17}{Physics Department, Graduate School of Science, Ehime University, 2-5 Bunkyou, Matuyama, 790-8577, Japan}
\altaffiltext{18}{Large Binocular Telescope Observatory, University of Arizona, 933 N. Cherry Ave.
   Tucson, AZ  85721-0065,   USA}
\altaffiltext{19}{Dipartimento di Astronomia, Universitˆ di Padova, vicolo dell'Osservatorio 2, I-35122 Padua, Italy}

\altaffiltext{}{}

%
%
%

 \begin{abstract}
 
 We present the first identification of large-scale structures (LSS) at z $< 1.1$ in the Cosmic Evolution Survey (COSMOS).
 The structures are identified from adaptive smoothing of galaxy counts in 
 the pseudo-3d space ($\alpha,\delta$,z) using the COSMOS photometric redshift catalog.
 The technique is tested on a simulation including galaxies distributed
 in model clusters and a field galaxy population -- recovering structures on all scales from 1 
 to 20\arcmin~without {\it a priori} assumptions for the structure size or density profile. 
 Our procedure makes {\bf no} {\it a priori} selection on galaxy spectral energy distribution (SED,
 for example the Red Sequence), enabling an unbiased investigation of environmental 
 effects on galaxy evolution. 
 
 The COSMOS photometric redshift catalog yields a sample of $1.5\times10^5$ galaxies with  
 redshift accuracy, $\Delta z_{FWHM}/(1+z) \leq 0.1$ at z $< 1.1$ down to  I$_{AB} \leq 25$ mag. 
 Using this sample of galaxies, we identify  42 large-scale structures and clusters. 
 Projected surface-density maps for the structures indicate multiple 
 peaks and internal structure in many of the most massive LSS. 
 The stellar masses (determined from the galactic SEDs) for the LSS range 
 from M$_* \sim 10^{11}$  up to $\sim3\times10^{13}$ \msun. Five LSS have 
total stellar masses exceeding $10^{13}$ \msun. (Total masses including non-stellar 
baryons and dark matter are expected to be $\sim 50 \rightarrow 100$ times greater.) 
The derived mass function for the LSS is consistent (within the expected Poisson and cosmic variances) 
with those derived from optical and X-ray studies at lower redshift.  

To characterize structure evolution and for comparison 
with simulations, we compute a new statistic -- the area filling factor  as a function of the overdensity value compared to the 
mean at surface overdensity ($f_A(\Sigma/{\overline{\Sigma}(z)}$).  The observationally determined f$_A$ 
has less than 1\% of the surface area (in each redshift slice) with overdensities exceeding 10:1
and  evolution to 
higher overdensities is seen at later epochs (lower z) -- both characteristics are in  good agreement 
with what we find using similar processing on the Millennium Simulation. Although similar variations in the filling factors 
as a function of overdensity and redshift 
are seen in the observations and simulations, we do find that the observed distributions reach 
higher overdensity than the simulation, perhaps indicating  
over-merging in the simulation. 

 All of the LSS
 show a dramatic preference for earlier spectral energy distribution (SED type) galaxies in the denser regions of  the 
 structures, independent of redshift. The SED types in the central 1 Mpc and 1 -- 5 Mpc regions of each structure
 average about 1 SED type earlier than the mean type at the same 
 redshift, corresponding to a stellar population age difference of $\sim$2 -- 4 
 billion years at z = 0.3 to 1. 
   
We also investigate the evolution of key galactic properties -- mass, luminosity, SED 
and star formation rate (SFR) -- with redshift and environmental density as derived from  
 overdensities in the full pseudo 3-d cube. Both the maturity of the stellar populations and the 'downsizing' 
of SF in galaxies vary strongly with redshift (epoch) and environment. For a very broad mass range (10$^{10} -- 10^{12}$\msun), we find that galaxies in dense 
environments tend to be older -- this is not just restricted to the most massive galaxies. And in low density environments, 
the most massive galaxies appear to have also been formed very early (z $> 2$), compared 
to the lower mass galaxies there.
Over the range z $< 1.1$, we do not see evolution in the 
mass of galaxies by more than a factor $\sim 2$ separating  active and inactive star-forming galaxy populations. 

\end{abstract}

 

 
 \section{Introduction}\label{intro}
 
 The Cosmic Evolution Survey (COSMOS) 
 is intended to probe the evolution of 
 galaxies, AGN and dark matter in the context of their cosmic 
 environment (large-scale structure -- LSS). The survey area
 samples scales of LSS out to  $\sim$ 50 -- 100 Mpc at z $> 0.5$. This corresponds 
 to 9 times the area of GEMS and EGS \citep{rix04,dav06}, the next largest HST imaging surveys.
 The COSMOS area \citep{sco06} is expected 
 to have a 50\% probability of having one 10$^{14}$ \msun~halo (dark and luminous matter) within every 
 $\Delta$z $\sim$ 0.1 at z $\sim$ 1 -- 2 (based on  $\Lambda$CDM simulations; see \cite{ben01});
 lower mass halos ($\sim10^{13}$ \msun) are $\sim 20$ times more abundant and therefore will be seen in 
 every $\Delta$z $\sim$ 0.1. A major theme for COSMOS is the effect of cosmological environment
 on the evolution of galaxies and AGN. The identification and measurement of LSS are therefore 
 a prerequisite to many aspects of science with COSMOS since the large-scale structures 
 define the local environment.  In COSMOS, the local galaxy number and mass densities can be compared with the total mass densities determined from weak lensing tomography and hot X-ray 
 emitting gas in the virialized parts of LSS having clusters of galaxies \citep[see below,][]{mas06,fin06}.

The identification of LSS from the observed surface-density of galaxies requires 
some means of discriminating galaxies at different distances along the line of 
sight; otherwise, the increased shot noise in the galaxy counts reduces the signal-to-noise ratio
for the large-scale structure. The better the redshift or distance discrimination, the easier
it is to see low density, large-scale structures. (This is true down to the point that the internal 
velocity dispersion of the structure is resolved.) For LSS finding, line-of-sight 
discrimination is usually accomplished using:
1) color selection \citep[e.g.~using broadband colors to select red sequence galaxies,][]{gla05};  2) 
photometric redshifts based on the broadband spectral energy distribution (SED),
or 3) spectroscopic redshifts \citep[see Appendix \ref{appendix};][]{wey94,pos96, sch98,mar02}. Color selection is not used here
since it would preclude investigation of correlations between environmental density and 
galaxy SED or morphological type \citep[][]{dre97,smi05}. With color selection, the defined large-scale structures 
would be {\it a priori} biased to a particular galactic SED type (e.g. early type galaxies if the red-sequence method is used). 
High density structures may 
well be rich in red galaxies of early morphological type, but exploring the dependence of 
galaxy type on environmental density requires that the environment be defined or identified without bias toward specific galaxy types. 
 Ultimately, the use of spectroscopic redshifts to determine distances is more desirable, 
 provided a sufficiently large sample exists \citep{lef05,ger05,men06,cop06,coi06}. The z-COSMOS spectroscopic  
survey  
will yield  $\sim$ 30,000 galaxies when completed in $\sim$2008
\citep{lil06,imp06};
however, at this time the spectroscopy is much more limited and we must rely on the alternative 
approaches for line-of-sight discrimination. 

In this paper, we identify LSS in the 2\sq\deg COSMOS 
field using the extensive COSMOS photometric redshift catalog \citep{mob06} to analyze the galaxy surface 
density in redshift slices out to z = 1.2. The galaxy samples used for this work and their completeness  
are discussed in Section \ref{dist}. We use an adaptive smoothing technique (see Appendix \ref{appendix}) to identify areas of significantly enhanced 
galaxy surface-density. For each significant peak in the smoothed surface-density pseudo 3-d cube, we find all 
connected pixels to delineate a sample of 42 structures (Section \ref{adapt_smo}). For each structure, we estimate the
dimensions, number of galaxies, and mass (from the broadband fluxes of the galaxies). The stellar mass distribution of identified  COSMOS LSS extends  from 10$^{11}$ up to $3\times10^{13}$ \msun . The relative amount of structure at different overdensities is analysed in Section \ref{sims} and compared 
with results from the Millennium Simulation. 

We then investigate the evolution of galaxies with respect to their location in the 42 LSS in Section \ref{lss_evol} 
and environmental density in  Section \ref{envir_evol}. Strong variation of the SED type and star formation activity 
is seen with both redshift and environmental density. The maturity of the stellar populations and the 'downsizing' 
of SF in galaxies 
is also strongly varying with epoch and environment (Section \ref{mature}). (Adopted cosmological parameters, used 
throughout, are: H$_0$ = 70 km s$^{-1}$ Mpc$^{-1}$, $\Omega_M$ = 0.3 and  $\Omega_\Lambda$ = 0.7)

\section{Photometric Redshifts and Sample Selection}\label{dist}

For this investigation we use photometric redshifts to separate the galaxy population
along the line-of-sight. It is vital for the analysis that the binning in redshift 
be matched to the accuracy of the redshifts. Using binning that is finer
than the redshift uncertainties distributes the galaxies in a single structure
over multiple redshift slices and thus reduces the signal in each slice. Conversely,
 bins of width larger than the redshift uncertainties will increase the shot noise associated
with the background surface-density of galaxies, relative to the large-scale structure signal.

The COSMOS photometry catalogs were generated from deep ground-based 
optical imaging at Subaru \citep{tan06} and  CFHT; they are combined with shallower near infrared imaging
from NOAO (KPNO and CTIO) \citep{cap06}. 
The resulting photometric redshift catalog contains 1.2 million objects 
at I$_{AB} < $26 \citep{mob06}. For approximately 900 objects (all with I$_{AB} \leq 24$ mag
from zCOSMOS \cite{lil06}), there 
exist spectroscopic redshifts; after removing 'catastrophic' outliers ($\sim$1\% of the
sample), the offsets between the photometric and spectroscopic redshifts
have $\sigma_z / (1 + z) \simeq 0.03$  or  $\Delta z_{FWHM}/(1+z) \leq 0.1$ \citep{mob06}. 
Since the spectroscopic --
photometric redshift comparison is limited to a small sample of mostly 
brighter objects, we will instead use the 'goodness' of fit from the 
photometric redshift determination for a more general assessment of the 
redshift accuracies. This represents an internal dispersion and hence is likely to be 
lower than the true uncertainty which includes systematic effects; nevertheless,
it does provide a characterization of the dependence on redshift and magnitude 
cutoff for any sample selection.

\begin{figure}[ht]
\epsscale{1.} 
\plottwo{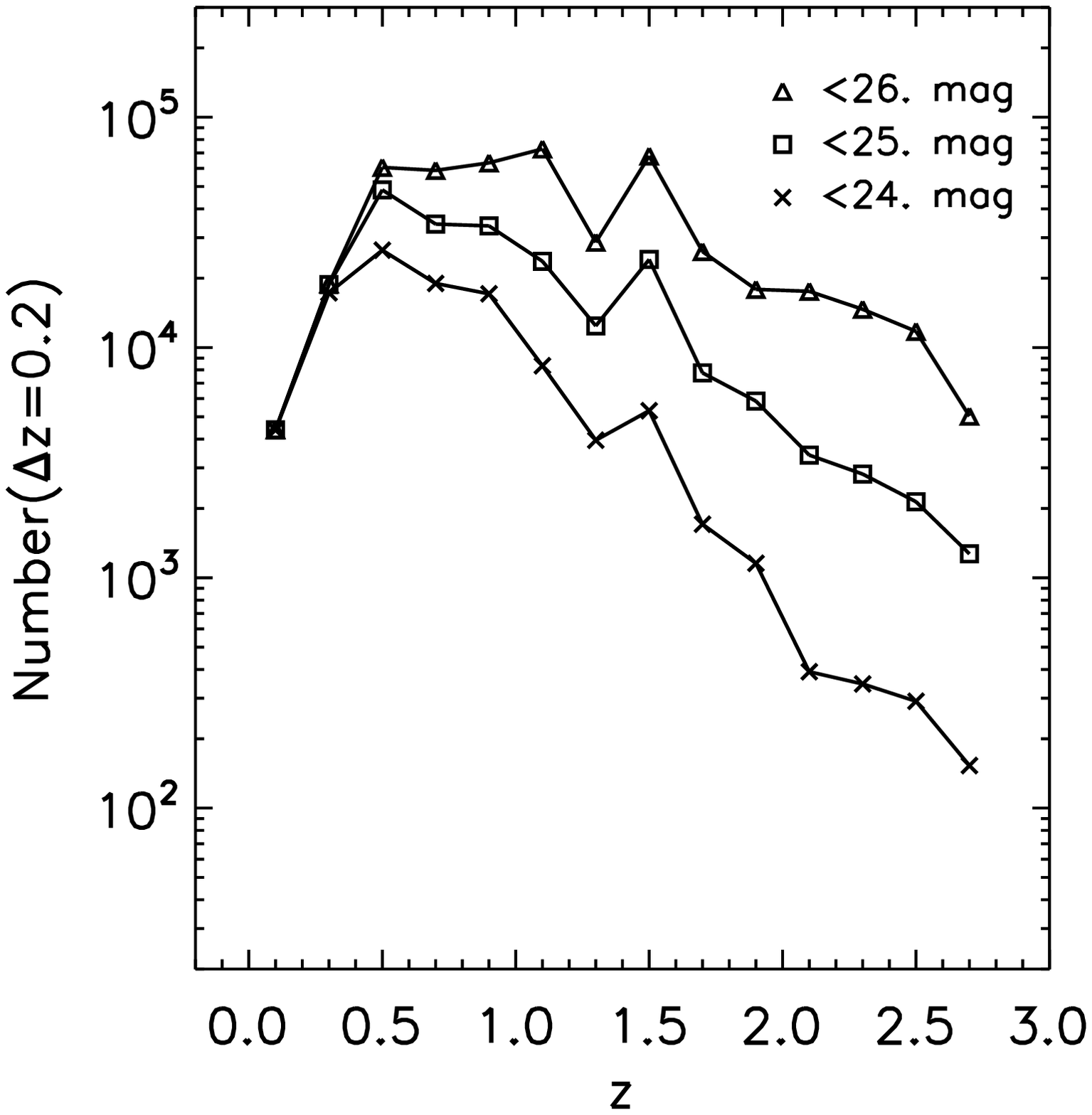}{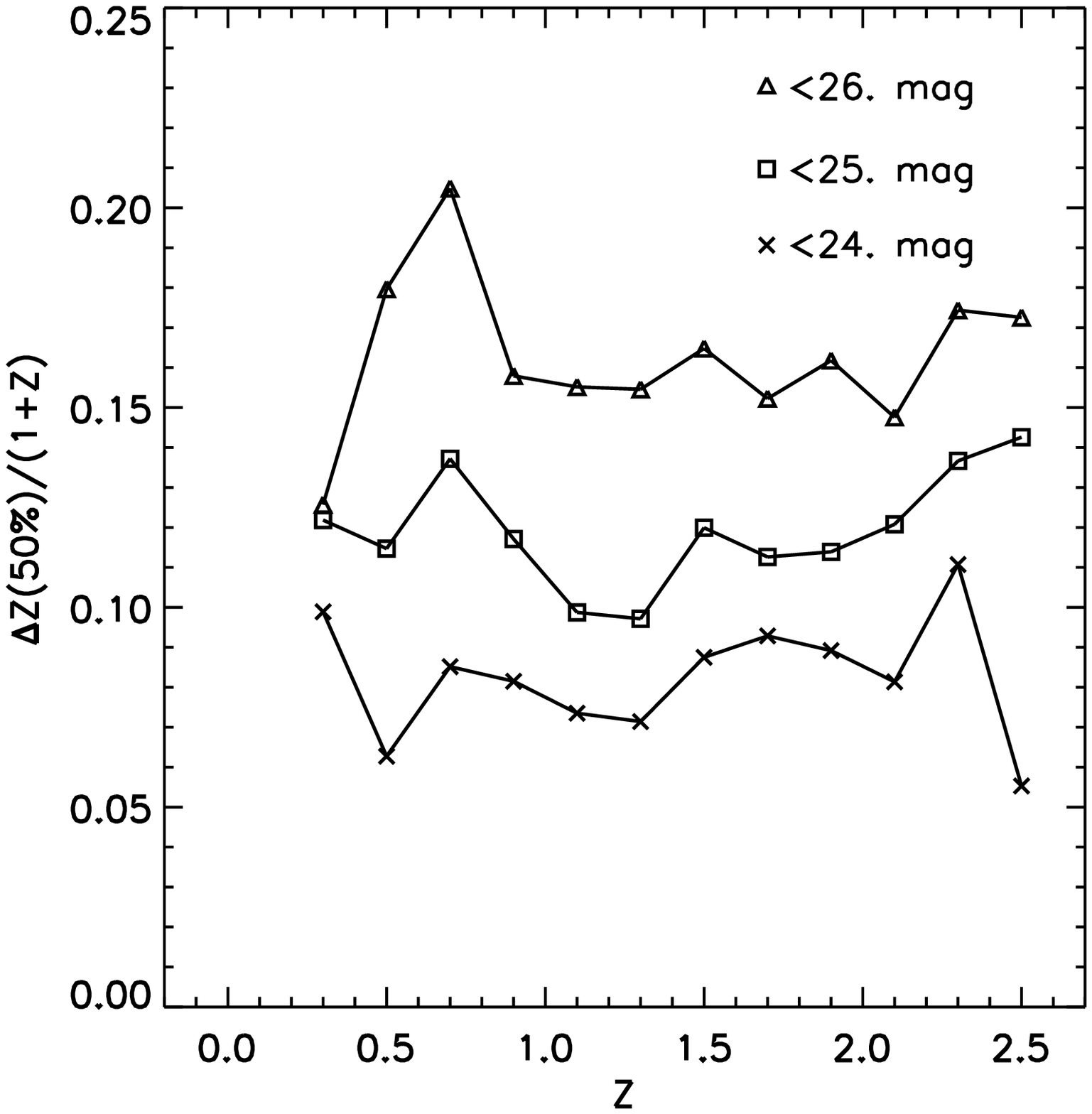}

\caption{a) Left : The number distribution of galaxies from the COSMOS photometric redshift catalog \citep{mob06} is shown as a function of redshift in bins of width $\Delta z =0.2$ for different I-band magnitude cutoffs. b) Right : The widths of the derived photometric redshift probability distributions are shown 
as a function of z. The quantity plotted is the mean of 1.3$\times\sigma_z/(1+z)$ for the photo-z likelihood distributions for all galaxies 
in each redshift
bin of $\Delta z =0.2$. For a Gaussian uncertainty distribution, 50\% of the 
probability is within a full width of 1.3$\times\sigma_z$  of the peak. Galaxies used in this 
plot were selected to have M$_V \leq -18$ ~mag and be detected in at least 4 bands.} 
\label{z_sigmas}
\end{figure}

The photometric redshifts were derived using the Bayesian Photometric Redshift
method 
\citep[BPZ:][]{ben00a,mob06}. The fitting presumes 6 basic spectral energy distributions (SEDs) and for the photometric redshift
catalog used here, there was no assumed 'prior distribution' for the galaxy magnitudes. The dust extinction within each galaxy was also a free parameter in the redshift fitting.  For SED types 0 to 2, a Galactic extinction law was assumed: for SED types greater than 2, a Calzetti law was used \citep{mob06}. The fitting outputs the most probable redshift,  68 and 95\% confidence intervals, the intrinsic SED
type and the absolute magnitude (M$_V$). In Figure \ref{z_sigmas}, the redshift 
distribution and uncertainty in the redshift fits are shown as a function 
of redshift and i-band magnitude cutoff in the sample. 
The SED classifications used here are : 1 (E/S0), 2 (Sa/Sb), 3(Sc), 4(Im), and 5-6(starburst). The quantity $\Delta$ z(50\%)/(1+z) shown in Figure \ref{z_sigmas}b is the 
mean value (at each z) of the width in z containing 
50\% of the probability distribution. (This 
full-width is 1.3$\times\sigma$ for the derived fits, assuming a Gaussian uncertainty distribution.) 
Also shown is the photometric redshift uncertainty as a function of  
apparent magnitude cutoff  -- based on these curves, we adopt a cutoff
$I_{AB} = 25$ mag. 
The uncertainties plotted in Figure \ref{z_sigmas} indicate that the bin width for identifying large-scale structures should be 
approximately $\Delta z \simeq 0.1 - 0.15$ up to z $\simeq 1.2$ 
and $\Delta z \simeq 0.25$ up to z $\simeq 2.5$ for the chosen magnitude limit. 

Throughout the investigation here, we use the galaxy rest frame SEDs to characterize 
the galaxy type, rather than the observed morphologies. 
\cite{cap06a} find a tight correlation between 
SED and morphology as measured by the Gini and Compactness measures.  We have also 
correlated the Sersic indices measured using GALFIT for 5,000 bright (I $<$ 22 mag)
galaxies in COSMOS at z = 0.2 to 0.8. The SED type 1 (E/S0) is strongly correlated with 
an R$^{1/4}$ law (Sersic n  = 4); however, there exists a large dispersion in the 
Sersic indices for the later SED types. This large dispersion probably reflects both the real 
dispersion in bulge to disk ratios of later galaxy types and the difficulties of measuring 
the morphologies for faint galaxies at high redshift. Throughout this investigation we will  
use the SED types, derived from the photometric redshift fitting, to classify the galaxies
since the SEDs are more readily classified for faint galaxies than the morphology. 

The COSMOS photometric redshift catalog also includes galactic stellar masses,
derived from the absolute magnitude, SED type and redshift  of each galaxy \citep{mob06}. 
Approximate estimates for the star formation rates  
(SFR) were also derived using  the SED type, absolute magnitude and fitted extinction.  
The intrinsic UV continuum (corrected for extinction at $\lambda = 1500$ \AA) was used to estimate the SFR 
 \citep{mob06a}.

\subsection{Galaxy Samples\label{sample}}

Although the COSMOS photometric redshift catalog contains over a million objects, 
we impose selection criteria to yield a more reliable galaxy sample for 
structure identification -- i.e. galaxies with the best photometric redshifts,  detected in several bands, and with significant intrinsic luminosity. 
We thus restrict the  analysis to :
\begin{mathletters}
\begin{equation}
 25 ~~mag \geq I_{AB} \geq 19 ~~mag 
\end{equation}
\begin{equation}
M_V \leq -18 ~~mag
\end{equation}
\end{mathletters}
\noindent We also require that each object be detected in at least 4 bands and the SExtractor 
stellarity parameter be less than 0.95. The former (in addition to the 25 mag cutoff -- Eq (1a)) limits the sample to galaxies with accurate photometry; the latter removes
objects which are likely to be stars or QSOs. Similarly very bright objects are also excluded by 
condition (1a) since they are likely to be stars. Since the fraction of galaxies with dominant AGN is 
probably not large, their exclusion should not significantly affect the large-scale structure definition. 
Condition (1b) removes galaxies 
which have low absolute luminosities and presumably low mass. These criteria yield the galaxy samples listed in Table \ref{tbl-1} along with the adopted redshift binning for the LSS identifications presented below. In Table \ref{tbl-1}, we also
provide a breakdown of the samples with respect to SED type from the 
photometric-redshift fitting. The majority of the analysis in this paper refers to the 
low-z sample in Table \ref{tbl-1}.

\subsection{Galaxy Selection and Completeness with Redshift}\label{select}

At larger redshifts, the galaxy sample used to define LSS will be incomplete at 
low luminosities (masses) and we evaluate here the severity of this effect
with two approaches : evaluating the cutoff $L_*$ (the characteristic luminosity at the knee in the Schecter luminosity function) as a function of redshift and comparing the galaxy mass functions 
as a function of redshift for the sample described in Section \ref{sample}.

Figure \ref{selection} shows the distributions of absolute magnitude (left) and stellar masses (right)
for galaxies in our sample as a function of redshift.  
On the left panel, the lines indicate the expected absolute magnitudes 
for L$_*$ and L$_*/10$ galaxies assuming a typical, passive evolution brightening of 1.2 mag from 
z = 0 to 1.2.  Here one sees that the sample easily goes down to L$_*/10$ out to 
z = 1.2 with $I_{AB} < 25$ mag (the lower envelope in the grayscale). Incompleteness does 
set in at less than L$_*/10$, but typically less than 30\% of the total luminosity is contained 
in these galaxies for a Schecter luminosity function. 

Alternatively, one can compare the distribution functions of galactic stellar masses as a function 
of redshift to assess the incompleteness (assuming, to first order, that the  
mass function is not strongly varying over this redshift interval -- e.g. $\leq 2\times$; see \cite{bor06} and below).  The derived mass functions for 
galaxies entering the sample (Section \ref{sample}) used here for LSS definition are shown on the 
right panel of Figure \ref{selection}. These mass functions are in agreement in both 
shape and absolute value with previously determined mass functions for this redshift 
range \citep{dro05,bor06,bun06}. The higher noise seen in 
the low z mass functions is due to the much smaller volume and hence smaller number of galaxies sampled.

The total number of galaxies and total mass of galaxies per unit comoving 
 volume were evaluated by integrating the distribution functions shown in right panel of Figure \ref{selection} 
 at masses above 10$^9$ \msun. In columns 2 \& 3 of Table \ref{tbl-2}, the totals are divided by the 
comoving  volume in order to assess the 
 count and mass incompleteness relative to this maximum redshift bin (see Table \ref{tbl-2}). 
 The falloff seen in the mass functions for z $> 0.7$ at M$_{stellar} < 5 \times 10^9$\msun~
is probably due to incompleteness at the apparent magnitude limit $I_{AB} < 25$ mag for our sample. Incompleteness at this 
magnitude limit is quantified for COSMOS in \cite{sco06}.
 In terms of integrated stellar mass for galaxies above $10^9$\msun, our sample is 
 at most missing only 10\% (Table \ref{tbl-2}) of the total mass, relative to the lower redshift 
 bins which are more complete (e.g. z $\sim 0.5$). In the analysis below we will not correct for this 
 incompleteness unless noted explicitly, given the fact that it is probably not large and the 
 uncertain assumptions of constancy in either the mass- or luminosity- function which would be required.

\subsection{Pseudo 3-D Surface-Density and Noise Estimates}

The adaptive smoothing algorithm we employ here is designed to analyze 
redshift slices, each of which represents the surface-density of galaxies ($\Sigma$)
in a redshift bin. The custom-built algorithm is formally described along with test results in 
Appendix \ref{appendix}. As noted in Section \ref{dist}, the width of these slices will
be $\Delta z = 0.1$ and 0.25 for the low and high-z samples respectively. 
However, given that  the different galaxies may have quite different widths for the fitted 
redshift probability distribution, insertion of each galaxy
into the 3-d cube ($\alpha$,$\delta$,z) as a delta function ($\delta(z-z_{pk})$) at the most likely redshift
would not optimally weight the galaxies with the most accurate photometric redshifts. 
Instead, we populate the 3-d cube with a Gaussian distribution in z  for each galaxy. The Gaussian  dispersion was taken from the high and low-z 68\%-confidence limits from the photometric redshift fit -- specifically 
$\sigma_z = (z_{high~68\%} - z_{low~68\%})/2$). Thus, in the adaptive smoothing 
procedure, galaxies which have a large uncertainty in their derived redshifts
will have relatively low weight, because they will be spread over a larger range 
in the redshift dimension. And galaxies with tighter redshift fits will be treated
more significantly. One concern might be that this tends to prefer structures 
defined by early type galaxies which have a strong Balmer break and thus small 
redshift uncertainty. As a test, we also used the adaptive smoothing on a 3-d cube 
with the galaxies located as points (rather than a probability distribution) 
 at their most probable redshift, Since this test yielded structures similar to those shown here, we 
prefer employ the 
probability distributions to take account of the redshift uncertainties. The cube being analysed is therefore
the 3-d 'probability' surface-density of galaxies, not the 
galaxies as discrete points in 3-d space. For the adaptive smoothing, the required 
noise estimate ($\sigma$) is taken as the
counting uncertainty, i.e. the square root of the galaxy surface-density cube.

The square COSMOS field is 1.4$\times$1.4\deg ~in size;  the comoving volume out to 
z = 1.1 in the low-z sample is $\simeq 10^7$ Mpc$^3$ \citep{sco06a}. For the adaptive smoothing, 
we use a grid of 300$\times$300 in ($\alpha$,$\delta$).
The angular resolution in the smoothing is therefore $\sim17$\arcsec. 
A typical redshift slice with $\Delta z$ =0.1 
contains $\sim 1 - 3\times10^4$ galaxies (see Figure \ref{z_sigmas});  
each cell will therefore be populated by $\sim0.2$ galaxies, on average. Significantly
higher computational resolution is therefore not warranted. 

\section{Adaptively Smoothed Surface-Density}\label{adapt_smo}

 The  galaxy surface-density derived using 
 the procedure described above and in Appendix \ref{appendix} is shown in Figure \ref{smooth} for redshift slices with  
 $\Delta z$ = 0.1, spaced by $\Delta z$ = 0.1 for the Low-z sample of galaxies. In the
 last two panels of Figure \ref{smooth} we show two of the higher redshift 
slices with $\Delta z$ = 0.25. We leave further analysis of the high redshift 
galaxies to a later paper since deeper near infrared and Spitzer IRAC imaging \citep{san06}
is required for higher accuracy photometric redshifts. 

The surface-density plots show a large number of very significant large features -- 
especially at z = 0.35, 0.75 and 0.85. And at every redshift numerous small groupings
of galaxies are seen. There is a definite 
trend towards increasing complexity of structure (clumpiness) at higher z. 
This is to be expected since structures at high z (earlier epochs)
are dynamically younger and expected to be less relaxed.

\subsection{Structure Identification}

From the derived surface density in the 3-D cube ($\alpha$, $\delta$ and z), we define preliminary LSS 
starting from $> 10\sigma$ peaks, finding all connected pixels above  the $1\sigma$ noise. Using an 
algorithm developed by \cite{wil94}, approximately 140 local maxima
are identified and their connected pixels catalogued. When multiple local maxima are found 
in proximity, the neighboring pixels are associated with the 
nearest local peak -- this can result in subdivision of  structures which have multiple 
peaks (real or noise). The maps of the 140 preliminary structures are therefore checked 
for possible recombination into composite structures. The decision to recombine was based 
on : whether the individual components were touching in 3-d space; their 
borders meshed; and their proximity in the 3-d space was unlikely by chance. This is somewhat subjective but a more physically justified 
recombination would require
spectroscopic redshifts with accuracy similar to the virial velocities of the
groups.  This will become possible with the COSMOS spectroscopic surveys \citep{lil06,imp06}.

Forty-two recombined, independent LSS were found from the procedure described above (i.e.~identification of all local maxima 
and the subsequent recombination).  The 
surface density of each structure, integrated in the z-dimension, is shown in Figure \ref{lss}. Many of the
most populated structures show complex structure with 
multiple peaks in $\alpha$ and $\delta$. The structures are ordered in terms of decreasing number of 3-D pixels
so the most complex and massive structures are those with lowest LSS number.
In Table \ref{struct_param} measurements for the structures are tabulated, including 
the location of peak density $\alpha$ and $\delta$, centroid redshift, sizes, number of galaxies and mass. Figure \ref{lss_2d} shows the projection of all LSS on to the plane of the sky, i.e. integrating in 
redshift;  a finding chart 
for the LSS within the COSMOS field is provided in Figure  \ref{lss_find}. The galaxy 
stellar masses are obtained from the photometric redshift fit, which 
yields an absolute magnitude and an SED type for the galaxy from
which a stellar mass-to-light ratio can be inferred \citep{mob06}. An important 
consideration in measuring the structure parameters is possible contamination 
from the background galaxy population. This contamination is estimated from 
the mean surface-density in each redshift slice. Then for each galaxy, seen in 
projection within the area of a given LSS, the probability that it is in fact associated 
with the structure is given by the ratio $\Sigma_{LSS} / (\Sigma_{LSS} + \Sigma_{background})$.
In calculating the cluster mass, the mass of each galaxy within it is multiplied by 
this local probability. Thus, the derived 
masses are corrected for foreground/background contamination. In the first column 
of Table \ref{struct_param}, we list (in parenthesis) the ID numbers of possibly associated 
X-ray clusters from \cite{fin06}. The surface densities at the cores of these structures are similar 
to those seen in the Rich Cluster and Subaru/XMM Deep surveys \citep[$\Sigma \sim 10 -- 20$ Mpc$^{-2}$; ][]{ohe98,pim06,kod04} 
but their extents at lower density go out to $> 10$ Mpc. Spectroscopic redshifts from zCOSMOS \citep{lil06} will 
be needed to confirm the coherence of these more extended structures.

\subsection{Radial Profiles : Clusters vs. Structures}

Without spectroscopic redshifts it is impossible to determine which 
structures are in fact gravitationally bound. However, the spatial 
distributions of galaxies within the structures suggest that many of 
the LSS are 'relaxed' clusters. In Figure \ref{radial}, we plot the azimuthally
averaged projected radial distribution of galaxies in each structure. For each LSS, radii were calculated from the position of peak number density. In several of the structures with significant
secondary peaks the radial structure is not monotonically decreasing (e.g. \# 10, 13 and 18). 
Since the largest, most complex structures have the lowest LSS numbers, these are most 
likely large-scale structures with multiple clusters (LSS \# 1 -- 8).  All of these are at the higher redshifts; this is due to 
the greater volume sampled at z $> 0.5$. They may eventually relax to form a 
centrally-concentrated cluster.  Conversely, LSS \# 30 -- 42 all appear 
fairly symmetric in their radial distributions and with size 1 -- 2 Mpc, similar to those
of present day galaxy groups ($\lesssim 20$ members) or small clusters.

Most of the structures can be fit by a power law surface density of roughly 
$r^{-1}$, within the central few Mpc, implying that the physical density is $\sim r^{-2}$ -- similar  
to what is usually found for  local clusters such as Perseus. There are however some LSS where the density dependence steepens in 
the central regions; this could reflect the presence of an unrelaxed, outer `infall' 
region. A better understanding of the nature of these density profiles and their 
variations will require better kinematics from spectroscopic redshifts. 

\subsection{Richness}

The last column of Table \ref{struct_param} provides an estimate of the {\it Richness} of the structures using a measure  similar to that used for galaxy clusters \citep{abe58}. For each structure, the central 
surface density of galaxies at R $\leq 1$ Mpc  is listed with the background number 
counts subtracted off.  Radius is measured from the location of peak surface density as in 
Figure \ref{radial}. We first find the 3rd brightest (M3$_V$) cluster galaxy and 
then count all galaxies brighter than M3$_V$ + 2 mag. (M3$_V$ is always brighter 
than -20 so this procedure is not conflicting with the cutoff in Equation 1b.)  The standard procedure 
for local clusters  employs R $\leq 1.5$ Mpc, but  \cite{pos96} find that for a typical 
cluster profile N($R < 1$ Mpc) / N($R < 1.5$ Mpc) $\simeq 0.72\pm0.05$ , so the estimates 
given in Table \ref{struct_param} can be scaled up by approximately 1.39 to 
make them comparable to the standard Abell Richness criteria. The distribution of Richness 
parameters is shown in the right panel of Figure \ref{lss_mass}. Most of the cores of COSMOS LSS fall
in Richness classes 1--3. 

\subsection{Structure and Cluster Masses}\label{struct_mass}

Figure \ref{lss_mass} shows the distribution  of photometrically derived stellar masses \citep{mob06} for the LSS
between $5\times10^{11}$ and $\sim3\times10^{13}$ \msun.  This distribution is clearly
subject to significant Poisson and cosmic variances as discussed below (Section \ref{var}). The 
observed distribution increases toward
low mass, but there are 4 with masses exceeding 
10$^{13}$ \msun. On the high mass end of the dark matter halo mass spectrum, the 
expected number distribution is $N(m)dm \propto m^{-1.6}$ (e.g. \cite{ben01}). The 
distribution of total stellar masses shown in Figure \ref{lss_mass} is much less steep,
but at this point the ratio of stellar to dark matter mass as a function of halo mass and z 
is not known. 

The highest mass structure is 
LSS \#1 with a stellar mass of 2.3$\times10^{13}$ \msun ; clearly, this is a super-massive structure,
equivalent to that of the Coma cluster if allowance is made for a dark matter contribution. The mass in LSS \#1 at z $\simeq 0.74$ is 
distributed over scales $\sim 10$ Mpc. In fact,  the structure appears to be
aggregating around a central cluster \citep{guz06,cas06} and is therefore possibly forming a
super-massive cluster like Coma.  LSS \#1 is also detected in the weak lensing shear analysis \citep{mas06} and in the X-rays \citep{fin06}. LSS \# 17 seems to exhibit a very complex sub-structure as discussed
   in detail by \citep[][]{smo06}. Within the inner $\sim2$~Mpc of LSS \# 17
   there are at least $4$ X-ray luminous clusters and one X-ray quiet
   overdensity at the same redshift. One of the clusters
   hosts a wide-angle tail (WAT) radio galaxy which \cite{smo06} discuss as
   a tracer for assembly of this complex cluster. They argue that
   the structure is in the process of formation and
   estimate that the mass of the final cluster, after merging of all 
   sub-components, will be $\sim20$\% of the Coma cluster
   mass.

The LSS masses listed in Table \ref{struct_param} are for just the stellar masses as derived from 
the observed galaxy fluxes using a mass-to-light ratio, based on the best fit SED from the 
photometric redshift determination \citep{mob06}. The total masses, including non-stellar or 
non-luminous baryons and dark matter, are at least an order of magnitude greater -- 
for  $\Omega_B = 0.025h^{-2} $ and $\Omega_M = 0.3$ with H$_0 = 70$, $\Omega_M / \Omega_B = 6.1$  \citep{kol90}. \cite{hoe05}
analysed the weak-lensing maps for a sample of individual galaxies at $0.2 < z < 0.4$ to estimate the 
total virial masses and baryon fraction in the stars. They find a virial-to-stellar mass ratio 
$M_{vir} / {M_* } = 20 \rightarrow 40$, depending on the assumed stellar IMF   
and $M_{*} / M_{baryon}$ = 14\% (early type galaxies)  to 33\%  (late type galaxies) \citep{hoe05}.
Similar results were found for lower redshift SDSS galaxies by \cite{guz02} and in 
semi-analytic simulations \citep{kau99,ben00,van03}. Since the mass-to-light ratio (including dark matter)
is found to increase as $\sim$L$^{1.5}$ at low redshift \citep{hoe05} 
and presumably, the stellar baryon fraction is lower at higher redshift, we adopt 
$M_{vir} / {M_* } =  50 $ as a reasonable lower limit for the LSS listed in Table \ref{struct_param}
and a more likely value might be $\sim 100$. For LSS \#1 which has been analysed in 
detail using weak lensing \citep{mas06}, X-ray emission \citep{fin06} and 
optical \citep{guz06}, the apparent ratio of total mass to stellar mass 
is $\sim 50$ to 100 \citep{guz06}. The {\it total} masses for the LSS are therefore 
likely to be in the range $10^{13}$ to $\sim3\times10^{15}$ \msun.  

\subsection{Variances in Distributions}\label{var}

The mass distributions derived for the LSS are subject to both  
shot noise, due the small number of structures within each redshift 
slice, and to the {\it cosmic variance} that characterizes
the mass distribution on very large scales. To estimate 
the resulting uncertainties in our LSS mass distributions
we follow the method described by \cite{som04}. 
We first calculate the volume and total mass ($M(vol)$) contained 
in each redshift slice (an area of 2.5 deg$^2$ for the photometric 
redshift catalog used here). The cosmic variance on this scale is 
given by $\sigma_{M(vol)}$
(see Figure 3-right in \cite{som04}) and the {\it relative} variance  
in number counts for halos of mass M$_h$ in a redshift slice is then :

\begin{mathletters}
\begin{equation}\label{eq10}
\sigma (M_h,z) =  (\sigma_{cosmic}^2 + \sigma_{shot}^2)^{1/2}
\end{equation}
\begin{equation}\label{eq20}
\sigma (M_h,z) =    ( (b(M_h,z) ~\sigma_{M(vol)})^2 + 1/N_h )^{1/2}
\end{equation}
\end{mathletters}

\noindent where $b(M_h,z)$ is the bias for the halos of mass $M_h$ at 
redshift z, calculated as in \cite{she99}, and $N_h$ is the average
number of halos in the slice. As explained in \cite{som04}, 
$\sigma_{M(vol)}$ should be a slight {\it overestimate} of the true 
cosmic variance, since
the volume in each slice is much deeper than it is wide, and thus
along its $z$-axis the slice samples much larger scales where the 
variance is smaller.

As suggested by \cite{mo96,som04}, we can identify the appropriate mass range 
for halos corresponding to the LSS by integrating the halo mass function, 
normalized to the survey volume in the redshift slice, down to a 
threshold mass which yields a total number of halos matching  the 
observed number of LSS. 
This assumes that the detected LSS corresponds to the most massive 
halos on a roughly one-to-one basis, and thus that the overall 
abundance of LSS indicates the characteristic 
mass-scale of halos with which they are associated.

In Table \ref{variance}, we summarize the expected relative variances (shot, cosmic and total) in each slice 
for two different mass ranges of DM structures ($10^{13}$ -- $10^{14}$ and $10^{14}$ -- $10^{15}$ \msun). These variances were calculated for the 
cosmological parameters specified in Section \ref{intro} and 
with $\sigma_8 = 0.74$ (the WMAP 3-year value, \cite{spe06}). Comparing the expected numbers of 
halos for the two mass ranges with the observed numbers of LSS, it is most reasonable
to identify the observed LSS with the higher mass range halos, i.e. $10^{14}$ -- $10^{15}$ \msun, 
for which the expected number is $\sim 37$. (This identification is only approximate since  
clearly some of the observed structures are much less massive.) 

Based on the results shown in Table \ref{variance}, we expect that the shot or Poisson 
noise and the cosmic variance are quite comparable for the mass range of the LSS sampled
here at all redshifts. The total combined relative variance is expected to be in the range 
0.4 to 0.6; i.e.,  for the very small number of very high mass structures the derived 
mass and number distributions will have typical uncertainties of $\sim$50\% for 
each redshift bin.

\subsection{Redshift distributions of LSS}

Redshift distributions of the structures in two ranges of LSS stellar mass, M$_*$ are shown in Figure \ref{lss_mass_z}.
As discussed in Section \ref{var}, uncertainties due to Poisson and cosmic variance are comparable and the 
expected total {\it relative} variance in these number distributions is 40-60\% (i.e. $\sigma_N / N \sim 0.5$). 
Also shown is the relative comoving volume (dotted line) for the redshift bins. 
The redshift distribution of total mass ($\Sigma M_{LSS}$) within structures with 
stellar masses in the range M$_* = 10^{12} \rightarrow 10^{13}$ \msun ~is similar (within the 
expected 50\% variance) to the dotted curve showing the 
variation of comoving volume sampled. This suggests that we are recovering structures in this 
mass-range without a strong redshift-dependent selection bias. The higher mass LSS (M$_* > 10^{13}$ \msun) 
exhibit an apparently steeper falloff at low z but this is not statistically significant given 
the small volume sampled.
These results are {\it consistent with} a lack of 
 dramatic evolution in the overall mass fraction for the most masssive structures out to z = 1
(a lookback time of $\sim8$ Gyr) -- as is also seen in $\Lambda$CDM simulations (e.g. \cite{ben01}).
However, this result is certainly not strongly constraining given the large variances.

The derived mass function for the COSMOS LSS can be compared with previously derived  mass distributions for galaxy clusters, mostly for clusters at lower
redshift. The cumulative mass function ($n(>M) =  \int_M^{\infty} n(m) \,dm $) is shown in Figure \ref{num_mass_cum}
for the 42 COSMOS LSS. The error bars are taken from the Poisson 
noise in each bin of width $2\times10^{12}$\msun. The expected cosmic variance is comparable 
to the Poisson noise (see Section \ref{var} and Table \ref{variance}); it is not explicitly included 
here since it is dependent on the adopted cosmological parameters (in particular $\sigma_8$) and on correct identification of the associated DM halo mass range.  The full error bars are likely $\sim \sqrt(2)
\simeq 40\%$ bigger than shown when including cosmic variance.
For the COSMOS sample volume
we adopt 1.5 $\times 10^7$ Mpc$^{3}$ out to z = 1.1. Also shown are the mass functions derived from optical and X-ray 
studies, as summarized by \citep[][]{rei02}. The mass function for masses within the Abell radius of each 
cluster is M$_A$; masses within the regions with density exceeding  200$\rho_c$ and 500$\rho_c$  are M$_{200}$ and M$_{500}$ 
\citep[where 
$\rho_c$ is the critical density for the universe, see][]{rei02}. \cite{rei02} derive the total 
mass including the dark matter and we have scaled their masses down by a factor of 100,
i.e. assuming a stellar mass fraction of 1\% of the total mass (baryons plus dark matter). We have also scaled to h = 70 (used here throughout)
from their h = 50.

Figure \ref{num_mass_cum} shows reasonably consistent number densities (per co-moving volume)
between the COSMOS LSS and the previous studies as summarized by \citep[][]{rei02}, given 
the somewhat uncertain ratio of total to stellar masses (taken as 100 for Figure \ref{num_mass_cum}). 
It should be noted that the mass function within the Abell radii (labelled M$_A$ in Figure \ref{num_mass_cum})
also closely approximates the mass function derived by \citep[][]{bah93}.
As noted in Section \ref{struct_mass}, \cite{hoe05} determined a value of up to $40$ for this mass ratio 
based on lensing measure for clusters at z = 0.2 to 0.4. The somewhat higher value, found here 
in order to achieve agreement in the local mass function shape, might indicate that the fraction of baryons in stars is less  at the higher redshifts (z $\simeq 0.2$ to 1.1) sampled in the COSMOS LSS. 
Alternatively, the COSMOS LSS measurements refer to more extended, lower density structures and filaments than those sampled by  \cite{hoe05}  and the conversion efficiency of baryons into
star is very likely dependent on environment in the context of $\Lambda$CDM models.

\section{Comparison of Structures with $\Lambda$CDM Simulations}\label{sims}

$\Lambda$CDM simulations provide quite specific and relatively confident predictions for the 
growth structure in the dark matter (DM) as a 
function of redshift, given a specified set of cosmological parameters.
On the other hand, the formation and evolution of the visible galaxies within 
the DM structures has relied on semi-analytic models or prescriptions for star and AGN formation, 
stellar evolution and feedback processes. These semi-analytic models   
and the predicted distributions of galaxies in $\Lambda$CDM 
have been mostly constrained from low redshift galaxy surveys. Relatively little 
constraint or testing of the semi-analytics {\it vis-a-vis} the DM LSS has been 
done at high redshift (i.e. z $> 0.2$). In this section, we compare in some detail 
the distributions of galaxy overdensities seen in the COSMOS field out to z = 1.1 with 
those predicted in simulations. 

In particular, we will compare the relative volumes (or areas) 
occupied by observed structures of overdensity  with the simulation
predictions -- as a function of redshift.  As time progresses,
the fraction of volume with high overdensity will increase and the maximum overdensity 
should increase at lower redshift. This measure of structure evolution enables significant
comparison between the simulations and the observed universe, avoiding
the azimuthal averaging which is inherent is a correlation function 
analysis. The structures are expected to filamentary and therefore are not circularly symmetric; they 
may also have multiple characteristic scales. For the same reasons, we have employed the adaptive smoothing 
technique developed here rather than matched filter  algorithms \citep{pos96, sch98} which are 
obviously well-adapted to the central, high density core structures but less appropriate to extended filamentary 
structures. Angular correlation functions for the COSMOS field are 
presented in \cite{mcc06}.

Figure \ref{dens_avedens} shows 
the fractional cumulative area with galaxy surface density greater than $\Sigma$/$<\Sigma>$ where $<\Sigma>$ is the average in
each redshift slice. The three colored curves show overdensity filling factors for the
redshift ranges 0.2 to 0.5, 0.5 to 1.1 and 0.2 to 1.1. These curves were computed from the adaptively 
smoothed overdensities shown in Figure \ref{smooth} divided by the mean background surface density ($\Sigma_B$, given 
in the top legend for each redshift slice). The level of the background is dependent on the density of true 'field' galaxies
and on the redshift accuracy -- thus in comparing with simulation predictions below, we also convolve the redshifts
of galaxies in the simulation with a Gaussian of z-width matched to that of the observational photometric redshifts.
Figure \ref{dens_avedens} exhibits the basic characteristic expected for structure growth as a function of 
redshift -- higher overdensities occuring at later times (lower z) and a larger fraction of the area in overdense 
regions as time progresses. 

A quantitative comparison can be made with the Millennium Simulation. 
Mock catalogs were constructed using the Virgo Consortium's Millennium 
Simulation and the Galform semi-analytic model of galaxy formation. Dark 
matter and merger trees were extracted from the Millennium Simulation using the 
techniques of \cite{hel03} , utilizing all halos of $\geq 20$ 
particles. These merger trees are fed through the Galform 
semi-analytic model (using the parameter set of \cite{bow06}  to populate the simulation with galaxies at all redshifts.
We did not have access to proprietary lightcone data from the Millennium Simulation,
so the 
mock catalogs were constructed taking cubes from the Millennium 
Simulation at z=0.3, 0.5, 0.7, 0.9 and 1.1. Regions with 
two sides equivalent to 1.4$\deg$ and one side extending 500 Mpc/h were extracted at each redshift.  
Galaxies were selected to have M$_V  < -18$ mag and 19 $<$ i $< $25. 
The mock based on the Millennium-Virgo semi-analytic (black curve in Figure \ref{dens_avedens})
is in remarkably good agreement with the mean curve determined from the 
observations for z = 0.2 to 1.1 (solid green line). Jackknife tests were done, 
splitting the data in half and the variances are typically $< 20$\% for most 
values of the overdensity -- this provides a limited estimate of the uncertainties. The overall area filling 
factors in the observations and theory track each other within a factor of $\sim 2$.
It does appear that the theoretical curve does not reach as high overdensities
as the observational curve -- possibly indicating a significant discrepancy on small 
scales. Since what is being measured in both the observed and theoretical distributions 
is the number counts of galaxies, not the mass distributions, the discrepancy might indicate 
that the simulations have too much merging in the denser regions.

\section{Correlation of Galaxy SED and Luminosity with Structure Location}\label{lss_evol}

A number of recent investigations have found early-type galaxies more strongly clustered 
than the later types \citep{lef05,men06,coi06,cop06} at z = 0.5 to 2.
Variation of galaxy SEDs as a function of both redshift and structure location is dramatically shown in Figure \ref{mass_types}. 
Here we plot the mean SED of galaxies as a function of z (with no 
selection for structures or the field) and the mean for galaxies 
within R $\leq 1$ Mpc and R $= 1$ -- 5 Mpc from the center of each structure. 
This is on a scale with six types ranging from 1 for an E/S0 galaxy SED to 6 for a starburst galaxy \citep{mob06}. Figure \ref{mass_types} 
demonstrates dramatically, and in every case that the interior of the structures are populated 
with galaxies having a mean SED type lower by $\sim$0.5 -- 1   
compared to the average SED type at the same redshift. \cite{buc84} 
first showed the trend for an increasing fraction of blue galaxies within clusters out to 
z = 0.5 and the trend for earlier  morphological types in the highest density 
regions is well known as the T - $\Sigma$ relation  at z $< 0.5 ~$\citep{dre97}.
The sample shown here for the COSMOS survey is the most extensive and covers a large range of 
redshift ($0.1 < z < 1.1$) using the same technique.  An analysis of galaxy morphology and environmental density in the COSMOS field is presented by \cite{cap06a}. Their results are consistent with those 
shown here.  
Figure \ref{mass_types} also shows a systematic gradient in the mean SED for the field galaxies -- about +0.5 to later types from z =0.2  to 1.

Figure \ref{samp} shows the fields surrounding a sample of six of the LSS (\#1, 2, 8, 10, 25 and 26)
with the galaxies shown in color depending on their SED type determined in the 
photometric redshift fit. (Galaxies within the $\Delta z$ range given in Table \ref{struct_param} are plotted for each structure.) These figures show the enhancements in galaxy density associated
with the LSS; they also indicate the level of background contamination which any
identification procedure must deal with. However, the most interesting feature easily seen 
in Figure \ref{samp} is the preference of the early 
SED-type galaxies for the denser LSS. It is important to recall that the sample selection used to 
identify the structures was involved all galaxy types, not just red galaxies.

\section{Evolution of  Galaxy Properties with Environmental Density and Redshift}\label{envir_evol}

The dependence of galaxy properties on redshift and environment is one of the central 
themes of current cosmological evolution studies \citep[e.g.]{lef05,ger05,men06,cop06,coi06}. Here we use the overdensities derived as a function 
of the pseudo-3d space ($\alpha,\delta$ and z -- Section \ref{adapt_smo} and Figures \ref{smooth})
and galaxy properties (SED type, mass, luminosity and star formation rate -- SFR) derived 
from broadband photometry to investigate the environmental influences. Use of the density 
cube precludes the need to identify and delineate specific LSS (Section \ref{lss_evol}). 

Our environmental 
densities were derived from the surface density of {\bf all} galaxies (above specified mass or luminosity cuts) --
the densities were not derived from clusters of color selected galaxies -- thus the analysis 
below is presumably unbiased and without {\it a priori} correlations of environment  and galaxy properties. 

\subsection{Environmental Density}\label{env}

To characterize the local environment of each galaxy, we use a 'relative density'  measure, $\rho_{rel}$
defined as 
\begin{equation}\label{eqrho}
\rho(\alpha,\delta,z)_{rel} = \frac{ \Sigma(\alpha,\delta,z)} {\overline{\Sigma}(z)}
\end{equation}

\noindent where $\Sigma(\alpha,\delta,z)$ is the overdensity for each redshift slice as 
shown in Figures \ref{smooth} and $\overline{\Sigma}(z)$ is the mean of this
overdensity at each redshift. The mean value of the overdensity is used for normalization
to enable comparison of widely separated redshift slices with $\Delta z = 0.1$ which 
have somewhat different surface densities and overdensities of galaxies due to varying line-of-sight depths 
and comoving volumes.  (The relative densities $\rho$ 
may be translated back to $\Sigma$ (Mpc$^{-2}$ per 0.1 in z) using $\overline{\Sigma}(z)$ = 
1.2, 0.52, 0.29, 0.17, 0.14, 0.19, 0.18, 0.15, 0.13 and 0.06 for z = 0.15 to 1.05, sampled every 0.1 in z.) 
For each galaxy (Section \ref{sample}), the 
environmental density was obtained from the pseudo 3-d cube using its $\alpha$, $\delta$ and best fit photometric redshift. 

\subsection{Sample Selection and Completeness}

\begin{figure}[ht]
\epsscale{1.} 
\plottwo{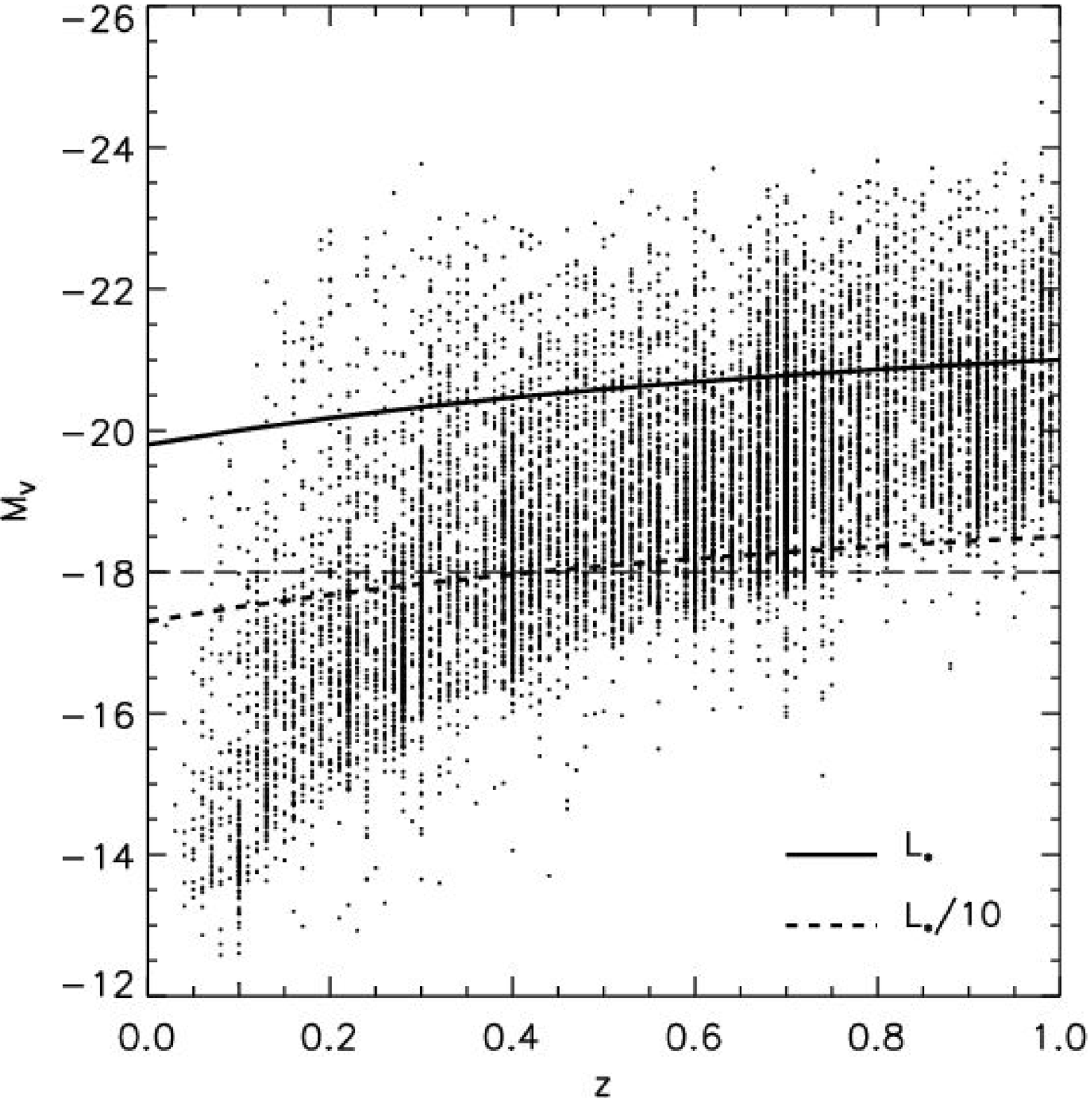}{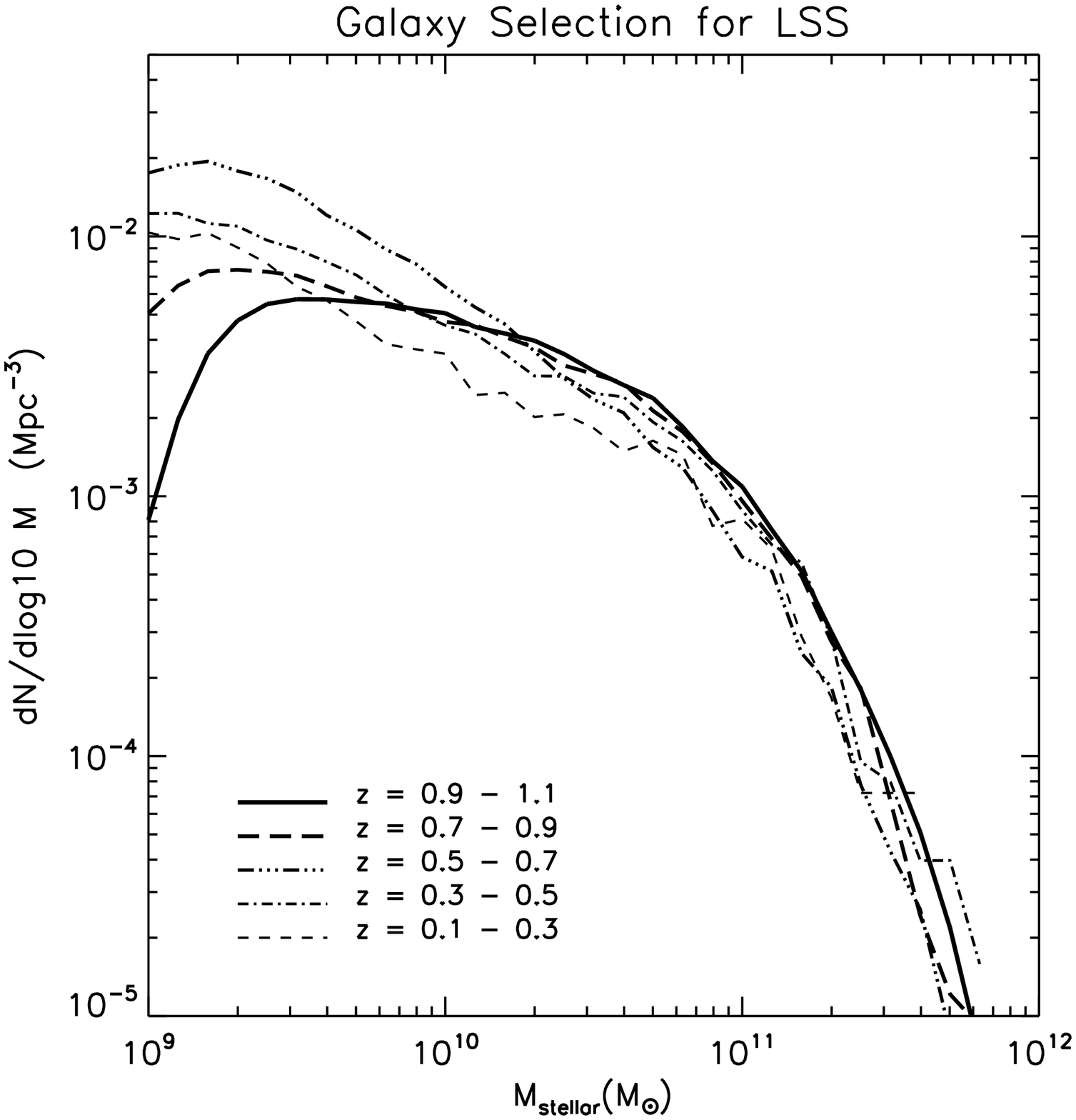}

\caption{a) Left : The distribution of rest frame M$_V$ for galaxies in our sample is shown as a 
function of redshift. The lines indicate the expected absolute magnitudes 
for L$_*$ and L$_*/10$ galaxies assuming passive evolution brightening of 1.2 mag from 
z = 0 to 1.  The horizontal line at M$_V = -18$ corresponds to the absolute magnitude cutoff used here. 
The lower envelope in the grayscale is imposed by our sample cutoff at
$I_{AB} < 25$ mag. (For clarity only 20,000 randomly sampled galaxies are plotted.) b) Right : The distribution functions of stellar masses for galaxies in our sample with $I_{AB} < 25$ mag and M$_V < -18$ are shown for redshift bins of width $\Delta z =0.2$. The higher noise seen in 
the low z mass functions is due to the much smaller volume and hence smaller number of galaxies sampled; the small falloff seen in the mass functions for z $> 0.7$ at M$_{stellar} < 5 \times 10^9$\msun 
~is due to the apparent magnitude limit $I_{AB} < 25$ mag for our sample. } 
\label{selection}
\end{figure}

It is of course vital that 
the sample selection function (see Sections \ref{sample} \& \ref{select}) not introduce biases as a function of redshift which 
masquerade as changes in the galaxy properties. We make use of two alternative samples with  
: \#1) a mass cutoff of $M_{*} > 3\times10^9$ \msun ~(82,274 galaxies)  ~and alternatively, 
\#2) a luminosity cutoff with $M_{V} < -19$ mag (101,018 galaxies).
Figures \ref{selection} show the observed distributions of $M_{V}$ and $M_{*}$ 
as a function of z. As discussed in Section \ref{select}, there is  little
change in the mass function at $z < 1$ and therefore most of the variations in the mass function 
at M $< 5\times10^9$ \msun~ are likely the result of incompleteness at $z > 0.7$ (see Figure \ref{selection}, right panel).
\cite{bor06} found a possible doubling of the integrated mass function of galaxies from 
z = 1 down to 0.2 (in the COMBO-17 survey). We take this as an upper limit since the sample used here shows no significant
variation aside from the aforementioned  incompleteness (see Figure \ref{selection}-right).
Similarly, the selection on $M_{V}$ is chosen to be close to the limit at which completeness starts to become 
an issue (see Figure \ref{selection}-left). 

We develop the two samples in parallel since one cannot assume 
{\it a priori} that the galactic masses and/or luminosities are invariant from z = 1 to 0. For example, one expects 
the luminosity of each galaxy to vary at z $< $1 (even in the absence of further star formation or merging)
due to dimming as the stellar population ages. For this reason,  adoption of a fixed M$_V$ cut  would 
yield a sample with larger surface density at z =1 than at z =0.2. The fixed mass-cut sample most likely comes closest to
generating equivalent galaxy samples at z $< 1.1$; however, at higher redshifts, it is likely the masses will 
be changing more rapidly. A later paper will explore various evolution scenarios for the 
luminosity-slected sample. 

More conservative higher mass and luminosity cutoffs would of course yield greater completeness at 
high z. On the other hand, since the various basic galaxy types have quite different masses and luminosities,
this would compromise one's ability to probe evolution between types. Specifically, a very high mass
cutoff (e.g. M  $>$few$\times10^{10}$ \msun) would largely limit the samples to just the most massive E's and spirals
and under-represent the lower mass, late type systems which have significant star formation activity.
This would severely compromise the dynamic range that could be 
investigated {\it vis-a-vis} the transformation from late to early type galaxies.

The distribution of $\rho$ for the sample of galaxies was also examined to select a 
lower cutoff in density for the analysis.  This was required since very low overdensities, compared to the mean background, are not
quantitatively meaningful -- in areas where the adaptive smoothing detects no significant overdensity exceeding 
3$\sigma$ (see Appendix \ref{appendix}), it smoothes the surface density down to a value detemined by the
largest spatial-smoothing width.  The adopted density cutoff reduced the final samples to 10,382 and 12,523 galaxies
for samples \#1 and 2, respectively. (The lowest overdensity to which one may carry this analysis is determined 
by the background counts of galaxies at each redshift. This is, in turn, largely a function of the photometric redshift accuracy. 
Higher accuracy photometric redshifts will enable extension of this investigation to lower density and the field.)

\subsection{Galaxy Properties : Mass (M$_*$), SED type, Early-Type Fraction, M$_V$, SFR and $\tau_{SF}$}

Galaxy  SED types and rest-frame luminosities (M$_V$) are by-products of  the 
photometric redshift fitting. Their masses were derived using the intrinsic SED to estimate the 
mass-to-light ratio together with the absolute V magnitude obtained from the observed fluxes
\citep{mob06}. The SED types range from 1 to 6 with : 1 = E,
2 = Sa/Sb, 3 = Sc, 4 = Im, 5,6 = two starburst populations \citep[defined by][]{kin96}.
The early-type galaxy fraction was calculated, taking all with SED type $< 1.9$ to 
be 'early-type'. For each galaxy, the SFR  was estimated  from the intrinsic
SED and observed fluxes, extrapolated into the UV. (As with the mass estimates, the SFRs have been 
aperture-corrected using the auto-magnitude parameter from SEXTRACTOR.) 
We use the SFR estimated from the extinction-corrected, 
rest-frame $\lambda = $1500\AA  ~continuum \citep{mob06a}. 

We also calculate the ratio of the
galaxy mass to the SFR, yielding a characteristic timescale to form the existing 
galactic mass of stars at the currently observed SFR (specifically, $\tau_{SF} = M_* / SFR$). 
For a starburst $\tau_{SF}$ will be significantly less than the Hubble time at the observed 
redshift whereas a galaxy for which the current star formation is relatively low, compared to 
that in the past, will have a long $\tau_{SF}$. $\tau_{SF}$ is equal to the inverse of 
the specific SFR per unit stellar mass of the 
galaxy, sometimes called the  'star formation efficiency' (SFE).

The galaxy samples were binned using 4 equal z bins of width $\Delta z = 0.23$ between z = 0.2 and 1.1
and 4 logarithmically spaced  bins in density $\rho$ from 8 to 215. For the adopted cosmology, the redshift bins are 
centered at lookback times of $\sim$ 3.5, 5.3, 6.6, and 7.7 Gyr.  Within each bin, the median values
of each galactic property  were determined. The median was used rather than the mean since it is less susceptible
to a few extreme values and hence the uncertainty in the median estimates can be 
small even for samples with a large intrinsic dispersion. To estimate the uncertainties 
in the median values, Monte Carlo simulations were done on the observed distributions,  
adding randomly sampled uncertainties from a normal distribution.  We adopted
uncertainties (1$\sigma$) in each of the bolometric quantities (M$_*$, M$_V$ and SFR) of a factor of 2 
from their nominal values for
each galaxy; for the  SED type, we assume an uncertainty of  $\pm1$ for the type. (The factor 
of 2 uncertainty is an approximation  to allow for uncertainties in 
photometric calibrations and the SED fitting.)
The median  was measured for each of 500 simulations, and the dispersion of the median distribution
was taken as the uncertainty in the median for the observed sample.  

\subsection{Galaxy Evolution with Redshift and Density}

In Figures \ref{mass_dens} to \ref{tau_dens}, the median stellar mass M$_*$, M$_V$, SED,  SFR and $\tau_{SF}$ are plotted 
for each redshift range as a function of density. The SEDs, SFR and $\tau_{SF}$ all exhibit very significant
variation as function of both redshift and density. The mass and luminosity distributions are partially affected 
by selection bias  but only in the highest z bin.

The median mass and M$_V$ distributions (Figures \ref{mass_dens} and \ref{mv_dens}) and comparison 
of  the mass- and luminosity- limited samples (left and right panels) may be used to assess 
the possible influence of incompleteness at the highest redshifts. The median masses show no systematic 
increase with z except in the z = 0.88 to 1.1 bin for which the masses appear systematically 
higher by a factor of 1.5 to 2 compared to lower z. This increase is very likely due to sample incompleteness 
or Malmquist bias since the sample is deficient in the galaxies with M$_*$ $<  5\times10^9$ \msun ~(see Figure \ref{selection}, right panel). 
$M_V$ exhibits a somewhat larger ($\Delta M_V \simeq 1$ mag) increase -- some of this is probably also 
due to incompleteness ~(see Figure \ref{selection}, left panel), but since it is larger than the mass shift, some of 
the M$_V$ variation is likely due to actual evolution of M$_V$ in the galaxies. Passive evolution of the stellar populations
from z = 1 to 0 is  $\Delta M_V \sim 1.2$ mag  \citep[e.g.][]{dah05}. To summarize, modest variations in the mass and luminosity 
medians between z = 0.8 and 1 are probably due to 
incompleteness; at lower redshifts, no such variations are seen and the samples are probably
complete to better than $\sim20\%$. 

The median masses clearly grow with increasing density, at each redshift. This cannot be a 
sample selection bias since that should be constant at each redshift. The doubling of the median mass 
in high density environments compared to lower densities is seen at all redshifts out to z = 1; this is undoubtedly 
reflecting the fact that the early type SEDs also are more prevalent in the denser environments (see below) and these 
are often massive galaxies.  Consistent with this interpretation is the fact that although M$_*$ (Figure \ref{mass_dens}) exhibits dependence on $\rho$, M$_V$ 
 does
not (Figure \ref{mv_dens}), implying that the median mass-to-light ratio is lower at high density. 

\subsubsection{Galaxy Spectral Type \& Early-Type Fraction}

The galaxy SED types shown in Figure \ref{sed_dens} exhibit very significant variations with both redshift and 
density --  in the sense that earlier type SEDs (E's) are seen at higher density, later types
in the lower density regions. And for all densities, the median galaxy type is later (i.e. bluer, star forming)
at higher redshifts. The major variation with z occurs between the  lowest two redshifts (from z $\sim 0.3 $ to 0.5 or lookback times 
less than 5.3 Gyr) -- all three high z bins have similar SEDs and their variations with density are the same. Numerous studies 
have noted the strong increase in the fraction of early type  galaxies (classified by both SEDs and morphologies) 
in dense enviroments out to z $> 1$ \citep[e.g.][]{dav76,pos84,dre97,kod04,kau04}. 
The results shown in Figure \ref{sed_dens} show very clearly that 
similar density correlations are seen over the entire redshift range. The mean SED shifts to earlier type  at lower z for both low and high
density environments.  The actual surface density for the breakpoint between the late and early 
SEDs shifts does not appear to shift more than a factor of 2 since the break occurs at approximately 
the same relative density $\rho$ and the density normalization from z $\sim 0.3$ to the higher z's changes by less than a factor of
2 (see Section \ref{env}). 

The percentage of galaxies with SED type $< 1.9$ is shown in Figure \ref{red_dens}, exhibiting variations 
like those in the median SED type (as it should, since they are closely related). We include the early type fraction since it is often used 
to characterize the galaxy populations in evolutionary studies. As with the SEDs, the major shift with z occurs beween 
the lowest two redshift bins, and at all redshifts an increased early-type fraction in seen above $\rho \sim 50$. 
Figure \ref{red_dens} clearly demonstrates that the galaxy-type correlation with density was clearly in place before 
z = 1 and we have extended this correlation to low densities, as well as the dense clusters. 

\subsubsection{Star Formation Rates and Timescales}

The median SFRs per galaxy (Figure \ref{sfr_dens}) rise systematically with redshift for all densities. The 
overall increase by a factor of 4.5 from z = 0.3 to 1 is similar to that found in many 
earlier studies \citep[e.g. see][]{mad96,hop04,jun05,bun06,bel05,sch05}. The observed increase at higher redshift
 is extremely well fit by a linear dependence on lookback time over this 
range, $\tau_{lookback} = 3.5$ to 7.7 Gyr. Figure \ref{sfr_dens} also shows evidence of a slight decrease in the 
median SFR at higher densities, with this decrease being steepest at low redshift (z$\sim0.2$). The 
steep decline in the SFR to lower redshift is possibly due to the depletion of ISM to fuel star formation
and AGN/SF feedback processes. Discriminating between these may be accomplished 
with future observations of the star forming gas content with the ALMA array.

Normalizing the SFRs  by the stellar mass of each galaxy, the SF timescale ($\tau_{SF}$, Figure \ref{tau_dens}) 
shows much stronger density correlation than the SFR. At all densities, the SF timescale 
is a factor of 2 -- 3 shorter in all three high redshift bins compared with z = 0.2 to 0.43.
And a factor of 4 -- 5 increase in the SF timescale occurs between the 
low and high density environments at all redshifts with the strongest density dependence occurring at the lowest redshift.  
These results imply that most of the stellar mass in dense environments must have formed much earlier
than  z $ = 1$ whereas a significant amount ($\sim$25\%) of the stellar mass in the low density environments 
must have formed at z = 1.1 to 0.4 (based on the measured SF timescales). 

\subsubsection{Downsizing of Star Forming Galaxies -- the Maturity Parameter ($\mu$) }\label{mature}

A number of investigations have suggested that star formation occurs earlier in the most massive galaxies
and as the universe ages the star formation progresses to less and less massive systems, a phenomenon 
often referred to as  'downsizing'  \citep{cow96,kod04,bun06}. However, this phenomenon can be  blurred and  
sometimes confused with the earlier formation times for galaxies in dense clusters  coupled with 
 the high abundance of  massive galaxies in clusters. Here, we attempt to separate these 
affects to investigate the relative formation times of high versus moderate mass galaxies as a function of 
both redshift and density. 

For this discussion, we define a parameter which we will call the Maturity ($\mu$), equal to the ratio of the star formation timescale ($\tau_{SF}$ used above) 
to the cosmic time ($\tau_{cosmic}$ = the age of the universe at each galaxy's redshift). With this definition, the Maturity is
unity if  the observed stellar mass could have formed at the observed star formation rate
within the age of the universe (at the redshift of the galaxy). The Maturity will be $< 1$ ({\it youth}) if it is forming stars at a sufficiently 
high rate that its mass could be produced in less than the cosmic time; the Maturity will be $> 1$ ({\it middle to old age}) if its current SF rate is low
and most of its stars must have been formed earlier (with $\mu$ $<$1, youth) at a star formation rate much higher than that presently measured. Obviously,
initial starburst systems would have $\mu$ $<< 1$ and old elliptical galaxies $\mu$ $>> 1$. The Maturity, defined in this manner, will continue to increase 
at later cosmic epochs if the star formation rate remains low. On the other hand,  if the aging galaxy undergoes a late-life starburst ({\it mid-life crisis}), it will be rejuvenated 
($\mu$ $\Downarrow$). But, if the the starburst is brief and not substantial, the galaxy  will return more or less to its prior state
of Maturity after the starburst. (As with humans, rejuvenation may be superficial and illusory ! Our use of medians for charting the 
overall evolution of the galaxy populations probes the typical Maturity, thus avoiding the 'noise' due to short starbursts.
[Anthropomorphizing this galactic parameter can actually help to visualize and 
track the galactic changes associated with evolution of $\mu$.] (This Maturity parameter 
is similar but not identical to the 'Birthrate' quantity discussed by \cite{bel05}, but 'maturity' 
more aptly connotes what this parameter characterizes. )

In Figure \ref{tau_dens_color} the Maturity is shown as a function of both redshift and density, separately for galaxies of 
high and low mass. The two samples were separated at M$_* = 5\times10^{10}$ \msun. \cite{kau04} and \cite{kod04}
found a 'break' between old, red galaxies and younger, blue galaxies at a  mass of $\sim3\times10^{10}$\msun ~at z = 0 and 1, respectively. 
We have adopted a somewhat higher value in order to minimize incompleteness in the low mass sample at z $\gtrsim 1$.
\cite{bun06} argue that the 
break mass varies with redshift, rising to $\sim 10^{11}$ \msun ~at z =1; however, we do not see 
such clear variation (see below).  In any case, we have found by experimenting with the mass cut 
that a factor of 2 variation in the vale mass cut (from $5\times10^{10}$ \msun)
did not change the behaviors discussed below. 

Figure \ref{tau_dens_color} shows that at all redshifts 
and densities probed here, the more massive galaxies are always more mature than the lower mass galaxies. 
At each redshift and environmental density, the lower mass 
galaxies are systematically 5 -- 10 times less 'mature' than the massive galaxies. Once again, we emphasize that 
these are not color-differentiated galaxy samples -- just mass-differentiated for which there is 
no {\it a priori} association with 'age'. Although, if the more massive galaxies tend to be more mature, 
obviously, they will appear redder. 

For both the high and low mass galaxies, 
the median Maturity is either constant or increases with time, i.e.  to lower redshift (Figure \ref{tau_dens_color}); it never decreases
with time -- as well it could if the star formation in either environment was delayed to commence at 
a late epoch (e.g. some dwarf galaxies). 
A constant Maturity implies a steady SFR over cosmic time, whereas an increasing maturity suggests diminishing 
SF with time. The more massive galaxies clearly must have had an early phase of rapid 
star formation at z $< 2$ \citep[c.f.][]{jun05} with relatively little star formation at z $< 1.1$ in order to
appear so mature ($\mu$ $\simeq 1$ to 2 at z $\sim 1$, Figure \ref{tau_dens_color}). By contrast, the lower mass
galaxies exhibit fairly constant immaturity down to z $\sim 0.43$, implying that on-going, fairly constant 
star formation has occurred from z = 1.1 to 0.43 and very likely also at the high z.  However, at the z $< 0.43$ the maturity of the lower mass 
galaxies rises in all environments, implying a significantly decreased star formation rate for 
lookback times $< 5$ Gyr. 

Figure \ref{tau_dens_color} suggests  that galaxies of {\bf all} masses (at z $\leq 1.1$) are
more mature in the dense environments, not just the high mass galaxies! The lowest redshift bins both 
show  $\tau_{sf}/\tau_{cosmic}$ rising quite 
significantly at the highest density while at the other redshifts, a factor $\sim 2$ increase 
in the maturity occurs between the 
lowest to highest densities. This clearly requires that the epoch of rapid star formation for galaxies of {\bf both} high and
lower mass must be earlier in the denser environments than in the field. 

In  Figure \ref{mtrans} the Maturity is shown as a function 
of galactic mass for high and low density environments, with separate plots for each redshift. 
The cut between high and low density was taken at $\rho = 45$, i.e. between the middle two bins of the 4 density bins
used earlier. In Figure \ref{mtrans}, the overall distribution of galaxies for all densities is shown by the colored shading while
the high and low density environments are shown in the red and blue contours, respectively. 
The separation of the old ($\mu$$ > 1$) and young ( $\mu$$ < 1$) galaxies is seen as a bimodal distribution
and their loci change systematically with redshift. Evidence of evolution in the 
mass separating starforming and non-starforming has been claimed by \cite{bun06} and \cite{bor06}. 
\cite{bun06} found  $M_{break} \propto (1+z)^4$; however, we find it difficult to identify a distinct mass which can be said to 
 divide the 
 mature and immature populations, since at most masses between $10^{10}$ and $10^{11}$ \msun~,  $\mu$ can range from $< 1$
  to $> 10$ (see Figure \ref{mtrans}).  Most of the mature galaxies occur above few$\times10^{10}$ \msun, but there is no sharp cutoff 
at most redshifts. In fact, for the lowest redshift bin,  two distinct, mature sequences can be seen 
at $\mu$$\sim 5$ and 20 -- 30. The latter could correspond to the maturation of the $\mu$$\sim 6$ to 8 sequence seen 
at higher z; the former might correspond to maturation of the $\mu$$< 1 $ galaxies seen at earlier epochs. 
Possibly,  the combination of  these two mature sequences at low redshifts 
account for the apparent evolution of the break mass as discussed by \cite{bun06}. Future work 
is planned using the COSMOS GALEX UV measurement to verify this second 
mature sequence. 

Galactic down-sizing with the most massive galaxies forming earliest is of course at variance with the expectation 
of the most simple hierarchical galaxy formation scenarios. However, \cite{kai84} and later \cite{cen93} suggested a model of biased galaxy formation 
with the most massive galaxies forming within the highest peaks of the initial density field, and this is a commonly accepted
explanation. In the highest peaks, there is more mass available for buildup of the most massive galaxies and 
the rate of growth is higher where the density of sub-halos is higher \citep[e.g.][]{del04}. The results shown here suggest 
that even at relatively low environmental densities, the more massive galaxies are formed earlier than
the low mass galaxies -- although not as early as the massive galaxies in very dense environments. 
This suggests that the formation of  massive galaxies occurs by two processes -- one local,  
responsible for the early growth of massive galaxies in low density regions,  the other associated with high overdensity regions 
where the growth occurs more rapidly and in some cases is carried to the very highest galactic masses.

\section{Summary and Conclusions}

The COSMOS photometric redshifts now have sufficient accuracy ($\sigma _z / (1+z) \simeq 0.03$) to enable identification of LSS at z = $ 0.1 $ to 1.1. We have developed an adaptive 
smoothing procedure to be applied to the galaxy density distributions in photometric redshift 
slices with $\Delta z = 0.1$ to identify LSS on scales less than 1 Mpc up to 30 Mpc with optimal 
signal-to-noise ratio across the range of spatial scales. This procedure has been tested with excellent results on mock redshift slices
and on the dark matter particle distribution from a $\Lambda$CDM simulation (see Appendices \ref{appendixa} and \ref{appendixb}).

The adaptive smoothing is applied to the COSMOS photometric redshift catalog with 
selection z $< 1.1$, 19 $< $I$_{AB} < $25 mag and M$_V < -18$ mag --  a sample of 150,000 galaxies. 
No color or SED selection is imposed, so that the defined structures are intrinsically 
unbiased with respect to galaxy type. From the galaxy over-densities derived from 
the adaptive smoothing, we have delineated 42 LSS and galaxy clusters in the 
pseudo 3-d space ($\alpha,\delta$,z). The surface density plots of the structures are 
shown in Figures \ref{lss} and \ref{lss_2d};  their measured properties are given in Table \ref{struct_param}. Five of the most massive structures have stellar masses (determined 
from the galaxy photometry) of M$_* > 10^{13}$ \msun. Several have extents which can be traced over 10 Mpc (comoving). Their total masses including dark matter are likely 
to be 50 -- 100 times greater. The Richness of the core regions of these structures is 
typical of Abell class 1 - 3. The derived mass function for the LSS is consistent 
with the {\it total} mass function for clusters derived by \citep[][]{bah93,rei02}
from optical and X-ray studies. 

The clusters at the center of the most massive LSS (\#1) are discussed in 
detail by \cite{guz06} and \cite{cas06}. The compact structures with diffuse X-ray emission, many of 
which are located within the LSS discussed here, are discussed by \cite{fin06}. These clusters 
are identified optically by wavelet analysis of the early type galaxies in the 
COSMOS photometric redshift catalog and from the diffuse X-ray emission.
We have compared the fractional areas seen at different overdensities and find 
general agreement to within $\sim 50$\%  with the predictions of $\Lambda$CDM simulations (processed similarly) -- 
with less than 1\% of the areas of the redshift slices having overdensities exceeding 10:1. However, the observed 
filling factor distribution does reach higher overdensity and this may indicate that the 
simulations have too high an efficiency for merging in dense regions., 

We have investigated the dependence of galaxy evolution on environment using the 
structures defined here and the SED types taken from the photometric redshift fitting. 
We find that in {\bf every} structure the mean galaxy SED type within the high density core
of the structures is earlier than the mean SED type at the same redshift. 
Our study thus confirms, with a sample of 42 structures/clusters, the correlation 
of galaxy evolution with environmental location \citep[e.g.][and references cited therein]{dre97,smi05,pos05,cop06}) over the full range
of redshift z = 0.1 to 1. \cite{cap06a} find a similar result using an entirely independent
measure of environmental density and using galaxy morphology instead of SED type. 

Extensive analysis was done to analyze the correlations of galaxy properties (SED, mass, luminosity and SFR)
with redshift and enviroment. The median  SED type and star formation activity 
varies strongly with both redshift and environmental density. The maturity of the stellar populations and the 'downsizing' 
of SF in galaxies are both strongly varying with epoch and environment. Although the more massive galaxies clearly tend to have
lower SFR per  unit galactic mass, we question whether it is possible to define a distinct 'break mass' separating  active and inactive star-forming galaxy populations.
And over the range z $< 1.1$, we don't see strong  evidence of evolution in the masses of galaxies undergoing  active star-formation (at the level of $< $ a factor 2).

 
 \acknowledgments
 
 The HST COSMOS Treasury program was supported through NASA grant
HST-GO-09822. 
 We gratefully acknowledge the contributions of the entire COSMOS collaboration
 consisting of more than 70 scientists. 
 More information on the COSMOS survey is available \\ at
  {\bf \url{http://www.astro.caltech.edu/$\sim$cosmos}}. 
 The COSMOS Science meeting in May 2005 in Kyoto, Japan was supported in part by 
 the NSF through grant OISE-0456439. Major work on this project was done
 while NZS was on sabbatical at the Institute for Astronomy at the University of Hawaii 
 and during a visit at the Aspen Center For Physics. We would also like to thank
 the referee for a number of suggestions which have greatly improved this 
 paper. 
 
 
 
 {\it Facilities:} \facility{HST (ACS)}, \facility{HST (NICMOS)}, \facility{HST (WFPC2), \facility{Subaru (SCAM)}}.
 
 
  \appendix
 
 \section{LSS Identification with Adaptive Smoothing}\label{appendix}
 
In the past, a number of algorithms or techniques have been used for automated 
identification and characterization of galaxy clustering, including
percolation and Voronoi tesselation techniques \citep{wey94,ebe93,mar02,ger05}, wavelet analysis \citep{esc95,fin06} and matched filter  \citep{pos96, sch98}.
An algorithm for the identification of structures must be capable of detecting structures on
multiple angular scales, and with only low order assumptions regarding the 
internal density profile of the structures. Techniques which search for 
a particular scale or assume, {\it a priori}, a density profile (or equivalently a 
spatial weighting function) will have highest sensitivity for structures with the
specified parameters, thereby biasing a derived distribution function for the recovered 
structures. It is also highly desirable that the algorithm be capable of 
displaying compact structures simultaneously with more extended, low density 
structures. Presumably, within large structures there will be high density 
substructures which one would not want to smooth out into low spatial frequencies. Conversely, if a high density structure is fully detected at
high spatial frequencies, one would not want its power to be carried out to 
low spatial frequencies as an extended halo. Multi-scale algorithms like wavelet and adaptive smoothing seem therefore 
most appropriate. 

For the structure identification, we have developed an adaptive kernel 
smoothing algorithm, specifically tailored  to have these characteristics.

\subsection{Algorithm}

The algorithm consists of a loop, starting at low smoothing width, going to 
successively larger smoothing kernels, removing power from the 
current 2-d residual 'image' if it exceeds a specified signal-to-noise ratio at the 
current level of smoothing. The 'image' being processed is the projected 
surface-density ($\Sigma$) of galaxies
in a redshift slice. Starting at the initial highest resolution (n=1), we  calculate the smoothed surface-density ($\Sigma_n$) and background surface-density (B) from 
\begin{mathletters}
\begin{equation}\label{eq1}
\Sigma_{n} = \Sigma^{\prime}_{n-1} ~\star ~ K_{n} 
\end{equation}
\begin{equation}\label{eq2}
B_{n} = (\Sigma^{\prime}_{n-1} -\Delta_{n} )~\star ~ K_{2n}
\end{equation}
\end{mathletters}
\noindent where '$\star$' is the convolution operator, K$_n$ is a 2-d smoothing kernel of width n, normalized such that 
its integral is unity and the Kernel K$_{2n}$ used to convolve the background has twice the width, ie. 2n. The
'power available' at resolution n is  calculated as  
\begin{equation}\label{eq3}
\Delta_{n} = \Sigma_{n} - B_{n} 
\end{equation}
Since Eq. \ref{eq2} depends on \ref{eq3}, these equations are iterated (typically 4 times)
to arrive at the 'best' estimates of the background (without the high frequency power included)
and the $\Delta_n$ with the most low frequency, background removed.

If $\sigma_{n}$ is the noise 'image' at resolution n, then 
the signal-to-noise ratio, S($\Delta$)$_n$,  on the delta residual-density is then 
\begin{equation}
S(\Delta)_{n} = \frac{ \Delta_{n} } {\sigma_{n}}
\end{equation}
and the signal-to-noise ratio, S($\Sigma$)$_n$,  on the original surface-density, smoothed to 
resolution n, is
\begin{equation}
S(\Sigma)_{n} = \frac{ \Sigma_{0} ~\star ~ K_{n}} {\sigma_{n}}
\end{equation}

The power to be removed at resolution n is then given by
\begin{equation}
P_{n} = \left\{H\left( S(\Delta)_{n} - snr_{\delta} \right)~\cdot~H\left( S(\Sigma)_{n} - snr_{total} \right)\right\}\cdot \Delta_{n} \\
\end{equation}
where H(x) is the Heavyside function (H=0 for x $\leq$ 0, H=1 for x $>$ 0).
'snr$_{\delta}$' is an adjustable parameter specifying the minimum signal-to-noise 
ratio in the residual image required before power is removed at width 'n'. Similarly 
snr$_{total}$ is a parameter specifying the minimum signal-to-noise ratio 
required when the original total-power surface-density is smoothed to resolution
'n'.  Having these two conditions is crucial to the excellent results obtained 
with this procedure -- allowing small values of
snr$_{\delta}$ to be used while avoiding the retention of 'noise' peaks. For any pixels
which do not satisfy this double criteria for signal-to-noise ratio, the residual power is retained 
to the next level of smoothing.  The residual image (with lower spatial frequency power) 
to be used as input on the next iteration at larger smoothing kernel (n+1) is therefore 
given by
\begin{equation}
\Sigma^{\prime}_{n} = \Sigma^{\prime}_{n-1} - P_{n} 
\end{equation}
Steps A1 to A4 are repeated with successively larger values of 'n' up to n$_{max}$. 

After reaching n$_{max}$, the adaptive smoothed surface-density ($\Sigma_{final}$) is then given by
\begin{equation}
\Sigma_{final} = \Sigma^{\prime}_{n} +  \sum_{n} P_{n} 
\end{equation}

The procedure described above has the following desirable features :

\noindent 1) It is conservative, i.e. the 2-d integral of the original and final surface-densities are equal.

\noindent 2) Power is retained at the highest spatial frequencies and not smoothed out to lower 
frequency as long as its signal-to-noise ratio is sufficient (i.e. greater than the specified  snr$_{\delta}$).

\noindent 3) High frequency power is removed first, extended haloing around high density 
regions is thus minimized.

\noindent 4) Features seen in the final adaptively smoothed surface-density have a well-determined 
significance and resolution.

One caution :  since the resolution is variable across the adaptively smoothed 'image', the
usual intuition that judges significance or signal-to-noise ratio by  
comparison with the amplitude of high-frequency noise is not reliable. 

There are several parameters which are important to results of the adaptive smoothing process  outlined 
above :

1) the signal-to-noise ratio,  snr$_{total}$, to be required in the total surface-density, $\Sigma$ (smoothed to the current resolution).  This parameter is set at snr$_{total}$ = 3 so that 
virtually all features seen in the final adaptively smoothed image will be 'statistically' significant.

2)  the signal-to-noise ratio,  snr$_{\delta}$, specifying whether the  $\Delta_n$ signal is 
removed before proceding to a lower resolution filter. This 
parameter should be set such that power is removed at the highest spatial frequency for which 
there is a 'reliable' signal, but avoiding removal of what is essentially
low-frequency power before 'its time has come'. Based on trial and error, we have adopted 
snr$_{\delta}$ =1 for the LSS identification. Although it might seem that 1 $\sigma$ would be 
risky, the 3 $\sigma$ condition (above) assures that most features will be significant. 

3) the maximum filter width, n$_{max}$. The maximum filter width was taken at 0.33\deg,
i.e. 23\% of the linear size of the COSMOS field. 

Two smoothing kernels were used : a boxcar and Gaussian. The boxcar was used 
for program development since it was faster, but all final results employ a 2-d, 
symmetric Gaussian filter (implemented with a Fourier transform for speed). It is well known
that boxcar filters can introduce high spatial frequency edges whereas the Gaussian 
is better behaved in this respect.

 \subsection{Simulation Tests}\label{appendixa}
 
 To test the algorithm described above with conditions similar to the galaxy 
 counts in the COSMOS photometric redshift catalog, we have simulated a
 single redshift slice with 10,000 galaxies. Approximately 
 half of the galaxies were distributed
 within Gaussian profile structures with a distribution of peak densities and sizes.
 The other half of the sample galaxies were distributed randomly across the field. 
 
 The parameters for the Gaussian-profile structures 
 in the simulation are listed in Table \ref{sim_param}.  Figure \ref{simulation}a
 shows the input surface-densities profiles. For the lowest 3 rows, the simulated structures 
 have increasing peak surface-densities toward the top and increasing in size going to the left.
 The top row simulates more complex structures with 
 three internal components having varying sizes and surface densities. Figure \ref{simulation}b shows the 
simulated distribution of galaxies, consisting of 5260 galaxies randomly placed and 
4740 galaxies 
 populated with probability given by Figure \ref{simulation}a. It is important 
 to realize that since the density profiles of the structures are sampled randomly, the
 simulation distribution, input to the adaptive smoothing algorithm,
 is not identical to that shown in Figure \ref{simulation}a, i.e. there is shot noise. Therefore, 
 the algorithm should not be expected to return the smooth 
 input distributions (Figure \ref{simulation}a) exactly. 
 
 Figure \ref{recover} shows the surface-density recovered from the 
 galaxy distribution shown in Figure \ref{simulation}b using the adaptive smoothing.
 In fact, the algorithm has done an excellent job of recovering all structures 
 which were statistically significant in Figure \ref{simulation}b, including the 
 top row with complex, internal structure. The three structures in the lower
 right of Figure \ref{simulation}a were not recovered but these were all 
 sufficiently low in surface-density and/or size that their total numbers of 
 galaxies were not statistically significant (2, 8 and 6 galaxies respectively -- see Table
 \ref{sim_param}). Lastly, it is worthwhile to emphasize that the algorithm did 
 not find structures which were not in the input simulation, i.e. noise in 
 the random galaxy population was not falsely detected using parameters for 
 the simulation distribution (numbers of galaxies and fraction in structures) and 
 for the detection algorithm similar to those used for the COSMOS structure detection.
 
 \subsection{Test on $\Lambda$CDM Simulations}\label{appendixb}

As an additional test we have applied the adaptive smoothing procedure 
to one of the Virgo consortium  $\Lambda$CDM simulations
\citep{ben01} and the more recent Millennium simulation COSMOS wedge \citep{spr05,cro06}. 

The Virgo simulation had dark matter particles of 
mass 1.4$\times10^{11}$/h \msun ~and for the purposes of our test,
we sampled the dark matter particles to obtain a surface density of 
particles in each redshift slice similar to that of galaxies in the COSMOS photo-z
catalog. This was done to keep the simulation shot noise characteristics similar 
to those of the observational data being analysed here. 
The results for adaptive smoothing of the $\Lambda$CDM Virgo simulation are shown 
for z $= 0.35$ and 0.93 in Figure \ref{lcdm}. The algorithm reliably recovers 
all significant structures seen in the simulation. It is noteworthy also 
that the scale of the structures seen here is qualitatively similar to 
that actually found in our application to the COSMOS photo-z 
catalog. Compare Figure \ref{lcdm} with the similar redshift frames of Figure \ref{lss}.

In our tests on the Millennium Simulation, the objective was to determine if similar 
structures were seen in this most up to date simulation as in the COSMOS data. Thus the 
galaxies in the MIllennium COSMOS light cone were each given a redshift 
uncertainty similar to that in the COSMOS photometric redshifts and then processed
identically to the observed galaxies. In Figure \ref{mil} the overdensities from the simulation are 
integrated along the line of sight from z = 0.25 to 1.05 as was done for Figure \ref{lss_2d}, including 
keeping the same contours for both plots. Extremely good correspondence is seen from the comparison 
indicating that the adaptive smoothing is recovering very similar structures in both and 
by implication, both theory and observations have similar intrinsic structure. In Section \ref{sims}, we make a 
more quantitative comparison by measuring the area filling factor as a function of overdensity
and redshift.

 
 



 
 \clearpage

 \begin{deluxetable}{llcccc}
 \tabletypesize{\scriptsize}
 \tablecaption{LSS Galaxy Samples\label{tbl-1}}
 \tablewidth{0pt}
 \tablehead{
 \colhead{Sample} &  \colhead{z} & \colhead{I cutoff (mag)} & \colhead{\# of galaxies} &  \colhead{\# (early type)\tablenotemark{a} } & \colhead{\# (late type)\tablenotemark{a} } } 
 \startdata
all-z & z $\leq 3. $ &  26 & 487973 & 59285 & 428688 \\
... &  ...    & 25 & 228487 & 34303 & 194184 \\
... & ...   & 24 & 106487 & 25278 & 81209 \\
low-z & $ z <$ 1.1 & 25 & 150162 & 25583 & 124579 \\
 \enddata
 \tablecomments{All samples have M$_V < -18$  and galaxies must  be detected in at least 4 bands.}
 \tablenotetext{a}{Early type : SED type $\leq$ 2.5; Late type : SED type $>$ 2.5}
  \end{deluxetable}

 \begin{deluxetable}{ccc}
 \tabletypesize{\scriptsize}
 \tablecaption{Galaxy Selection 'Completeness'\label{tbl-2}}
 \tablewidth{0pt}
 \tablehead{
 \colhead{Redshift interval} &  \colhead{Rel. \# per unit vol. } &  \colhead{Rel. Mass per unit vol. } } 
 \startdata
0.1 -- 0.3 & 0.52 & 0.69 \\
0.3 -- 0.5 &   0.69 & 0.94 \\
0.5 -- 0.7 & 1.0 & 0.92 \\
0.7 -- 0.9 & 0.56 & 0.96  \\
0.9 -- 1.1 &  0.47 & 1.0 \\

 \enddata
 \tablecomments{The total number of galaxies and total mass of galaxies per unit comoving 
 volume were evaluated by integrating the distribution functions shown in right panel of Figure \ref{selection} at masses above 10$^9$ \msun. These totals were normalized to the 
 values in the redshift interval with the largest values in order to assess the 
 count and mass incompleteness relative to this maximum bin.}
  \end{deluxetable}

\begin{table}

\begin{center}
\vskip -1cm
\caption{~~~~ Structures in the COSMOS field\label{struct_param}}
\smallskip
 \tiny
\begin{tabular}{ l | ccccccccccc}
\tableline\tableline
 &&& & \multicolumn{2}{c}{~~~~~~~~~~~~~~~FWHM\tablenotemark{b}}   \\ 
  \cline{5-7} 
  Structure \# &  RA-150\tablenotemark{a} & Dec-2\tablenotemark{a} & z\tablenotemark{a}
 & $\Delta$RA & $\Delta$DEC & $\Delta$z & size\tablenotemark{c} & \% LSS\tablenotemark{d} & \# galaxies\tablenotemark{e} &  M$_*$\tablenotemark{f} & Central $\Sigma_{1 Mpc}$ \tablenotemark{g} \\ 
  \cline{2-12}
 (X-ray id)\tablenotemark{h} & (\deg)  & (\deg)  &  & (\deg) & (\deg) & & Mpc & & & $10^{12}$\msun &  \\
\tableline
  1 (73,97,100,103,106) &  -0.09 &   0.51 &   0.73 &   0.22 &   0.17 &   0.27 &  12.69 &   35 &    1767 &  23.63 & 188 \\ 
  2 (62,68,84)&   0.15 &   0.20 &   0.88 &   0.26 &   0.21 &   0.25 &  17.41 &   30 &     815 &  15.72 & 99 \\ 
  3 (126,128)&  -0.33 &   0.27 &   0.93 &   0.18 &   0.17 &   0.20 &  13.31 &   36 &     875 &  17.85 & 48\\ 
  4 &   0.57 &   0.49 &   0.71 &   0.15 &   0.14 &   0.28 &   8.94 &   24 &     569 &   4.55 & 46\\ 
  5 &  -0.49 &  -0.13 &   0.62 &   0.24 &   0.22 &   0.11 &  12.97 &   31 &     939 &   6.25 & 63\\ 
  6 (40,45,53)&   0.42 &  -0.14 &   0.92 &   0.16 &   0.17 &   0.19 &  12.70 &   33 &     580 &  11.51 & 47\\ 
  7 &   0.30 &   0.40 &   0.73 &   0.14 &   0.13 &   0.21 &   8.70 &   25 &     384 &   3.97 & 84\\ 
  8 &   0.40 &   0.77 &   0.75 &   0.25 &   0.10 &   0.18 &  12.30 &   27 &     526 &   4.93 & 96\\ 
  9 (32) &   0.51 &   0.23 &   0.89 &   0.12 &   0.12 &   0.18 &   9.25 &   40 &     512 &  10.32 & 57 \\ 
 10 (66,72)&   0.16 &   0.60 &   0.87 &   0.25 &   0.15 &   0.26 &  15.48 &   32 &     394 &   6.76 & 59 \\ 
 11 &   0.27 &  -0.42 &   0.46 &   0.12 &   0.09 &   0.25 &   4.70 &   44 &     403 &   1.93 & 92\\ 
 12 &   0.15 &   0.70 &   0.54 &   0.12 &   0.10 &   0.37 &   5.49 &   28 &     255 &   2.05 & 49 \\ 
 13 (25,64) &   0.46 &   0.56 &   0.29 &   0.13 &   0.16 &   0.17 &   4.20 &   53 &     197 &   1.31 & 41\\ 
 14 &   0.06 &   0.30 &   0.74 &   0.08 &   0.08 &   0.22 &   5.13 &   26 &     141 &   1.70 & 105\\ 
 15 (111) &  -0.23 &   0.32 &   0.34 &   0.14 &   0.15 &   0.11 &   4.81 &   44 &     168 &   1.47 & 54\\ 
 16 &  -0.24 &   0.92 &   0.38 &   0.10 &   0.08 &   0.12 &   3.28 &   56 &     226 &   2.18 & 103\\ 
 17 (78,85) &   0.05 &   0.22 &   0.24 &   0.07 &   0.16 &   0.14 &   3.10 &   62 &     127 &   1.10 & 16\\ 
 18 &  -0.50 &   0.01 &   0.98 &   0.10 &   0.09 &   0.11 &   7.57 &   31 &     102 &   2.07 & 16\\ 
 19 &   0.53 &   0.73 &   0.22 &   0.18 &   0.11 &   0.10 &   3.39 &   60 &      77 &   0.59 & 12\\ 
 20 (34,39,41,44) &   0.49 &   0.07 &   0.45 &   0.05 &   0.05 &   0.18 &   2.07 &   53 &     133 &   1.19 & 104\\ 
 21 &   0.20 &  -0.37 &   0.32 &   0.06 &   0.05 &   0.39 &   1.71 &   52 &      95 &   0.38 & 64 \\ 
 22 (89,105)&  -0.08 &   0.60 &   0.26 &   0.06 &   0.14 &   0.11 &   2.81 &   62 &      67 &   0.71 & 42\\ 
 23 (15) &   0.45 &   0.05 &   0.33 &   0.05 &   0.08 &   0.10 &   2.19 &   57 &      82 &   0.56 & 58\\ 
 24 (80) &   0.11 &   0.56 &   0.61 &   0.03 &   0.04 &   0.26 &   1.97 &   43 &      85 &   0.82 & 131\\ 
 25 &  -0.56 &   0.43 &   0.46 &   0.06 &   0.05 &   0.09 &   2.42 &   58 &      87 &   1.34 & 90\\ 
 26 (145) &  -0.60 &   0.57 &   0.39 &   0.05 &   0.03 &   0.12 &   1.49 &   67 &      92 &   0.78 & 131\\ 
 27 (54,57,59) &   0.33 &  -0.40 &   0.37 &   0.03 &   0.04 &   0.22 &   1.29 &   55 &      37 &   0.17 & 44\\ 
 28 (70) &   0.18 &  -0.24 &   0.30 &   0.03 &   0.06 &   0.24 &   1.40 &   59 &      43 &   0.37 & 39\\ 
 29 &  -0.34 &  -0.44 &   0.45 &   0.07 &   0.03 &   0.21 &   2.27 &   35 &      31 &   0.05 & 22\\ 
 30 (67) &   0.34 &  -0.39 &   0.19 &   0.03 &   0.03 &   0.08 &   0.63 &   76 &      17 &   0.24 & 21\\ 
 31 (132) &  -0.41 &   0.82 &   0.32 &   0.03 &   0.02 &   0.06 &   0.77 &   72 &      22 &   0.49 & 31\\ 
 32 (56) &   0.10 &  -0.00 &   0.30 &   0.03 &   0.03 &   0.03 &   0.93 &   49 &      12 &   0.08 & 22 \\ 
 33 &  -0.31 &   0.35 &   0.20 &   0.03 &   0.03 &   0.03 &   0.57 &   91 &       9 &   0.01 & --\\ 
 34 &   0.39 &   0.87 &   0.16 &   0.02 &   0.01 &   0.05 &   0.24 &   86 &       3 &   0.09 & -- \\ 
 35 (29,42) &   0.53 &   0.49 &   0.16 &   0.01 &   0.01 &   0.05 &   0.23 &   85 &       2 &   0.03 & -- \\ 
 36 (140) &  -0.59 &   0.57 &   0.17 &   0.01 &   0.01 &   0.05 &   0.23 &   85 &       2 &   0.00 & -- \\ 
 37 &   0.51 &  -0.45 &   0.20 &   0.01 &   0.01 &   0.00 &   0.29 &   86 &       3 &   0.02 & -- \\ 
 38 &  -0.16 &   0.68 &   0.20 &   0.01 &   0.02 &   0.00 &   0.29 &   70 &       ... &   0.01 & -- \\ 
 39 &   0.33 &  -0.36 &   0.60 &   0.01 &   0.01 &   0.00 &   0.63 &   36 &       ... &   0.01 & 10 \\ 
 40 &   0.74 &   0.67 &   0.90 &   0.01 &   0.01 &   0.03 &   0.77 &   29 &       ... &   0.02 & -- \\ 
 41 &   0.65 &   0.31 &   0.30 &   0.01 &   0.01 &   0.00 &   0.28 &   38 &       ... &   0.02 & -- \\ 
 42 &  -0.30 &   0.51 &   0.50 &   0.01 &   0.01 &   0.00 &   0.38 &   30 &       ... &   0.01 & --\\ 
 \tableline
\end{tabular}
\tiny
 \tablenotetext{a}{RA and DEC (J2000) of peak galaxy surface-density and the centroid z.}
  \tablenotetext{b}{Full-width at half-maximum evaluated from the 2.3$\times\sigma$ where $\sigma$ is the
  dispersion (from the calculated 2nd moment).}
   \tablenotetext{c}{Estimated as ($\Delta RA^2$ + $\Delta DEC^2$)$^{1/2}$ converted to co-moving Mpc.}
  \tablenotetext{d}{Mean probability that a galaxy within the structure is within the structure rather 
    than being in the projected background, estimated as $\Sigma_{LSS} / (\Sigma_{LSS} + \Sigma_{Background})$.}
    \tablenotetext{e} {Total number of galaxies estimated within structure, corrected for 'field or background'
  contamination by comparing the surface-density in the structure with the background surface-density for that redshift slice -- see note (d)}
      \tablenotetext{f}{The total stellar mass (M$_*$), estimated from the absolute magnitude of 
      each galaxy and using a mass-to-light ratio appropriate to the galaxy SED. For each galaxy,
      this photometric mass is multiplied by the probability that it is within the structure rather 
      than being a projected field/background galaxy (see note (e)).}
 \tablenotetext{g}{The central surface density used to evaluate the structure/cluster 
            Richness -- the number of galaxies within radius 1 Mpc brighter than 2 magnitudes 
            below the 3rd brightest galaxy. This column is blank if there are too few galaxies 
            to estimate a Richness (i.e $< 10$)}
 \tablenotetext{h}{In parenthesis we give the cross-reference to the ID for the X-ray 
 clusters from \cite{fin06}. The wavelet technique used by \cite{fin06} on the X-ray emission and on early SED type galaxies  in the photometric redshift catalog is selective toward 
 more compact structures than the adaptive smoothing technique used here; therefore, in many cases, these 
 cross-identifications should be viewed only as 'possible', based on close proximity in $\alpha, \delta, $
 and z. }
            
      \end{center}
\end{table}

\begin{deluxetable}{cccccccccc}
\tabletypesize{\scriptsize}
\tablecaption{Relative Variances for 2.5 deg$^2$ field \label{variance}}
\tablewidth{0pt}
\tablehead{
\colhead{Redshift} &  \colhead{Comoving Volume} & \colhead{M(vol) } &  \colhead{$\sigma$(M(vol))} &  \colhead{Halo mass} & \colhead{Number} & \colhead{$<b>$} &  \colhead{$\sigma_{shot}$ }&  \colhead{$\sigma_{cosmic}$ }&  \colhead{$\sigma (M,z)$ }\\
\colhead{range}   & \colhead{$10^{6}$ present-day Mpc$^3$} & \colhead{  10$^{17}$ \msun} & & \colhead{range (\msun)} & \colhead{expected} & & & & }
\startdata
0.2 -- 0.4  &  0.80  &  0.327  &  0.149   &   $10^{13 }$--$10^{ 14 }$ &  142  &  1.62  &  0.084  &  0.240  &  0.254 \\ 
0.2 -- 0.4  &  0.80  &  0.327  &  0.149   &   $10^{14 }$--$10^{ 15 }$ &  6  &  3.03  &  0.397  &  0.451  &  0.601 \\ 
0.4 -- 0.6  &  1.78  &  0.727  &  0.099   &   $10^{13 }$--$10^{ 14 }$ &  279  &  1.84  &  0.060  &  0.182  &  0.192 \\ 
0.4 -- 0.6  &  1.78  &  0.727  &  0.099   &   $10^{14 }$--$10^{ 15 }$ &  10  &  3.52  &  0.324  &  0.349  &  0.476 \\ 
0.6 -- 0.8  &  2.79  &  1.14  &  0.075   &   $10^{13 }$--$10^{ 14 }$ &  372  &  2.09  &  0.052  &  0.157  &  0.165 \\ 
0.6 -- 0.8  &  2.79  &  1.14  &  0.075   &   $10^{14 }$--$10^{ 15 }$ &  9  &  4.08  &  0.327  &  0.307  &  0.448 \\ 
0.8 -- 1  &  3.67  &  1.50  &  0.061   &   $10^{13 }$--$10^{ 14 }$ &  405  &  2.37  &  0.050  &  0.145  &  0.153 \\ 
0.8 -- 1  &  3.67  &  1.50  &  0.061   &   $10^{14 }$--$10^{ 15 }$ &  7  &  4.71  &  0.371  &  0.288  &  0.470 \\ 
1 -- 1.2  &  4.40  &  1.79  &  0.052   &   $10^{13 }$--$10^{ 14 }$ &  387  &  2.69  &  0.051  &  0.140  &  0.149 \\ 
1 -- 1.2  &  4.40  &  1.79  &  0.052   &   $10^{14 }$--$10^{ 15 }$ &  5  &  5.41  &  0.458  &  0.282  &  0.538 \\ 

\enddata
\tablecomments{The field size of 2.5 deg$^2$ was taken to match the area of the sample taken 
from the COSMOS photometric redshift catalog used here. The table includes the expected 
number of halos in two mass ranges, together with the relative variances due to 
shot noise, cosmic variance and the combined total variance. Values calculated for cosmological parameters : H$_0$ = 70 km s$^{-1}$ Mpc$^{-1}$, $\Omega_M$ = 0.3,  $\Omega_\Lambda$ = 0.7 and  $\sigma_8 = 0.74$.}
   \end{deluxetable}

\begin{deluxetable}{cccccc}
 \tabletypesize{\scriptsize}
 \tablecaption{Parameters for Structures in Simulation\label{sim_param}}
 \tablewidth{0pt}
 \tablehead{
 \colhead{Structure \#} &  \colhead{RA} &  \colhead{Dec} & \colhead{FWHM(\arcmin)\tablenotemark{a}} &  \colhead{Peak (\#/\sq\deg)\tablenotemark{b} } & \colhead{\# galaxies\tablenotemark{c} } } 
 \startdata
  1 &  -0.48 &  -0.64 &   1.0 &   6148 &     2 \\ 
  2 &  -0.16 &  -0.64 &   1.9 &   6148 &     8 \\ 
  3 &   0.16 &  -0.64 &   5.8 &   6148 &    65 \\ 
  4 &   0.48 &  -0.64 &  11.5 &   6148 &   252 \\ 
  5 &  -0.48 &  -0.24 &   1.0 &  18446 &     6 \\ 
  6 &  -0.16 &  -0.24 &   1.9 &  18446 &    22 \\ 
  7 &   0.16 &  -0.24 &   5.8 &  18446 &   194 \\ 
  8 &   0.48 &  -0.24 &  11.5 &  18446 &   773 \\ 
  9 &  -0.48 &   0.16 &   1.0 &  61489 &    18 \\ 
 10 &  -0.16 &   0.16 &   1.9 &  30744 &    36 \\ 
 11 &   0.16 &   0.16 &   5.8 &  30744 &   322 \\ 
 12 &   0.48 &   0.16 &  11.5 &  30744 &  1288 \\ 
 13 &  -0.42 &   0.58 &   1.0 &  43042 &    13 \\ 
 14 &  -0.48 &   0.64 &   1.9 &  24595 &    29 \\ 
 15 &  -0.48 &   0.58 &   5.8 &  24595 &   258 \\ 
 16 &  -0.09 &   0.58 &   1.9 &  24595 &    29 \\ 
 17 &  -0.16 &   0.67 &   1.9 &  24595 &    29 \\ 
 18 &  -0.16 &   0.58 &   7.7 &  12297 &   229 \\ 
 19 &   0.23 &   0.58 &   1.9 &  30744 &    36 \\ 
 20 &   0.16 &   0.64 &   5.8 &  18446 &   194 \\ 
 21 &   0.16 &   0.58 &  11.5 &  12297 &   514 \\ 
 22 &   0.48 &   0.64 &   1.9 &  24595 &    29 \\ 
 23 &   0.58 &   0.58 &   3.8 &  24595 &   115 \\ 
 24 &   0.48 &   0.58 &   8.7 &  12297 &   290 \\ 
 \enddata
 \tablecomments{Total number of galaxies in all the stuctures is 4740. }
 \tablenotetext{a}{Full-width at half maximum (arcmin) for the Gaussian galaxy distribution.}
  \tablenotetext{b}{Peak surface density of galaxies.}
   \tablenotetext{c}{Total number of galaxies in structure.}
  \end{deluxetable}


\clearpage
 
 


\begin{figure}[ht]
\epsscale{1} 
\plottwo{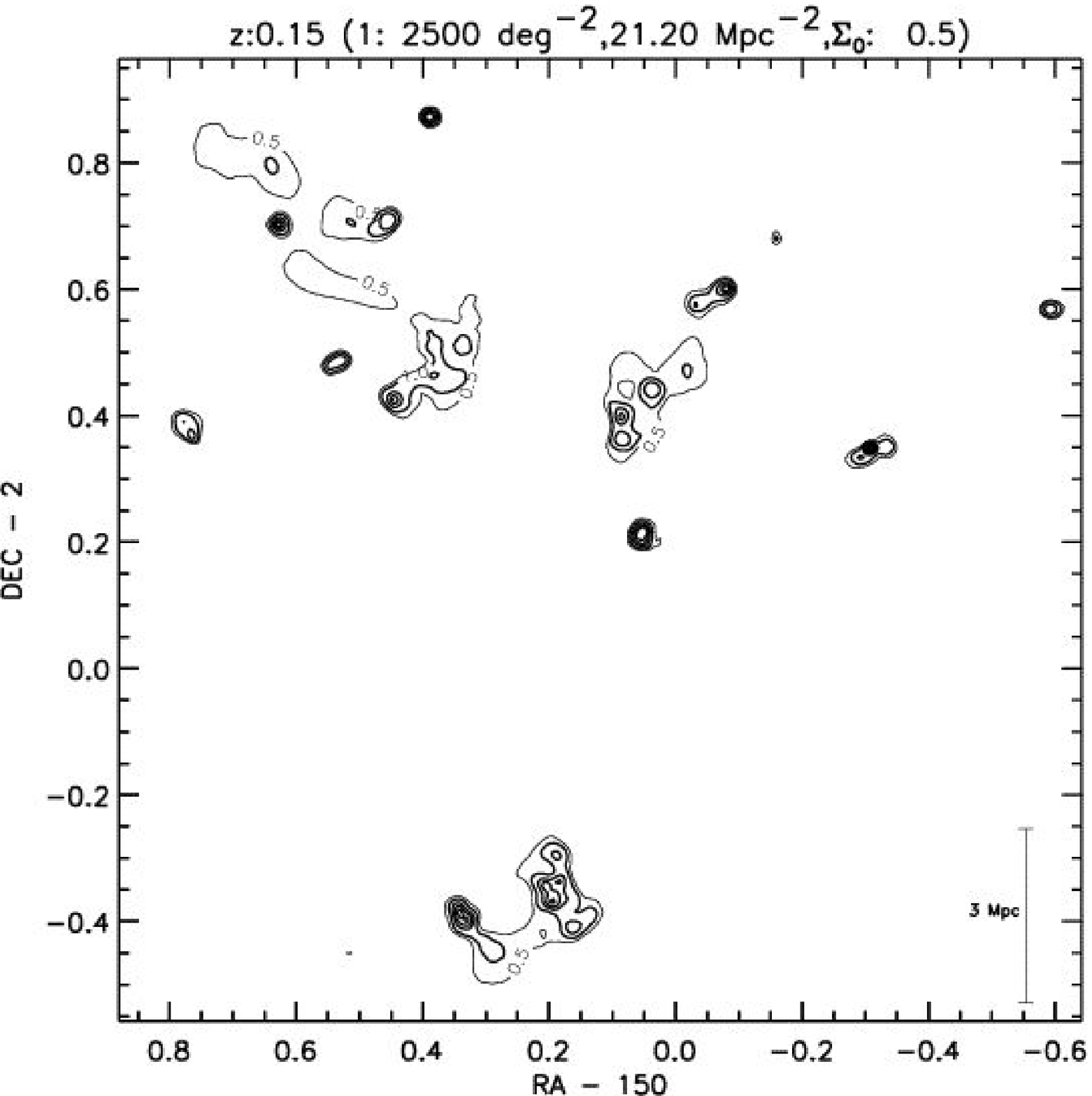}{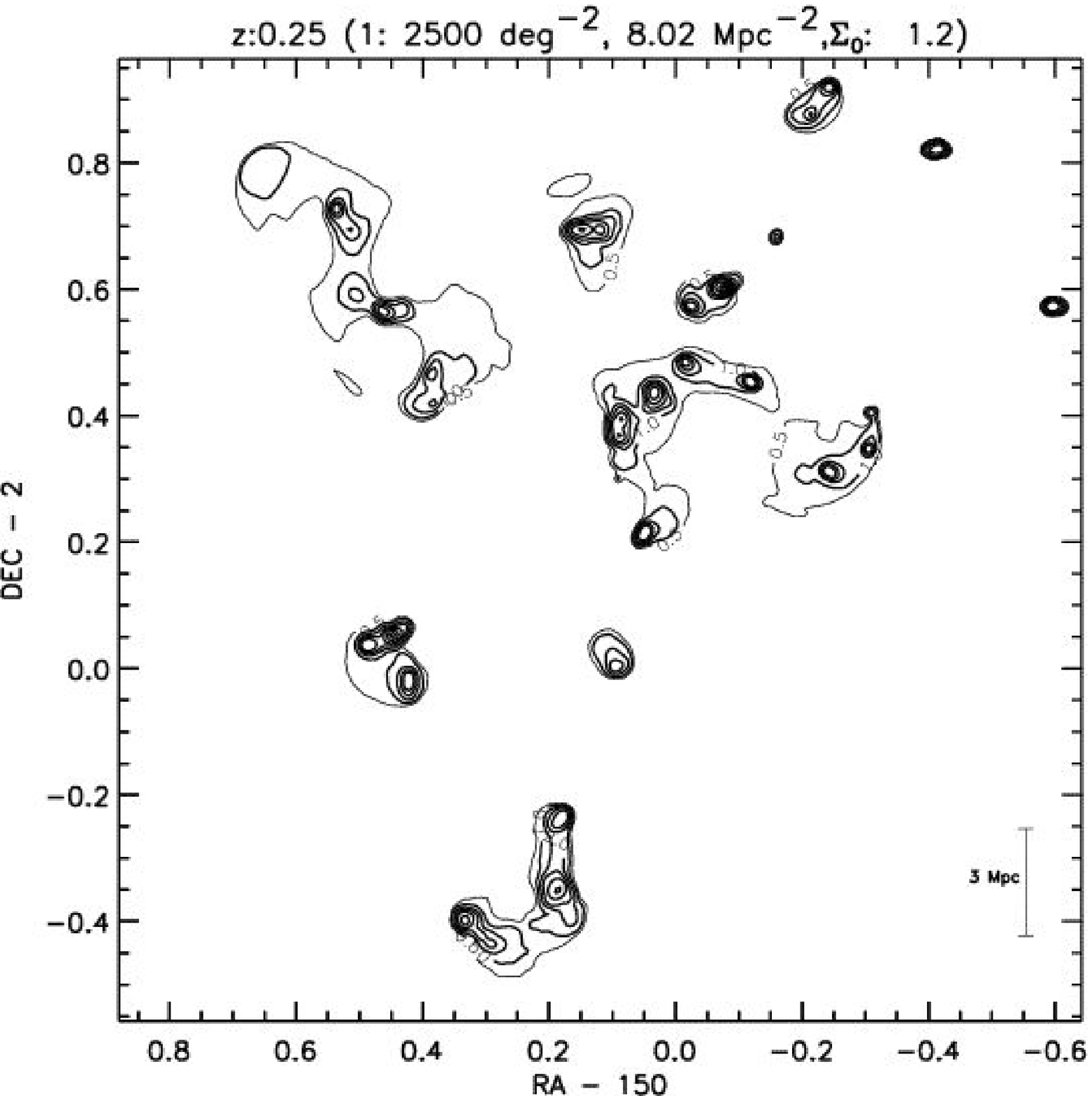}

\vskip 1cm
\plottwo{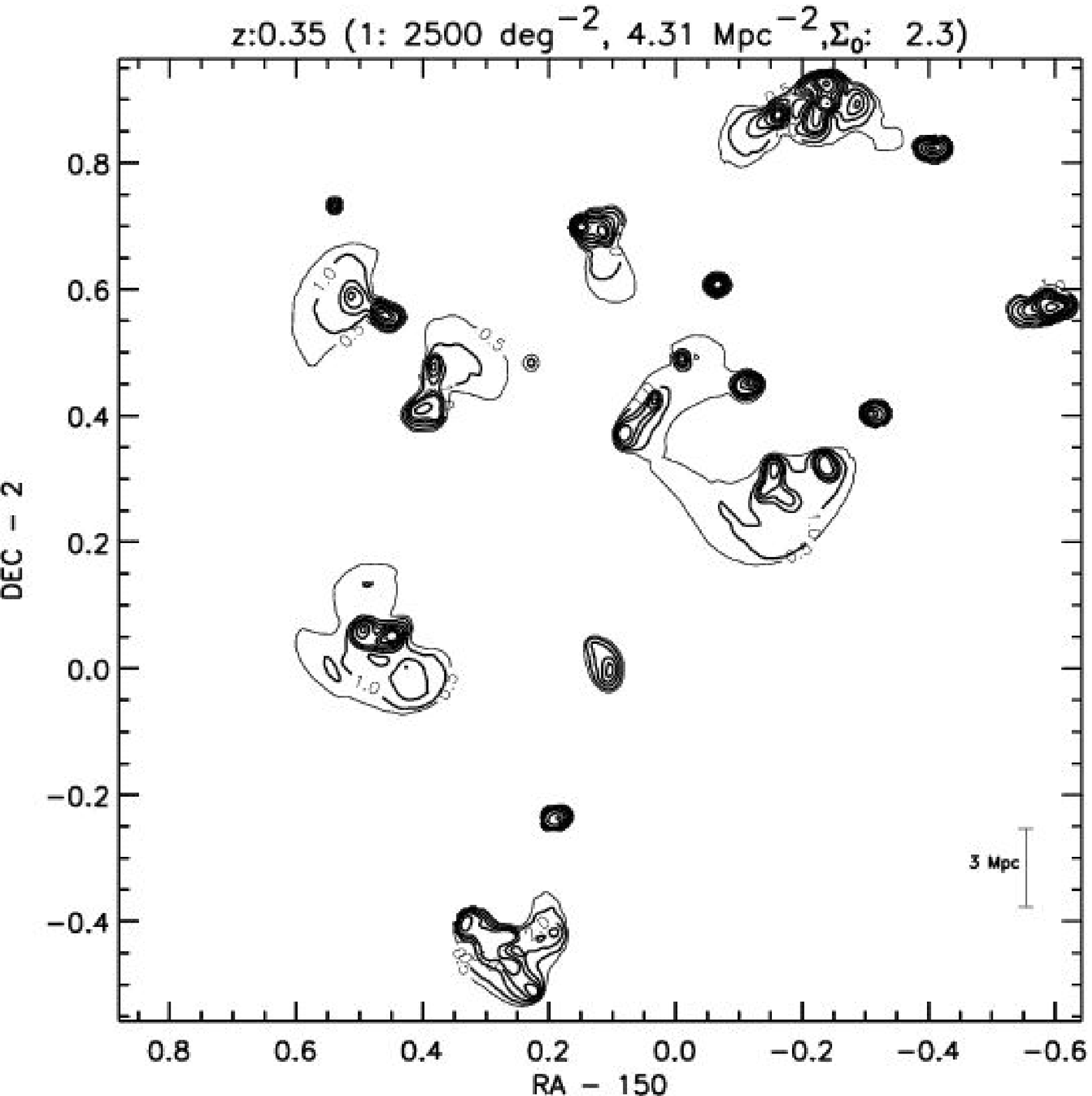}{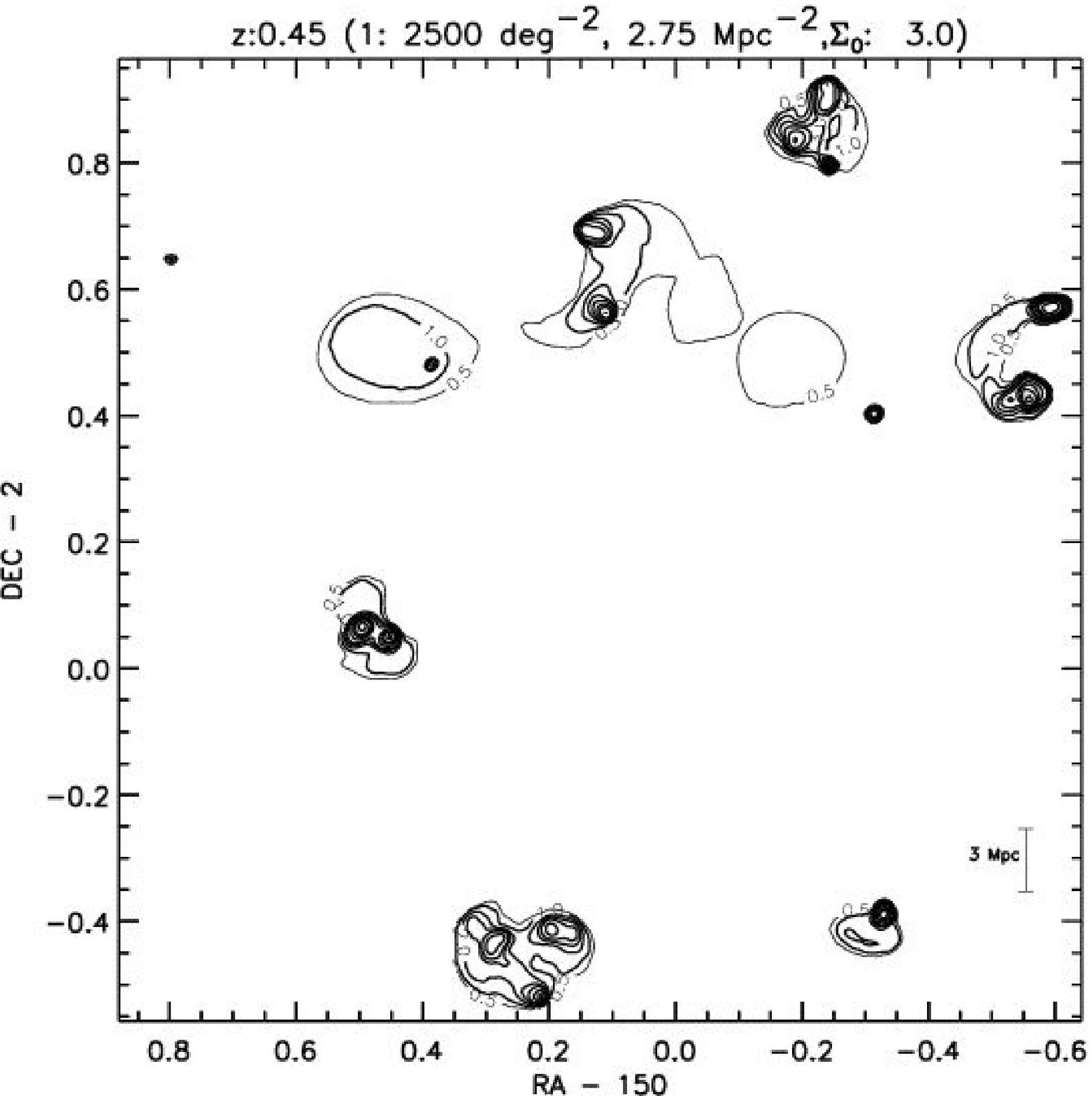}
\vskip 1cm
\caption{The surface-density of galaxies (with background subtracted out) computed using the adaptive smoothing 
algorithm described in Appendix \ref{appendix} is shown for slices at different redshifts. For z $\leq 1.05$, the width of the redshift slice is  $\Delta z = 0.1$; for the last three 
plots with z $\geq 1.3$, we use $\Delta z = 0.25$. On each plot, the legend at the top gives the 
surface density (galaxies per deg$^{-2}$ and per Mpc$^{-2}$) corresponding to 
1 contour unit.  The number of contour units corresponding to the background surface 
density ($\Sigma_0 =  B_0$) is also given. The scale bar indicates the size in co-moving Mpc. At higher z, we show 
only those redshifts with possibly significant structures.} 
\label{smooth}
\end{figure}

\clearpage
\epsscale{1} 
\plottwo{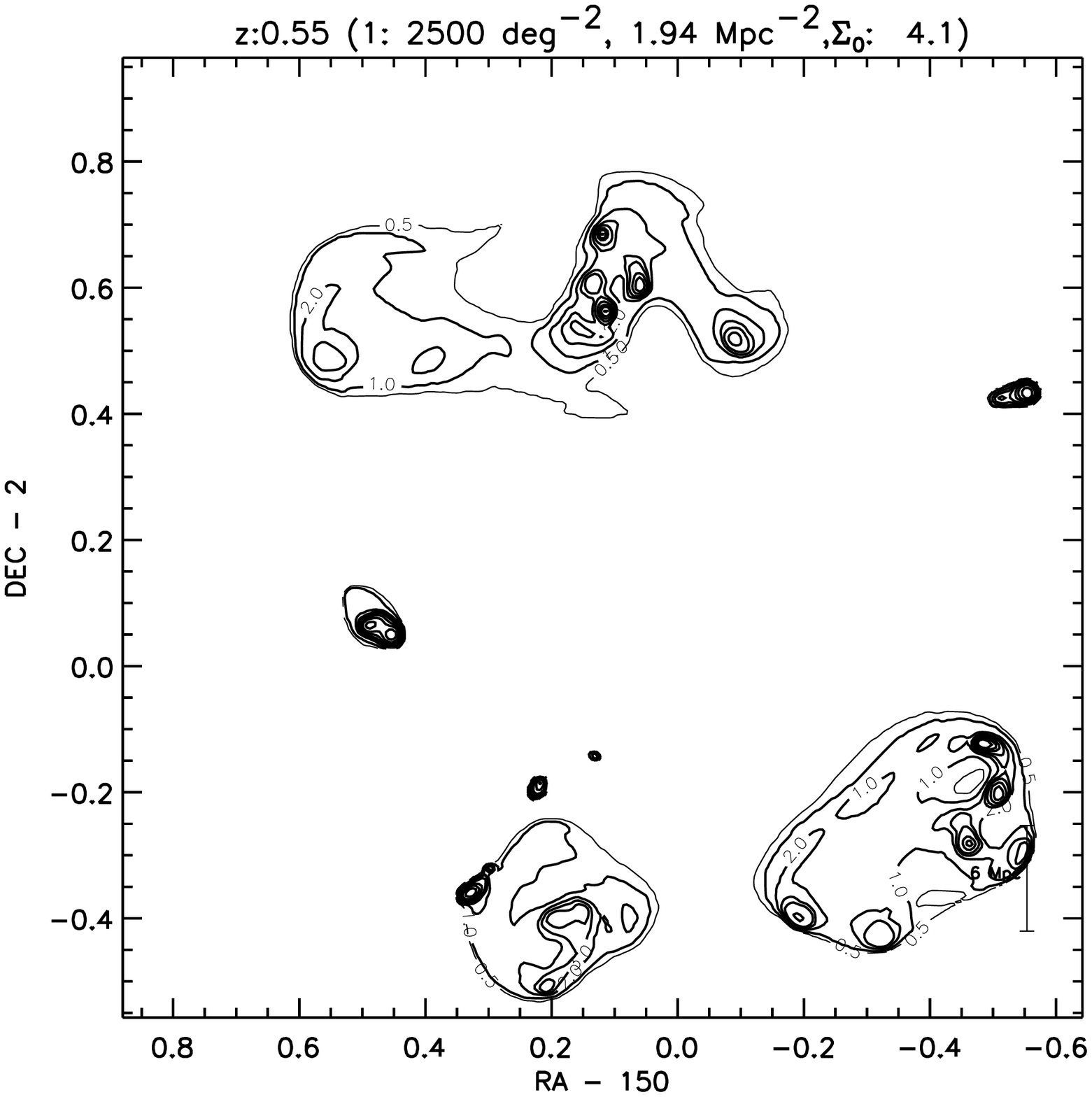}{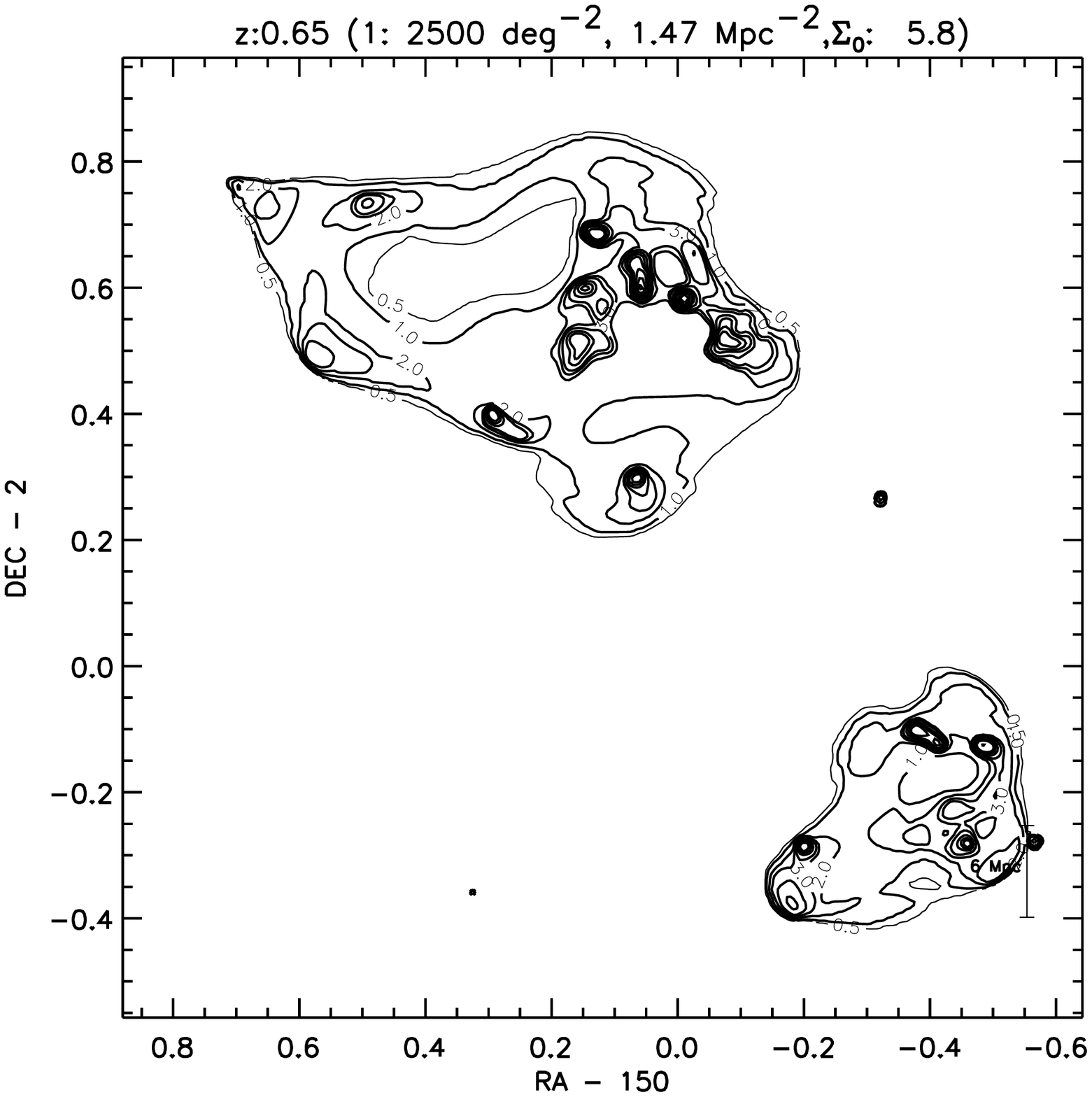}

\vskip 1cm
\plottwo{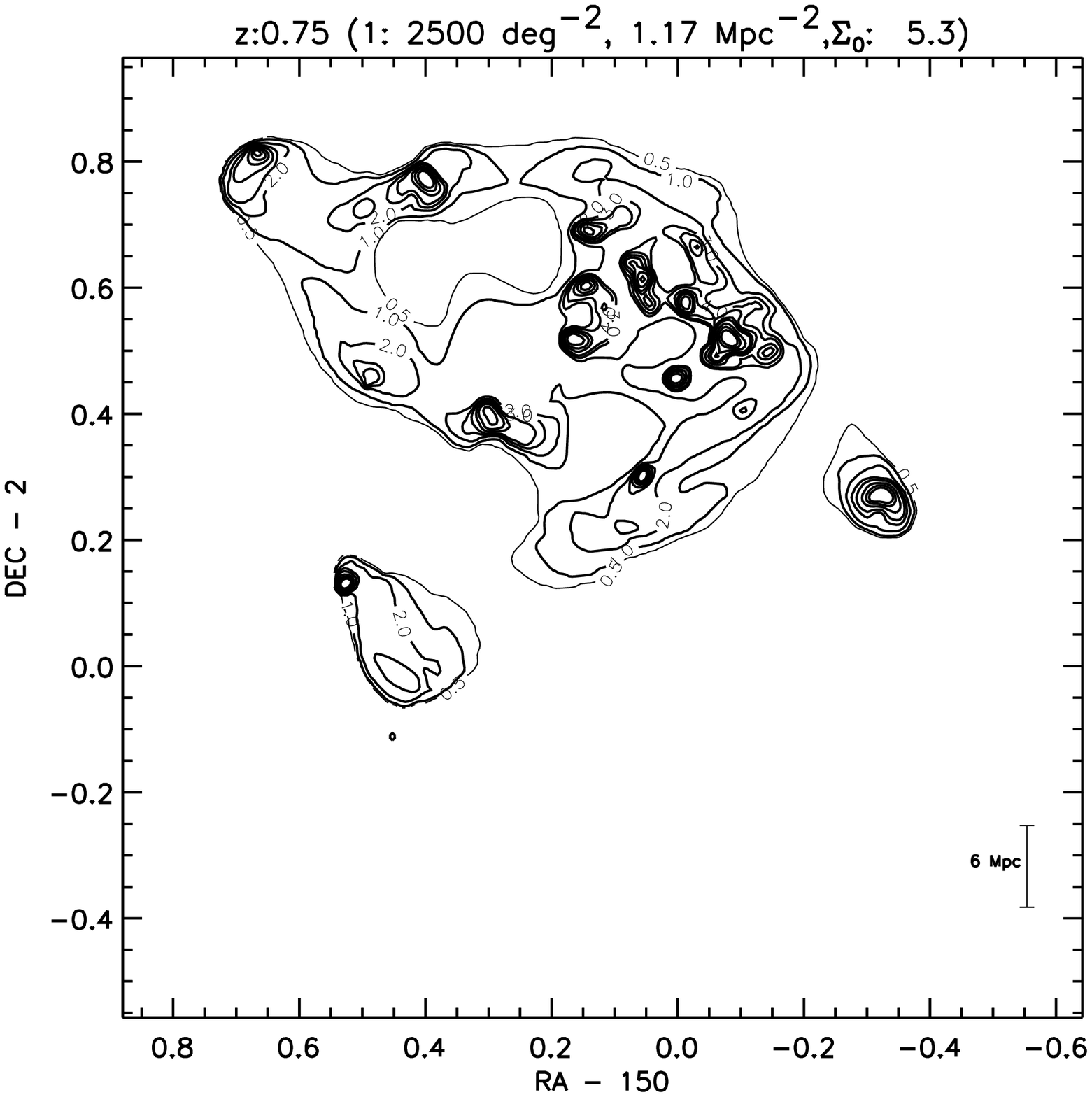}{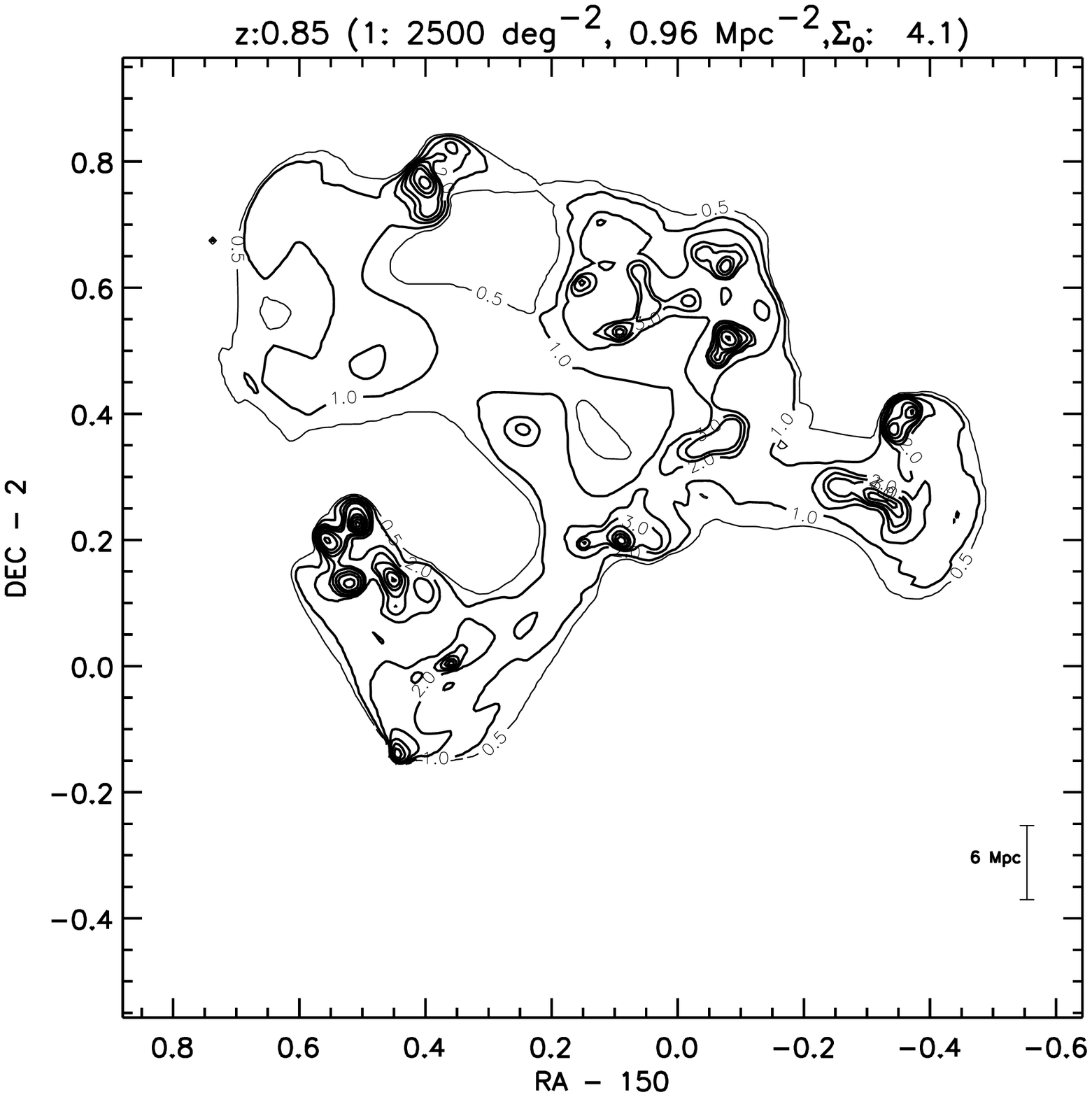}
\vskip 1cm
\centerline{Fig. 3. --- cont'd}

\clearpage
\epsscale{1} 
\plottwo{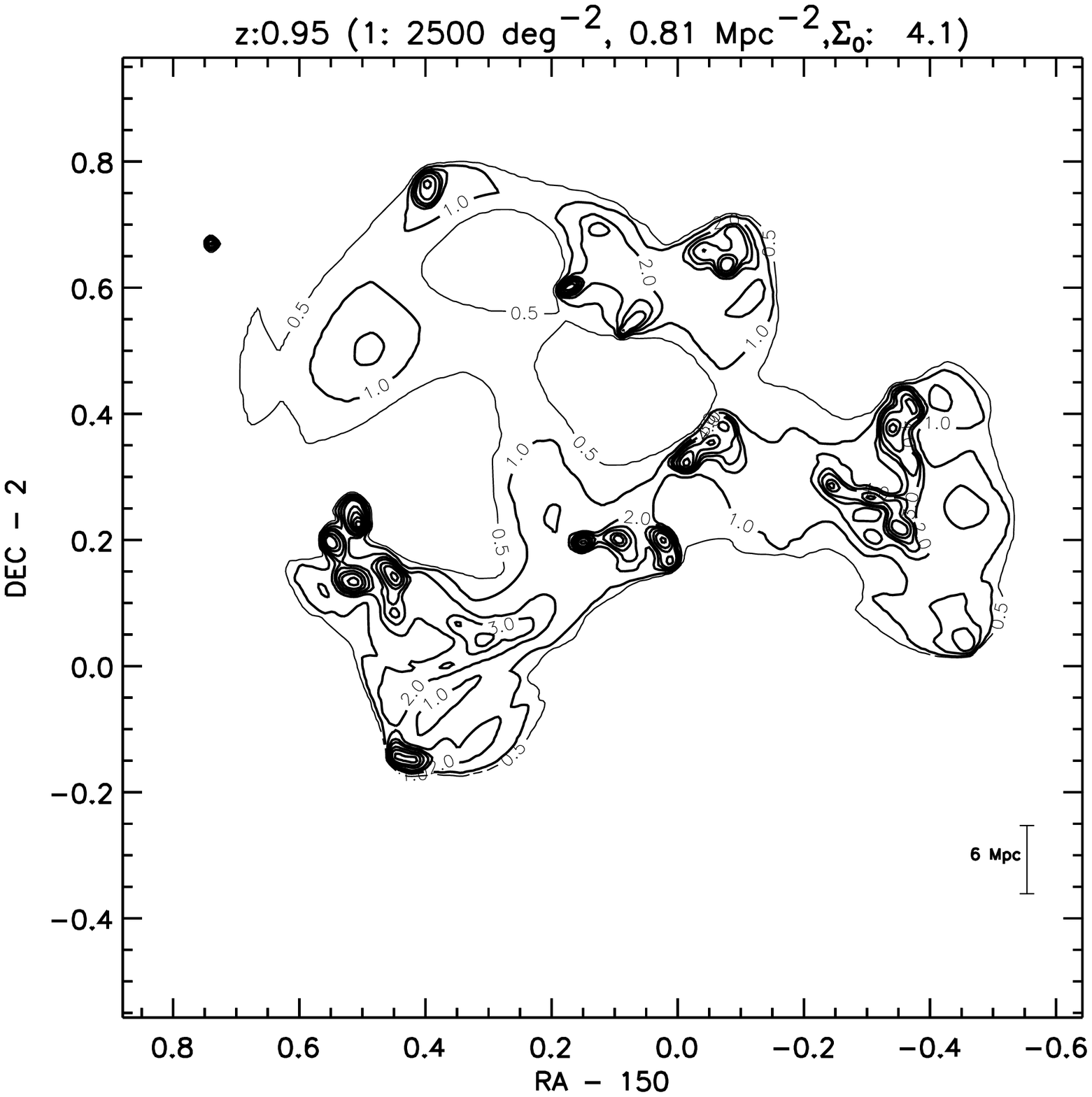}{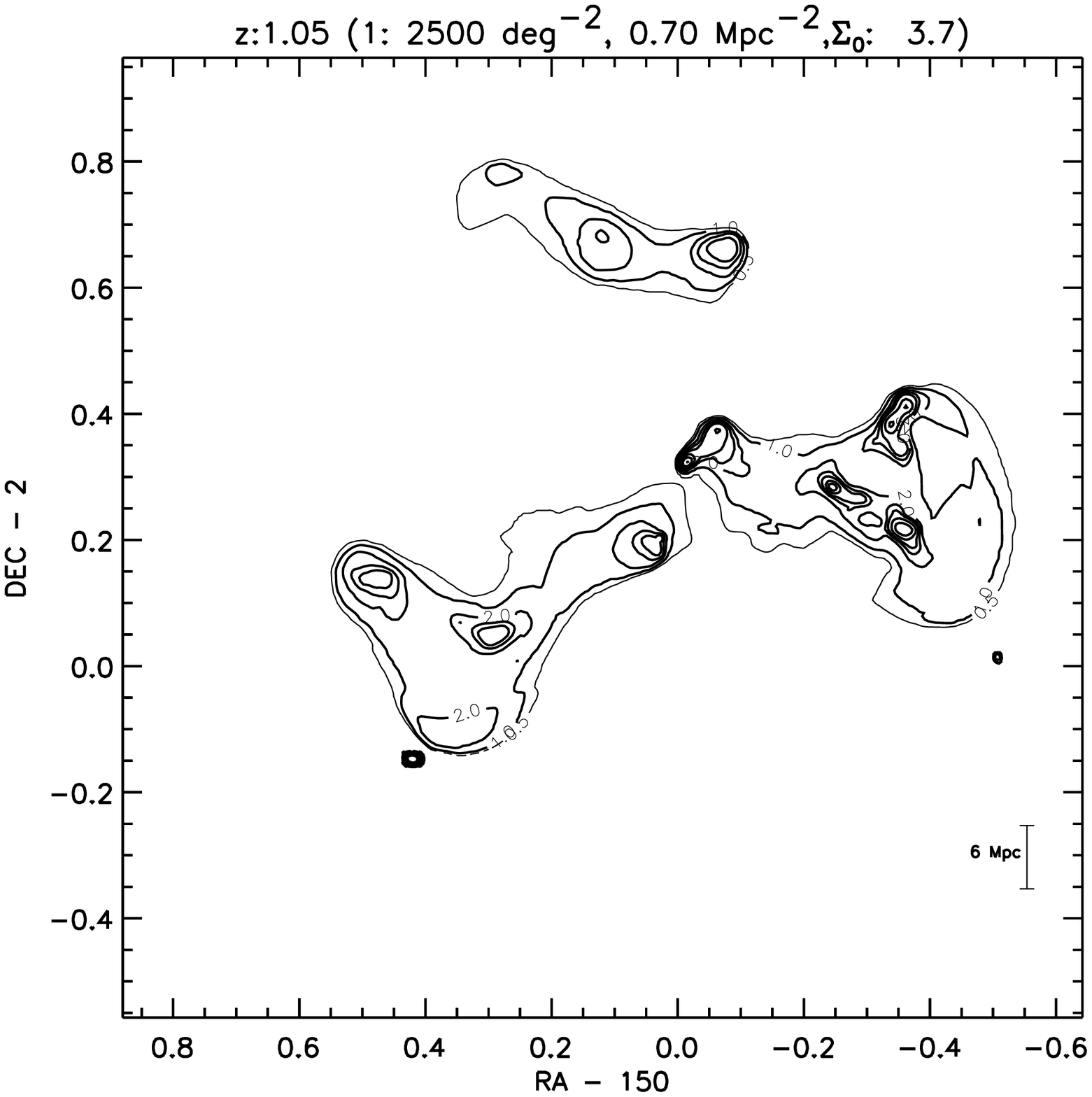}

\vskip 1cm
\plottwo{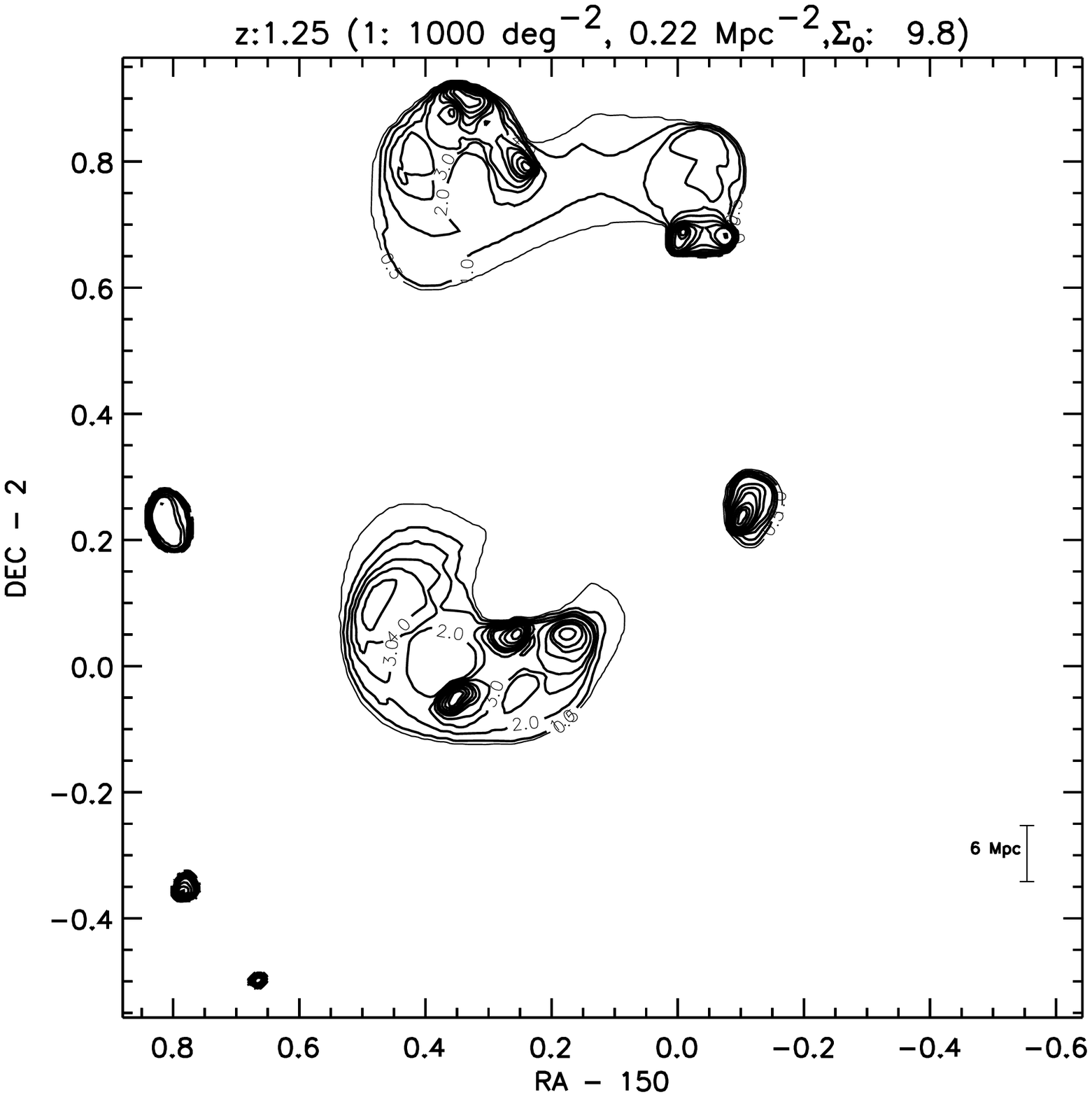}{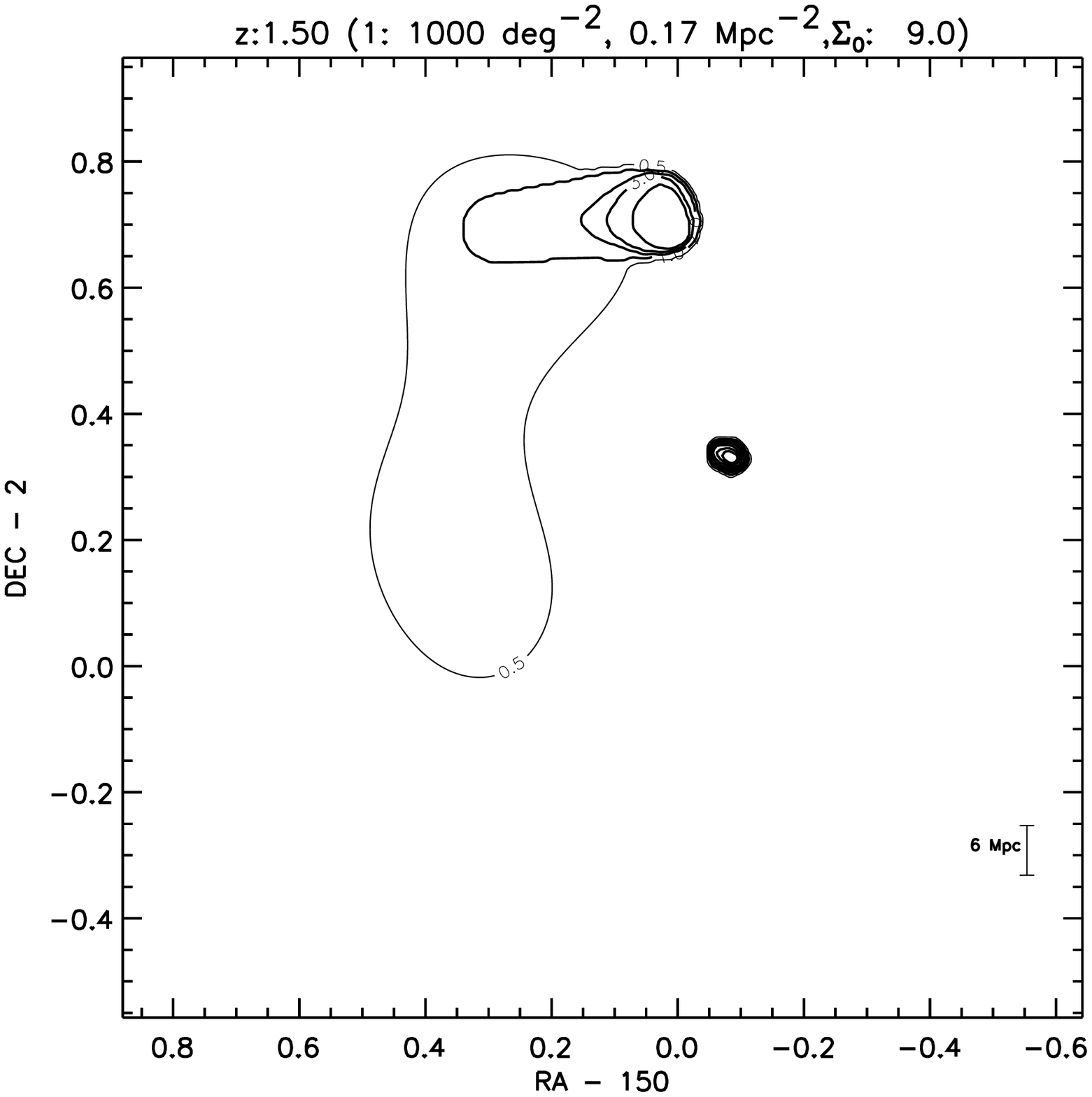}

\vskip 1cm
\centerline{Fig. 3. --- cont'd}

\clearpage
\begin{figure}[ht]
\epsscale{1}
\vskip -2cm 
\plotone{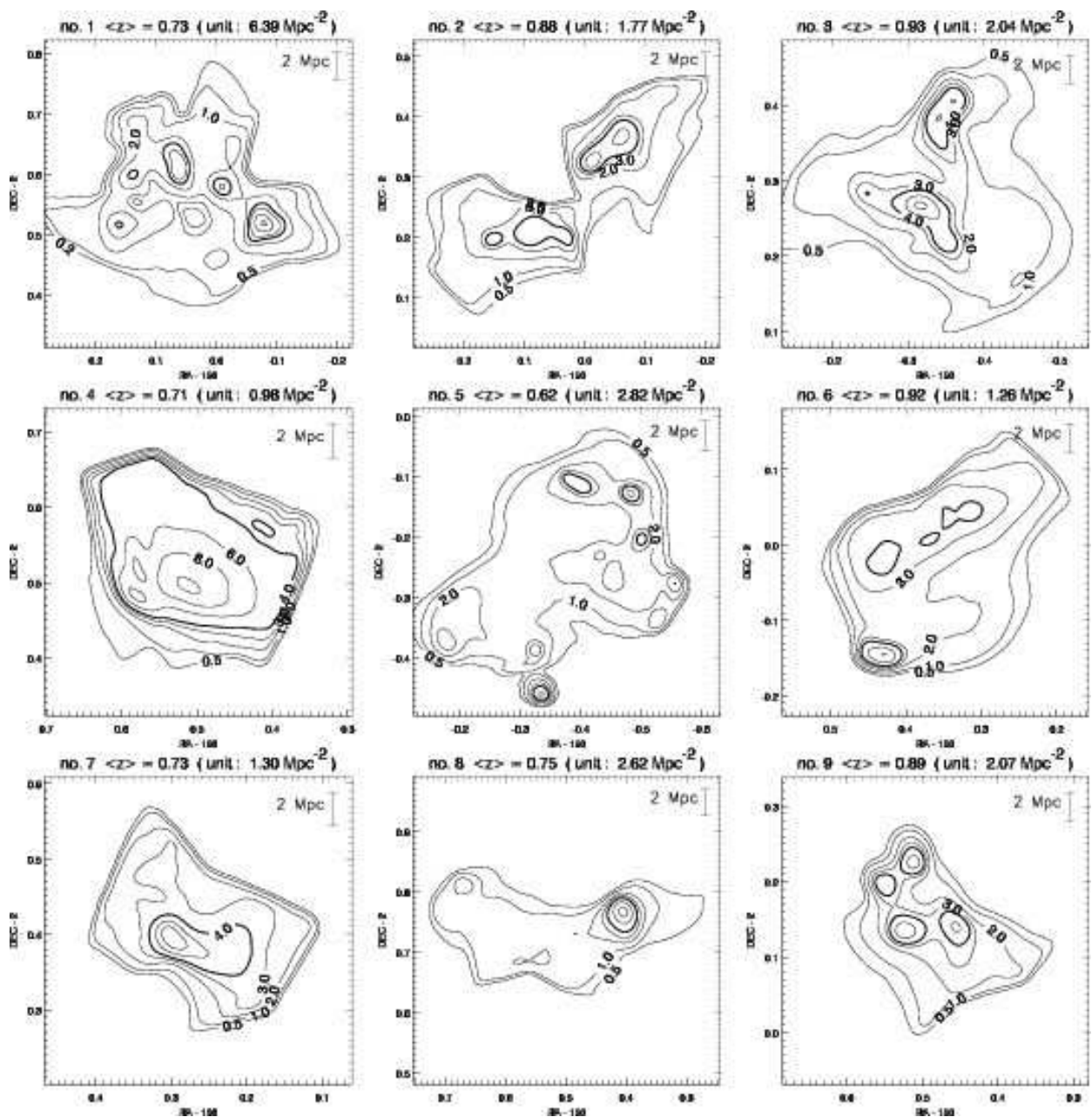}

\caption{The surface-density of galaxies in each structure found in the adaptive smoothing 
procedure. The galaxy densities are integrated in redshift over all connected 3-D pixels
for each structure. A scale 
bar on each plot indicates 2 Mpc (comoving). The top legend gives the surface-density 
of galaxies corresponding to 1 contour unit and the mean redshift $<z>$ of all galaxies within 
each structure is given.} 
\label{lss}
\end{figure}

\clearpage
\epsscale{1} 
\vskip -2cm 
\plotone{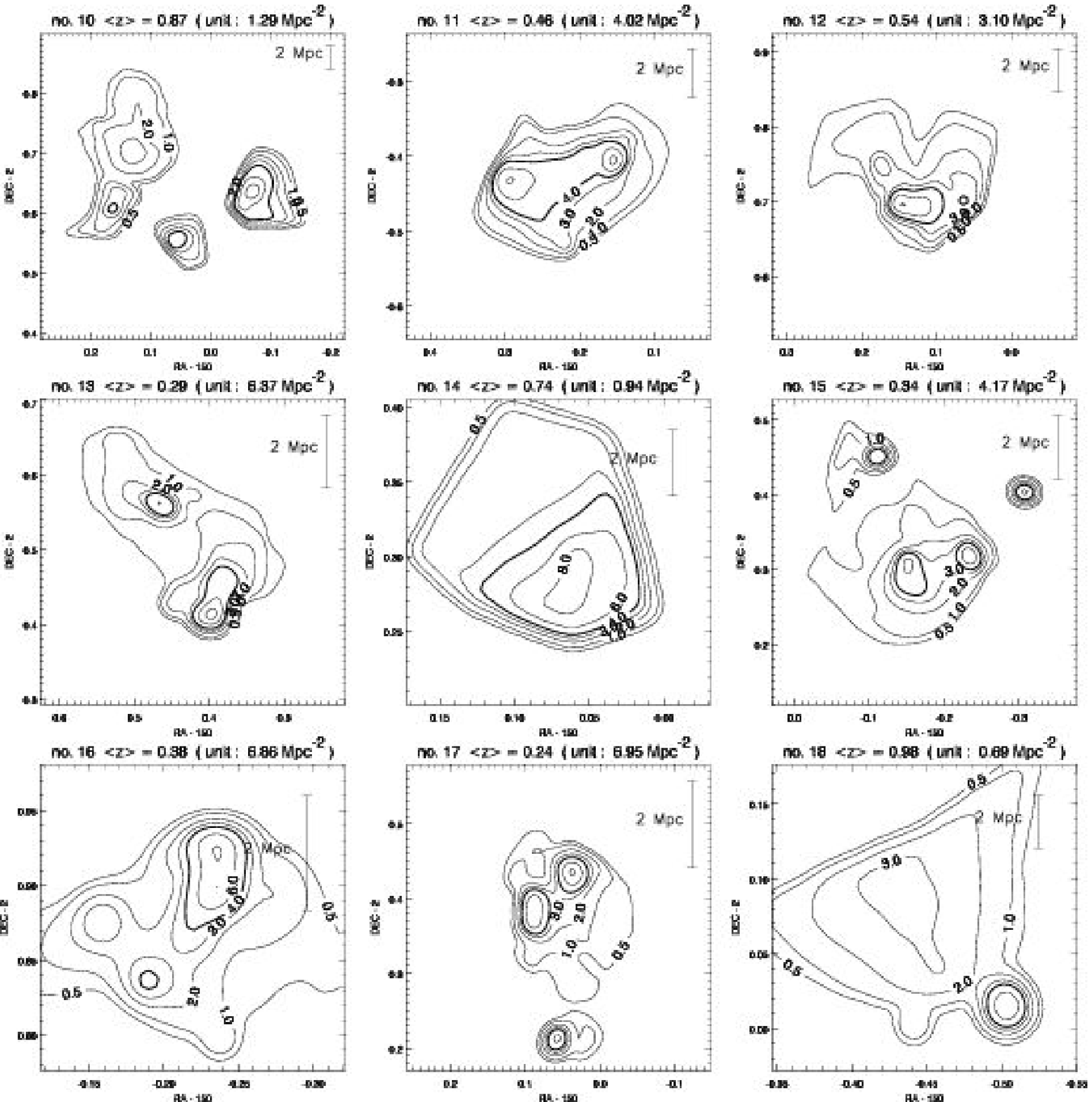}

\centerline{Fig. 4. --- cont'd}


\clearpage
\epsscale{1} 
\vskip -2cm 
\plotone{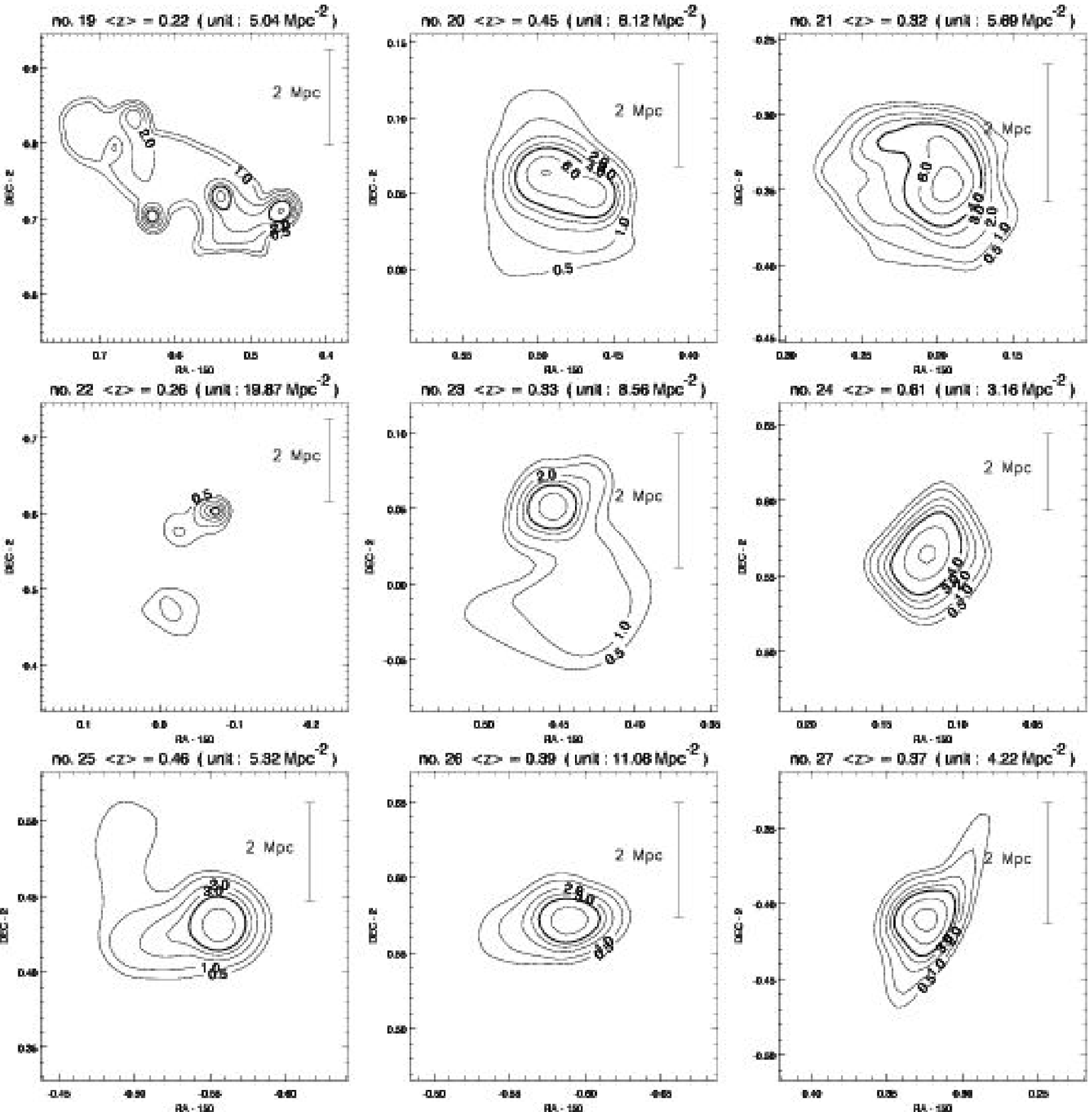}

\centerline{Fig. 4. --- cont'd}

\clearpage
\epsscale{1} 
\vskip -2cm 
\plotone{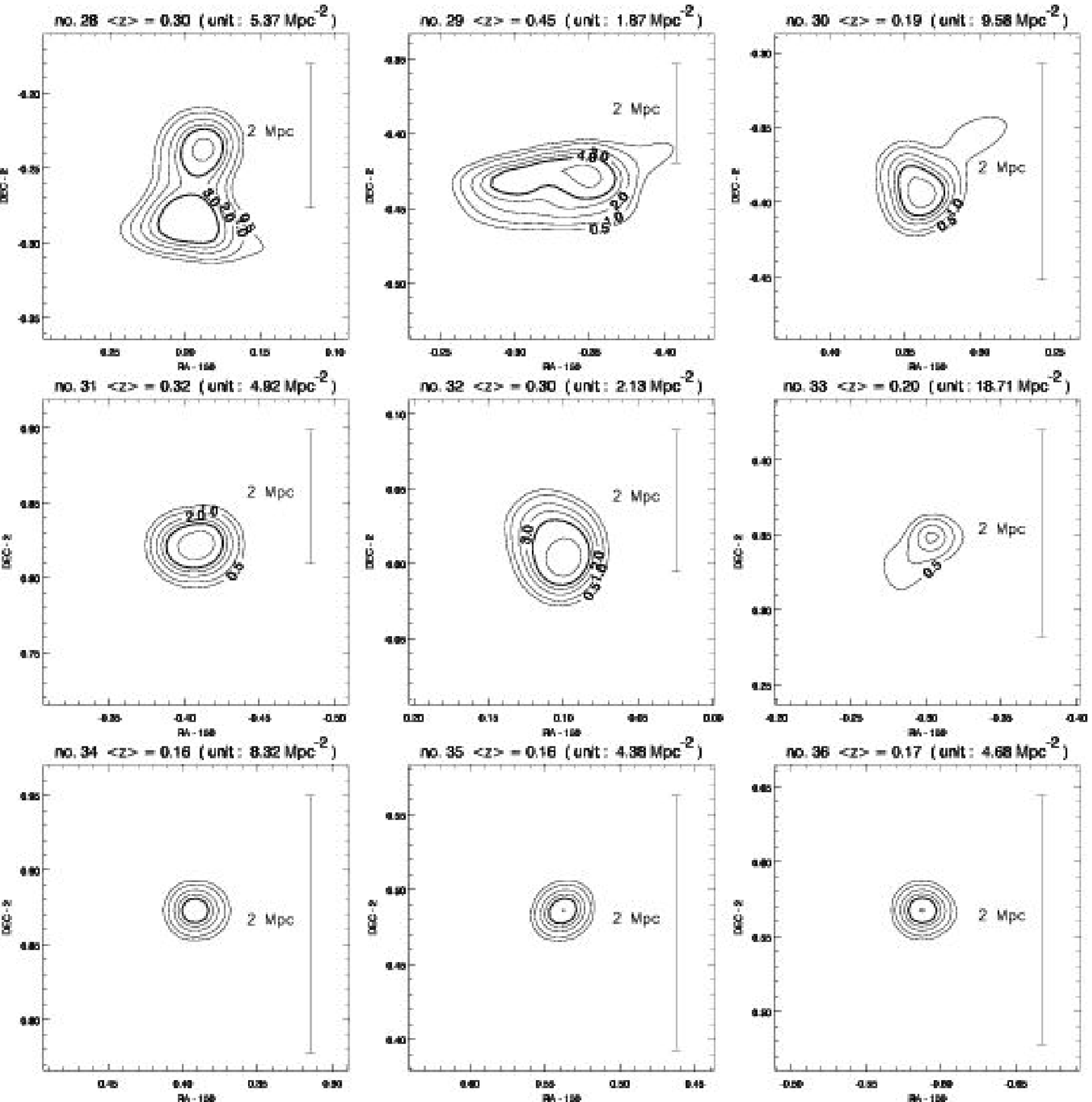}

\centerline{Fig. 4. --- cont'd}

\clearpage
\epsscale{1} 
\vskip -2cm 
\plotone{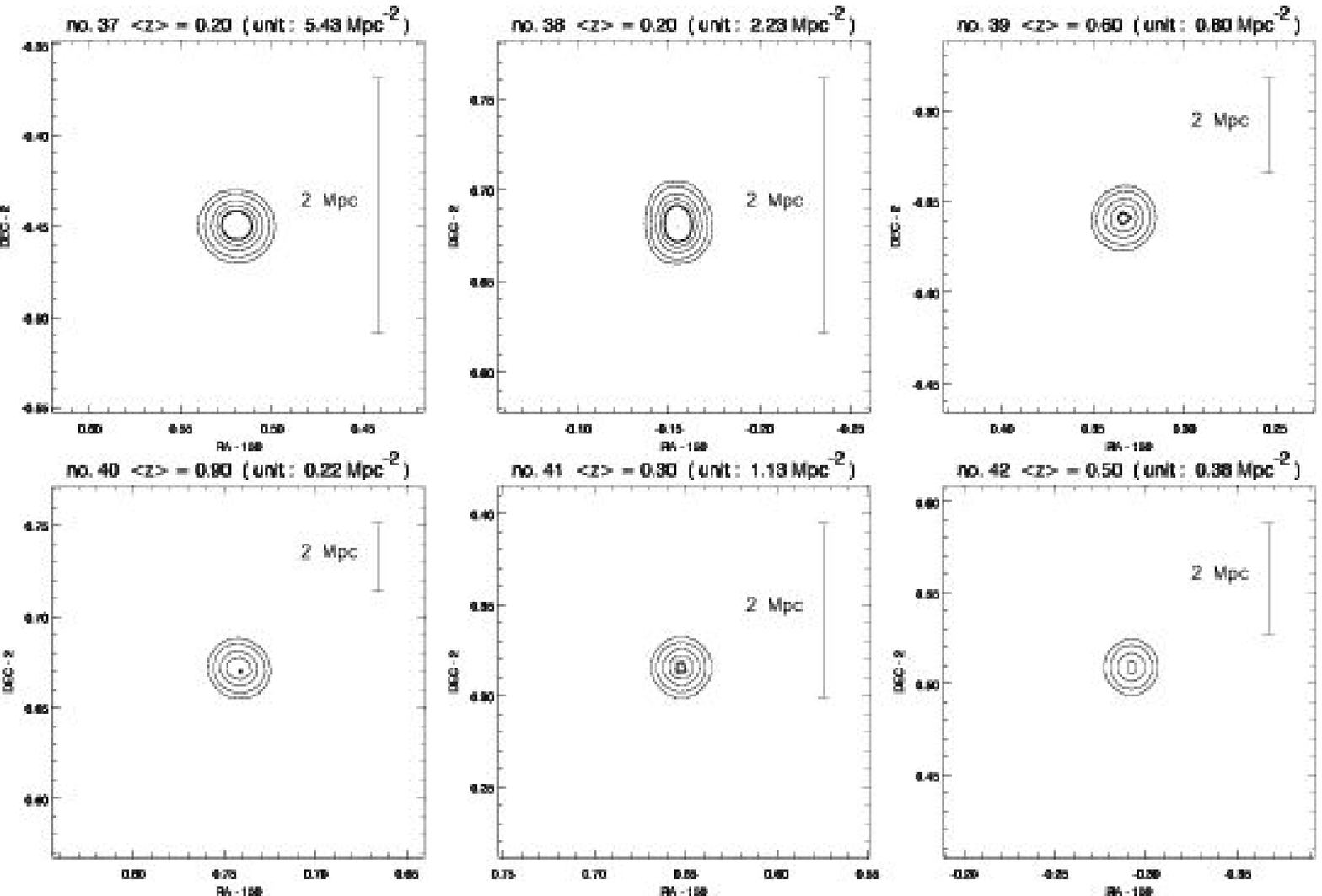}

\centerline{Fig. 4. --- cont'd}

\clearpage

\begin{figure}[ht]
\epsscale{1}
\vskip -1cm 
\plotone{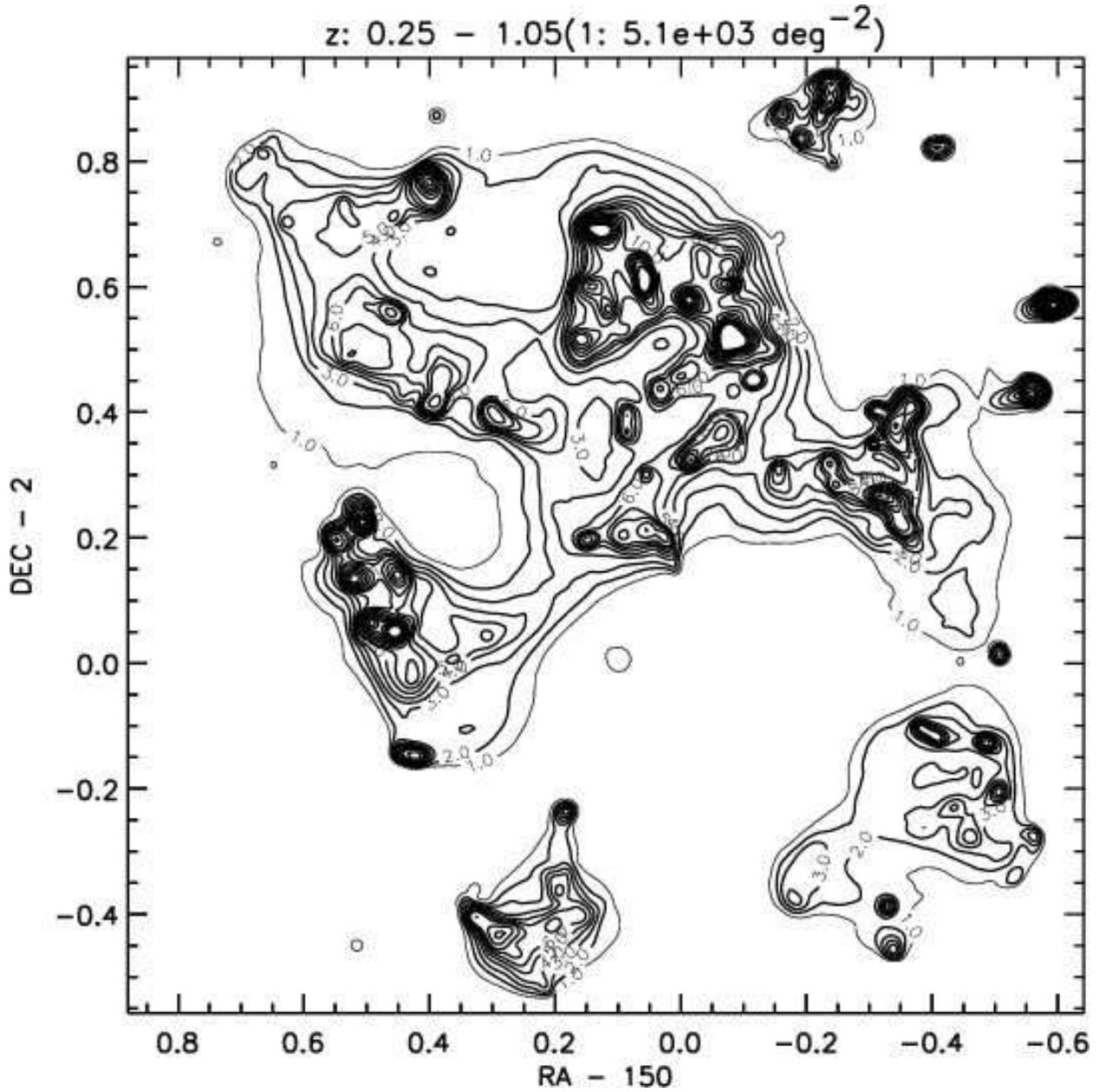}

\caption{The galaxy overdensities derived from the adaptive smoothing results, integrated in z from z = 0.25 to 1.05. The contour units are $5.1\times10^3$ galaxies  deg$^{-2}$
 and the contours  are at 1,2,3,4,5,6,7,8,10,12,14,16,18,20,22 and 24 units.} 
\label{lss_2d}
\end{figure}

\begin{figure}[ht]
\epsscale{1}
\vskip -1cm 
\plotone{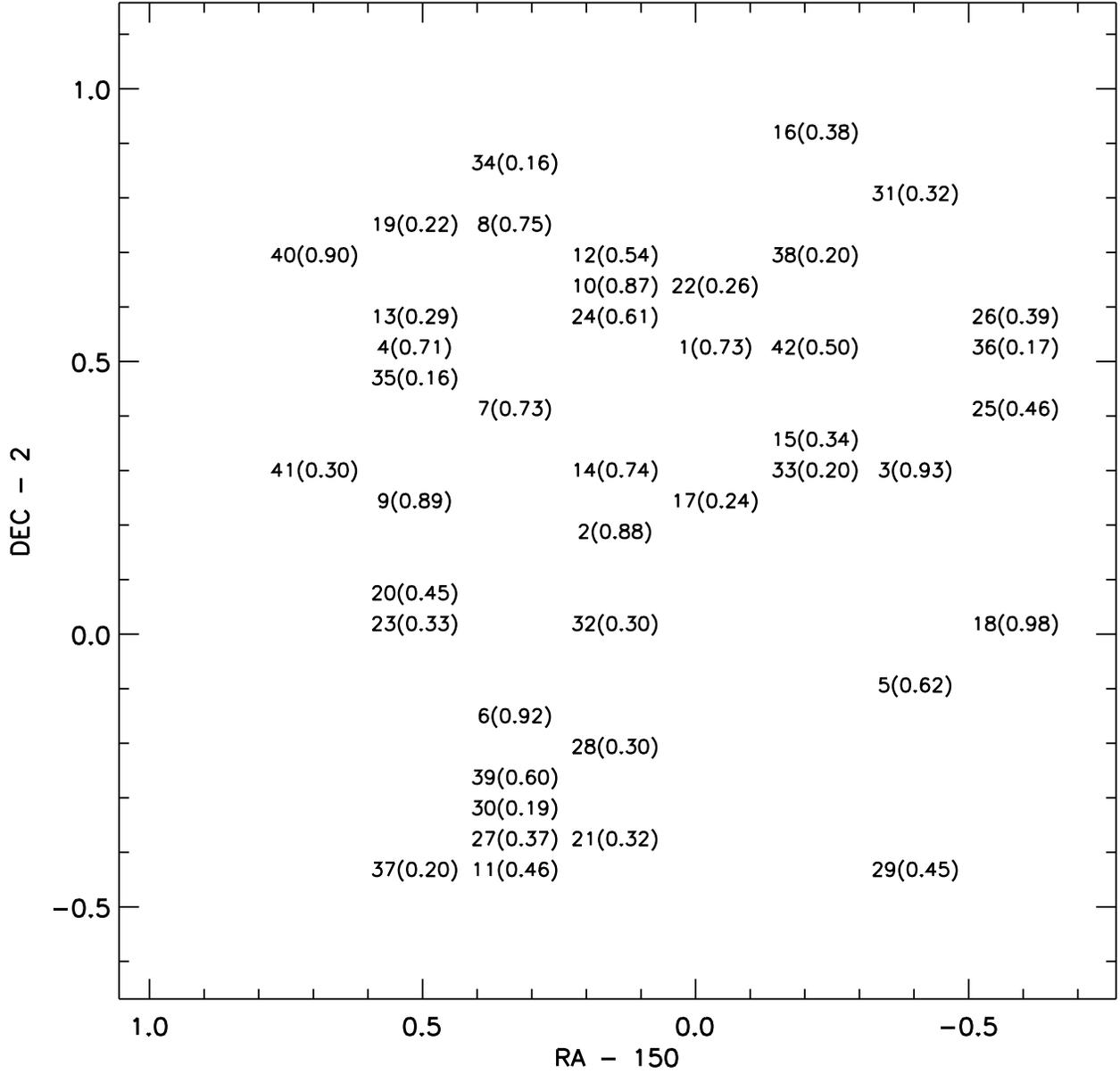}

\caption{The relative locations of each of the LSS are shown with their centroid redshifts.} 
\label{lss_find}

\end{figure}

\begin{figure}[ht]
\epsscale{0.8}
\plotone{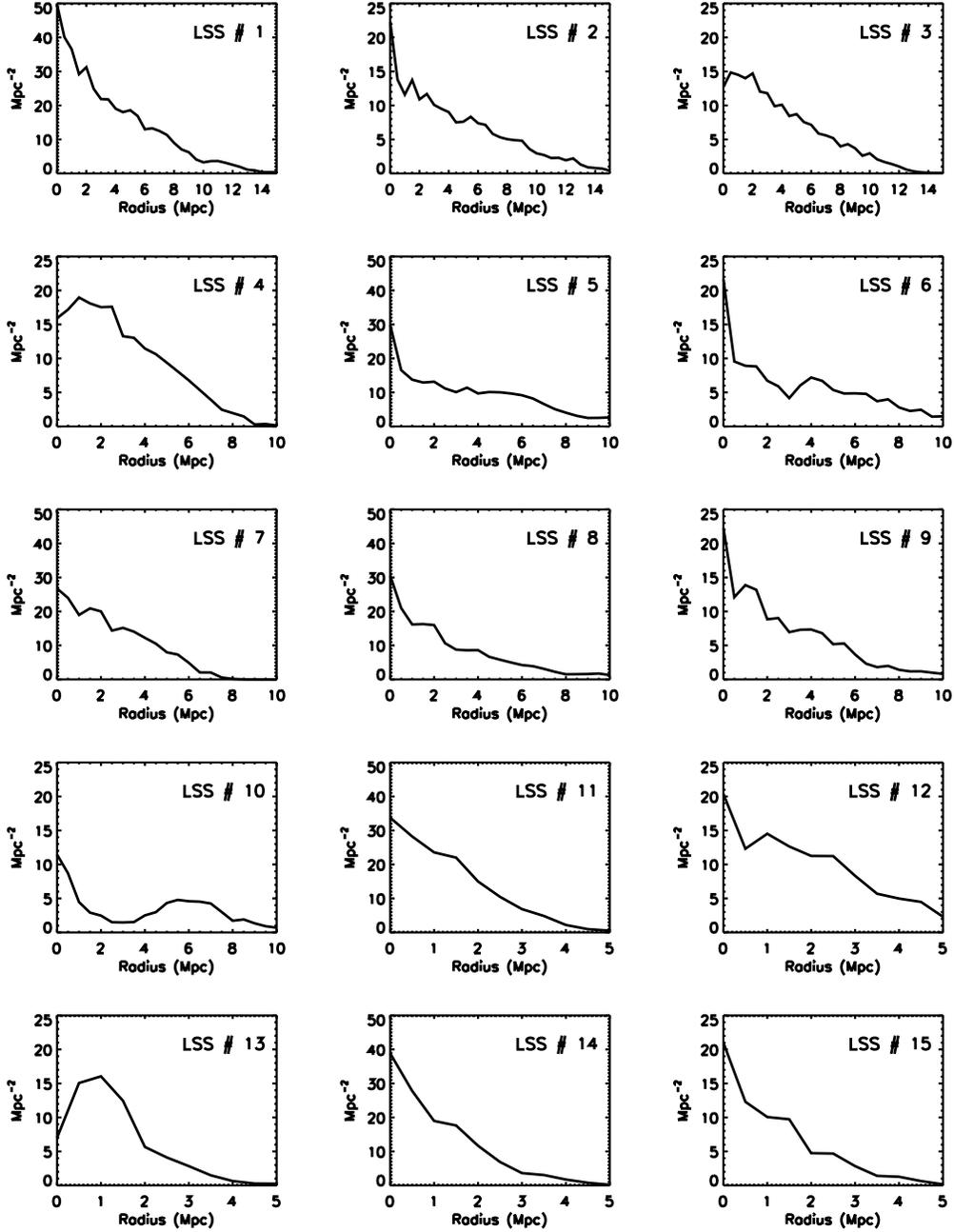}

\caption{The radial distribution of galaxies is shown for each structure using radius determined from the peak position of the number count distribution. The structures are ordered (as before) in decreasing numbers 
of pseudo 3-d space pixels. LSS above \# 26 are not plotted since they have too few galaxies 
for a meaningful radial distribution. In all cases, the
background galaxy distribution is removed. Many of the structures exhibit centrally 
peaked, well-behaved radial profiles, indicating candidate galaxy clusters with 
typical radii 1 -- 2 Mpc. In cases where the LSS has  multiple peaks of comparable amplitude, the radial distribution appears non-monotonic. } 
\label{radial}
\end{figure}

\clearpage
\epsscale{1} 
\plotone{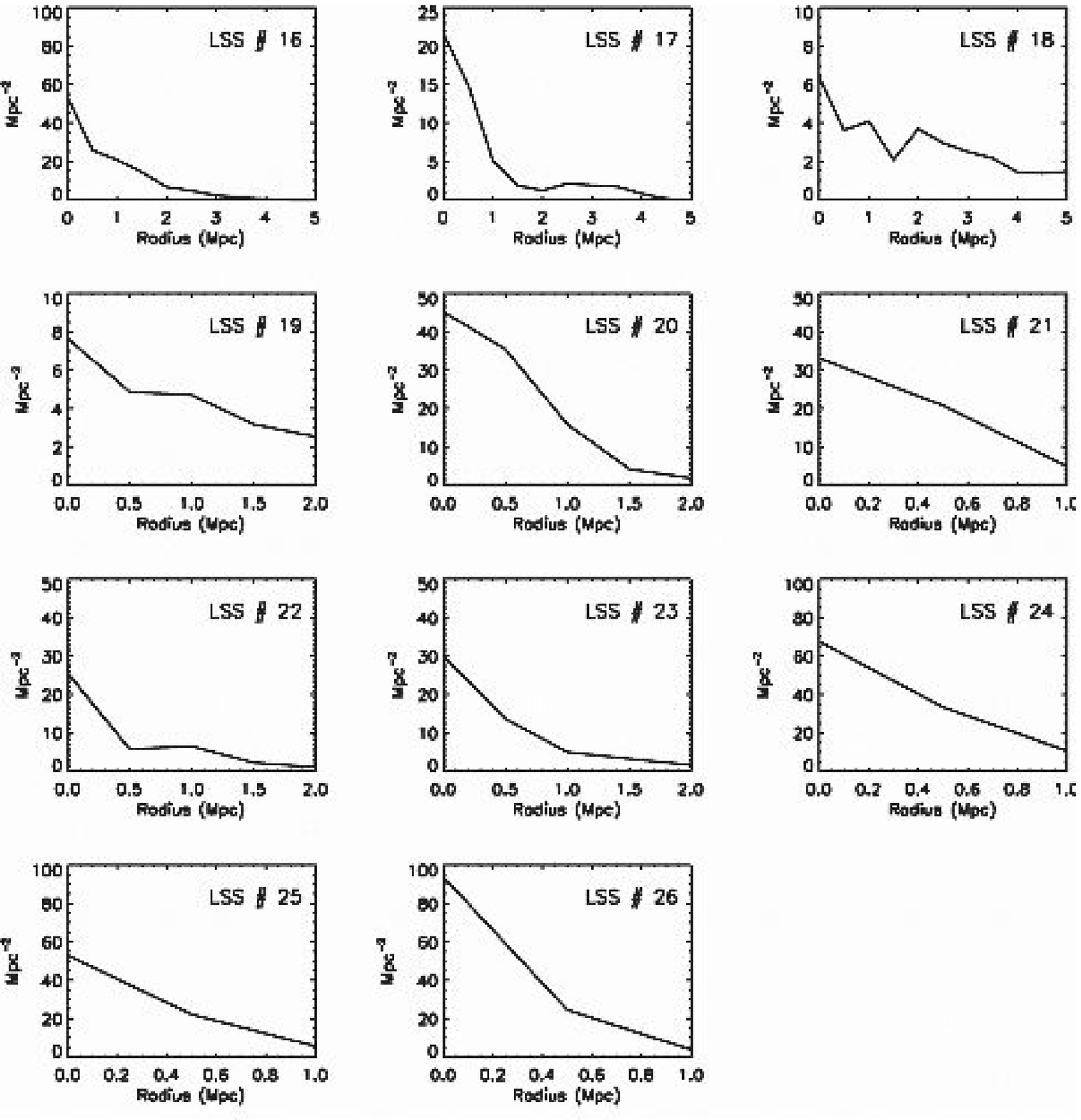}

\centerline{Fig. 7. --- cont'd}
\clearpage
\begin{figure}[ht]
\epsscale{1} 
\plottwo{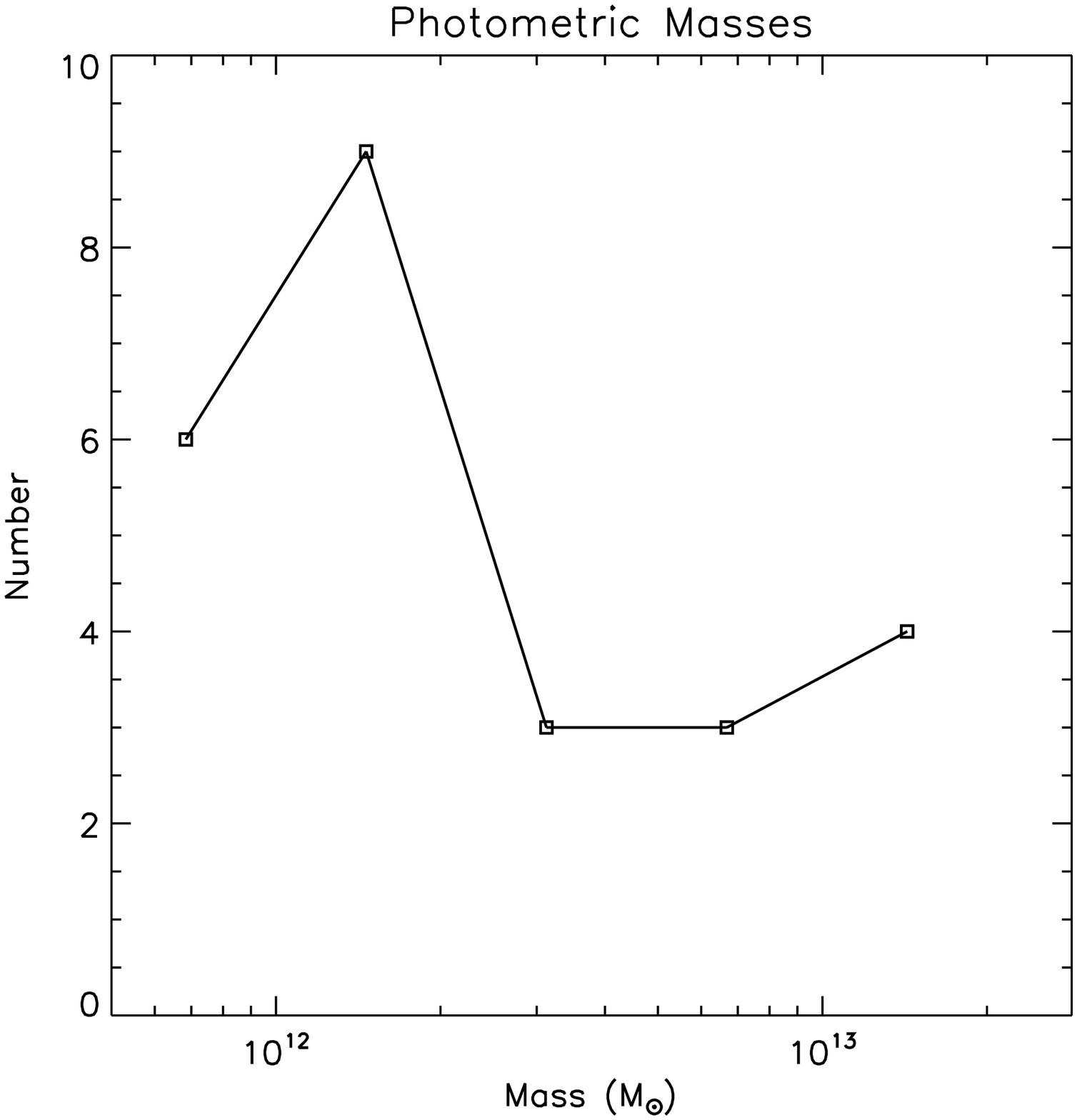}{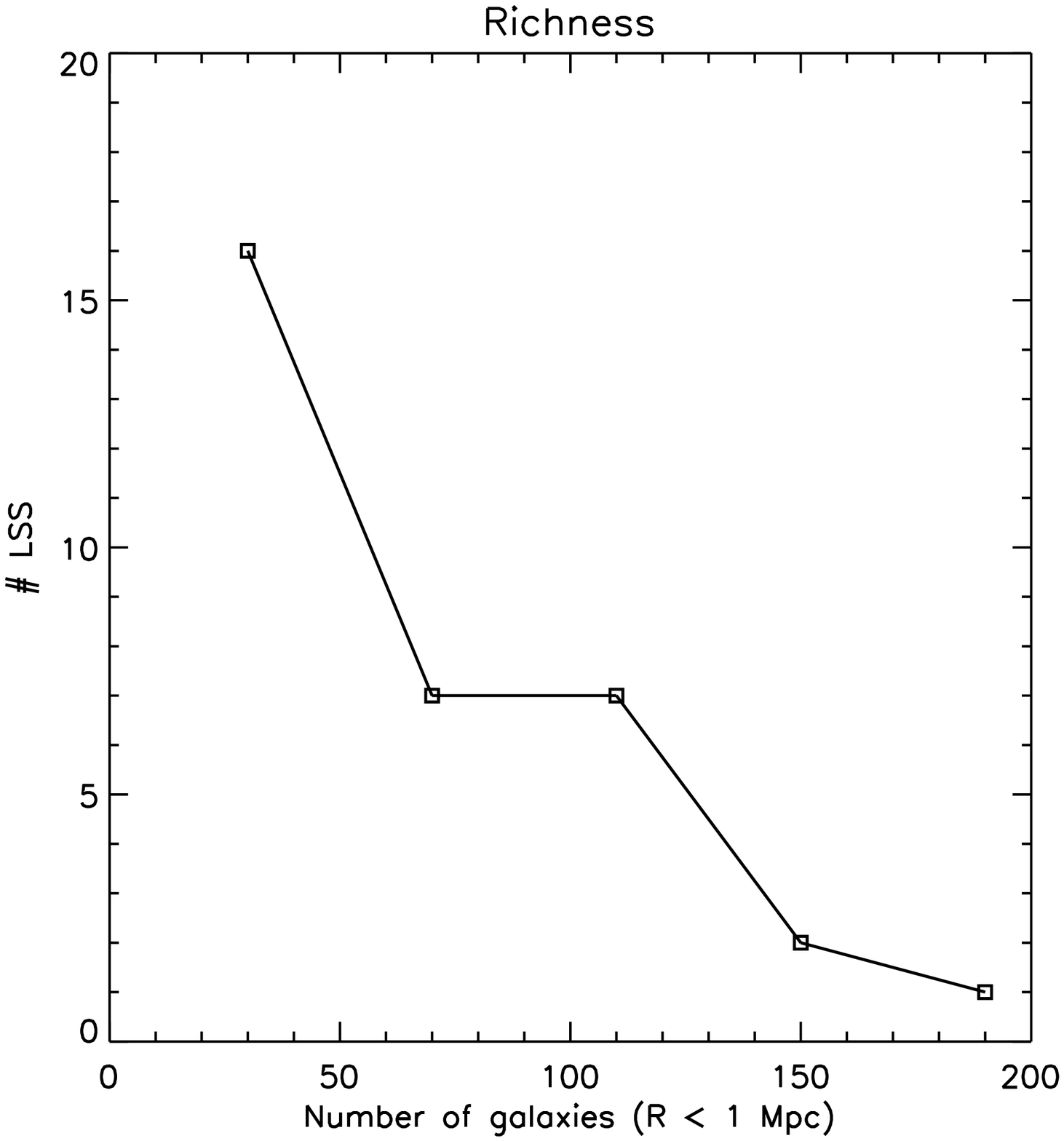}
\caption{a) Left :   The distribution of total stellar masses for the structures are shown for masses 
$5\times10^{11}$ to 10$^{14}$ \msun.  b) Right :   The distribution of Richness measures for the 
the structures is shown, calculated as the surface density of galaxies brighter than   
2 mag below the 3rd brightest galaxy (see text). Specifically, Richness is defined as the number of galaxies within 
the central R $\leq$ 1 Mpc brighter than 2 magnitudes fainter than
the 3rd brightest galaxy within the cluster (i.e. M$_V < $ M$_V$(3rd) + 2 ). Radius is measured from the 
location of peak surface density as in Figure \ref{radial}.} 
\label{lss_mass}
\end{figure}

\begin{figure}[ht]
\epsscale{1} 
\plotone{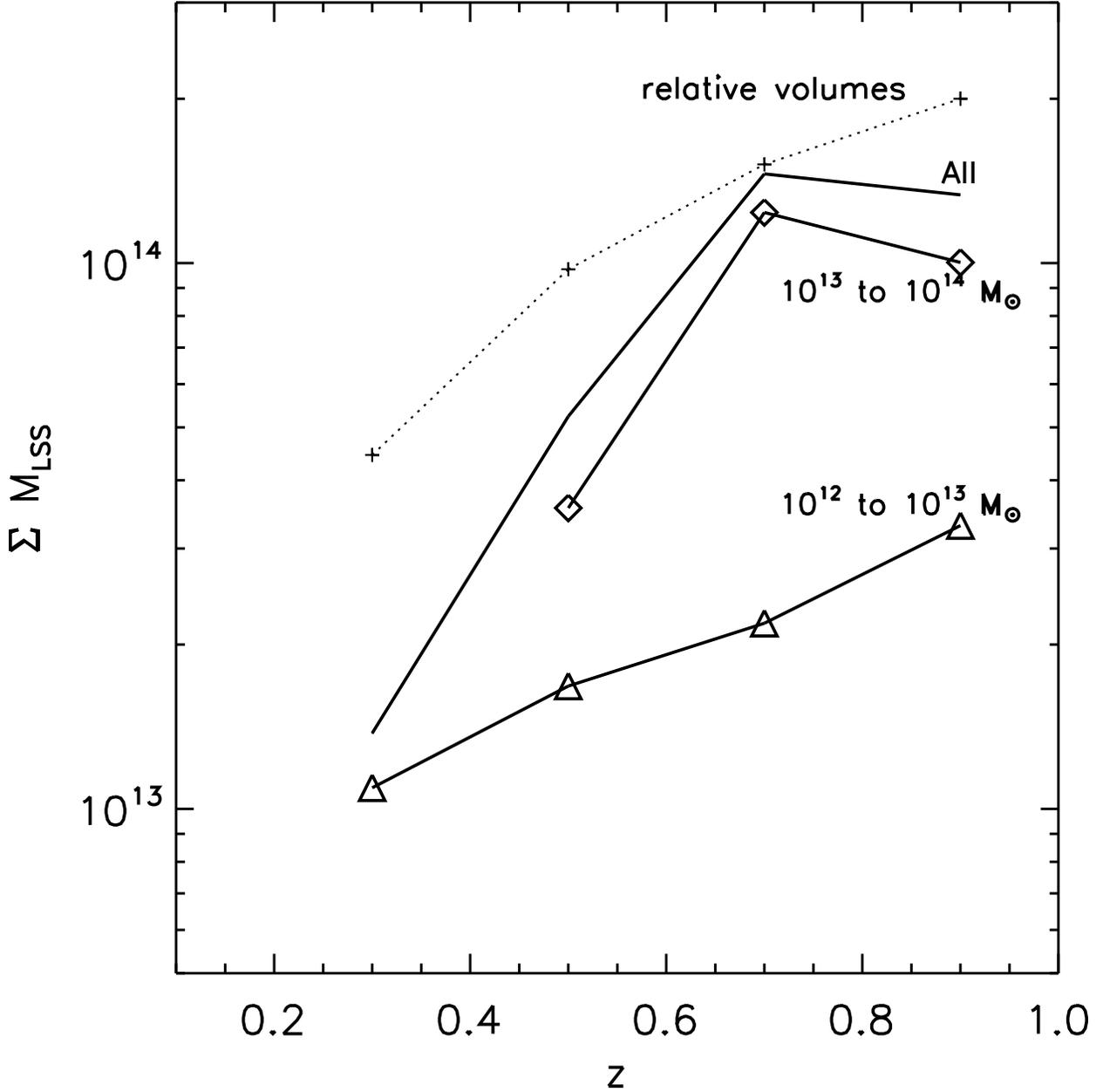}
\caption{The redshift distributions of summed structure masses are shown 
for redshift bins $\Delta z = 0.25$. Uncertainties due to Poisson and cosmic variance are comparable and the 
expected total {\it relative} variance in these number distributions is 40-60\% (i.e. $\sigma_N / N \sim 0.5$; see Section \ref{var} and Table \ref{variance}). The dotted line plots the relative comoving volumes for the 
selected redshift bins, with vertical scale arbitrarily normalized -- simply to indicate that the overall distribution of structure
mass with redshift is consistent with the expected mass conservation. The apparent discrepancy in the lowest redshift 
bin is probably not significant given the very large variances noted above. } 
\label{lss_mass_z}
\end{figure}

\begin{figure}[ht]
\epsscale{0.8} 
\plotone{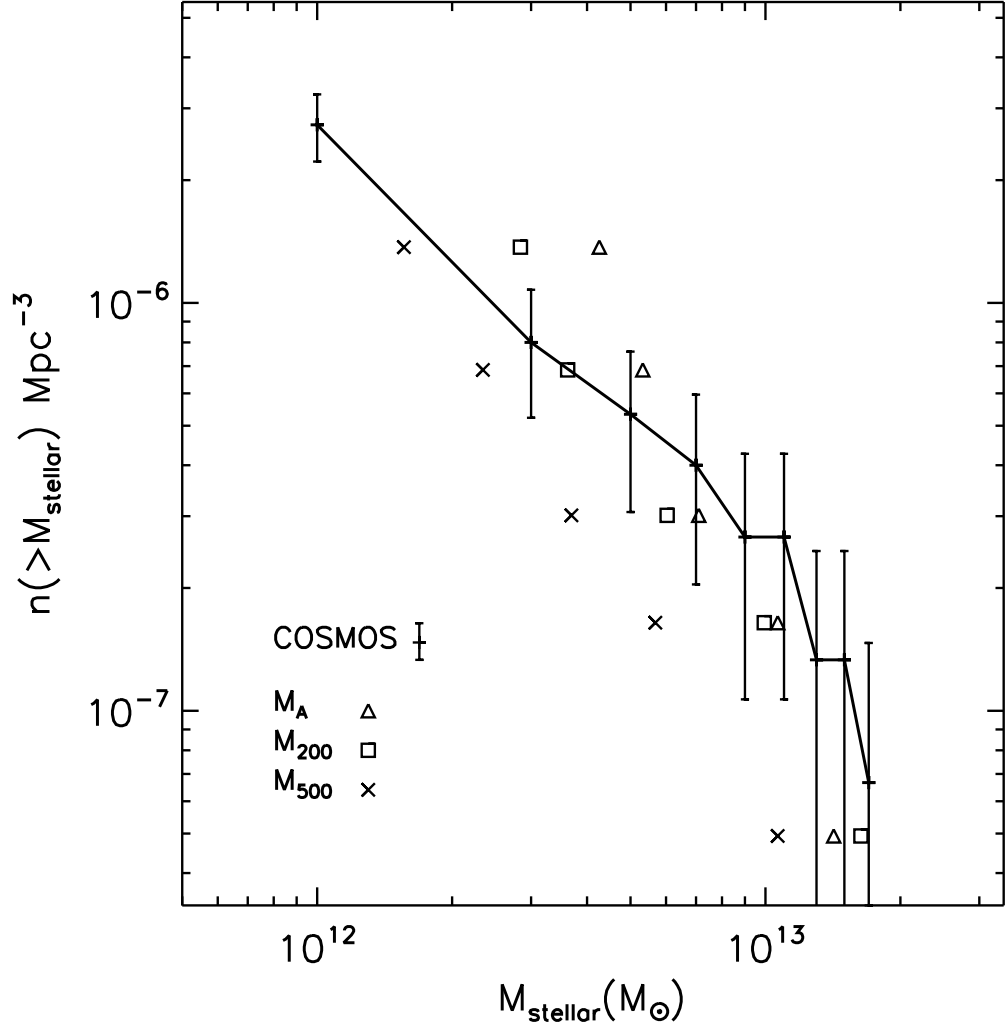}

\caption{The cumulative mass function ($n(>M) =  \int_M^{\infty} n(m) \,dm $) is shown for the 42 COSMOS LSS with linear mass bins 
of width $\Delta M_{*} = 2\times 10^{12}$ \msun. Also shown are the cumulative mass functions derived from optical and X-ray 
studies as summarized by \citep[][]{rei02}. The mass function for masses within the Abell radius of each 
cluster is shown by the triangle symbols (M$_A$); masses within the regions with density exceeding 200$\rho_c$ and 500$\rho_c$ (where 
$\rho_c$ is the critical density for the universe) are 
shown by the squares and crosses \citep[see][]{rei02}. The expected cosmic variance and Poisson noise 
are included in the error bars (see Section \ref{var} and Table \ref{variance}).
} 
\label{num_mass_cum}
\end{figure}

\begin{figure}[ht]
\epsscale{1} 
\plotone{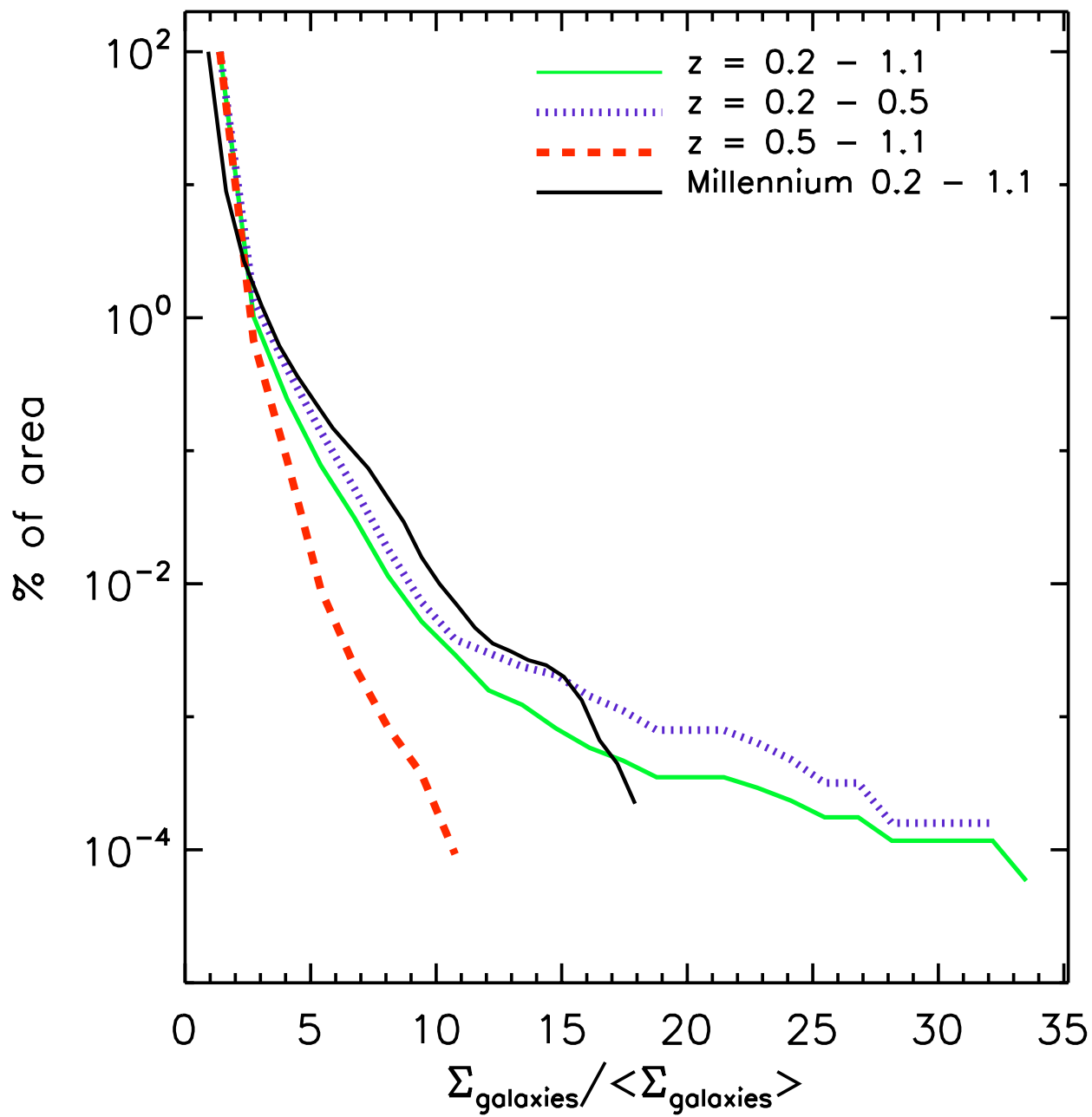}
\caption{The cumulative fraction of survey area with surface number density $\Sigma$ relative to  the average at 
each redshift ($<\Sigma>$) is shown for all redshifts and for low and high redshift ranges (colored curves). 
For comparison, the black line shows the average at z = 0.2 to 1.1 obtained from the Millennium Simulation (see text).  Jackknife tests, splitting the datasets in half, showed typical variances 
$< 20$\%. } 
\label{dens_avedens}
\end{figure}

\begin{figure}[ht]
\epsscale{.9} 
\plotone{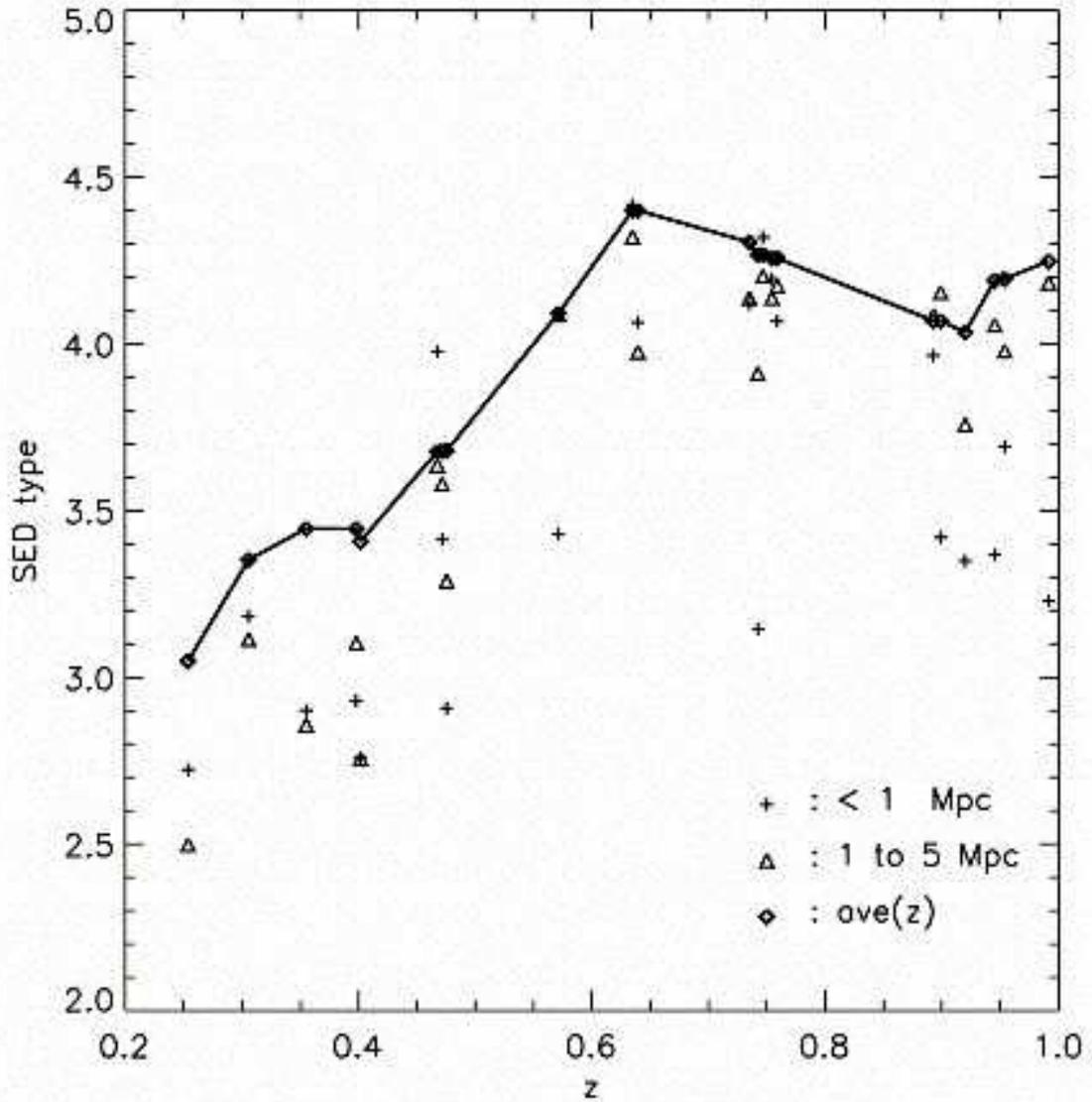}

\caption{For each structure, the mean SED types of galaxies at projected R $< 1$Mpc  and R = 1 to 5 Mpc are compared (solid line) with the mean type as a function of z (in a redshift bin 
$\Delta z = 0.2 $ centered on the same redshift).  One set of points is shown for each of the 
 structures plotted at the mean redshift of each structure. Note the very pronounced trend 
 for the central Mpc in the structures to have earlier mean SED type (low type) for the galaxies. } 
\label{mass_types}
\end{figure}

\begin{figure}[ht]
\vskip -0.5cm
\epsscale{.9} 
\plottwo{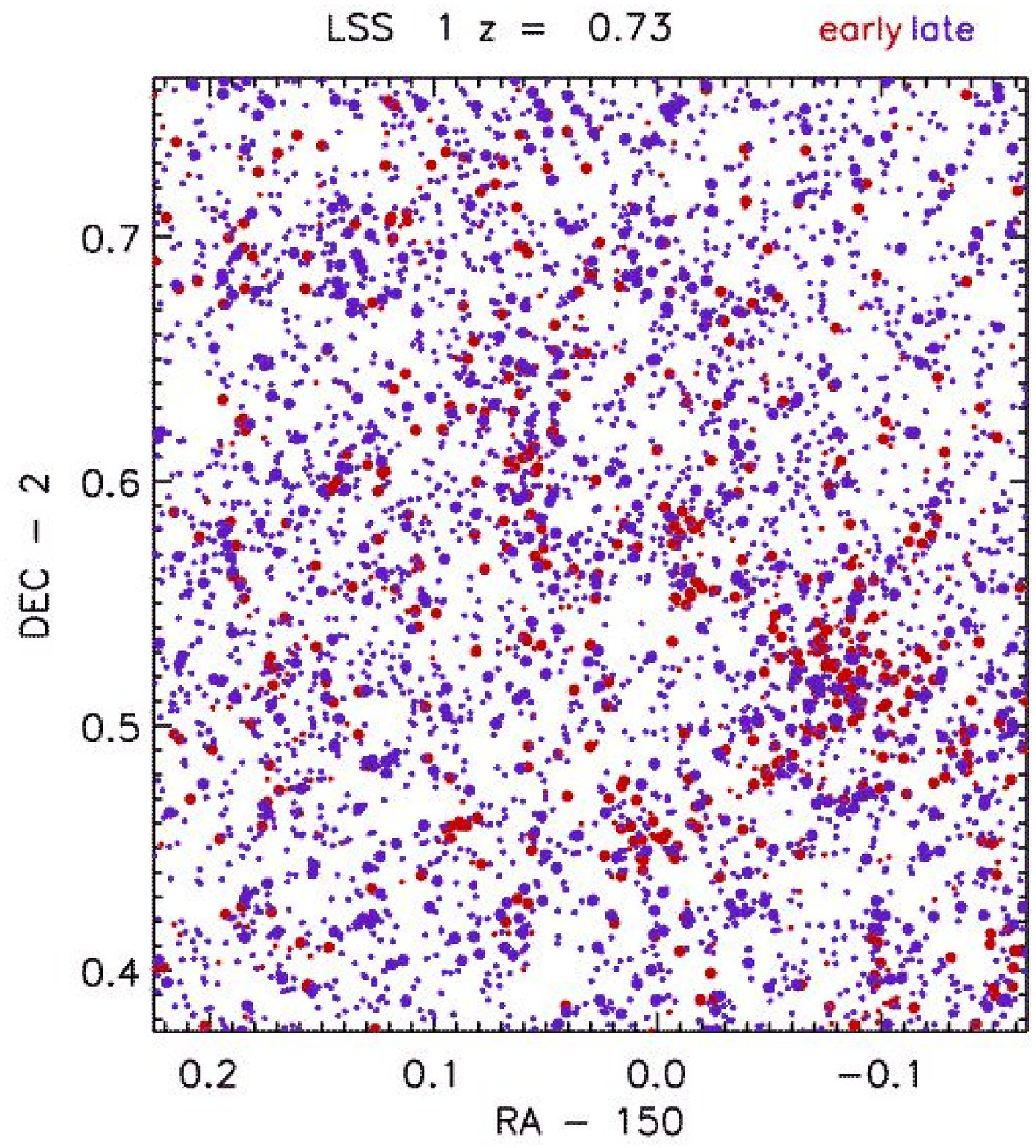} {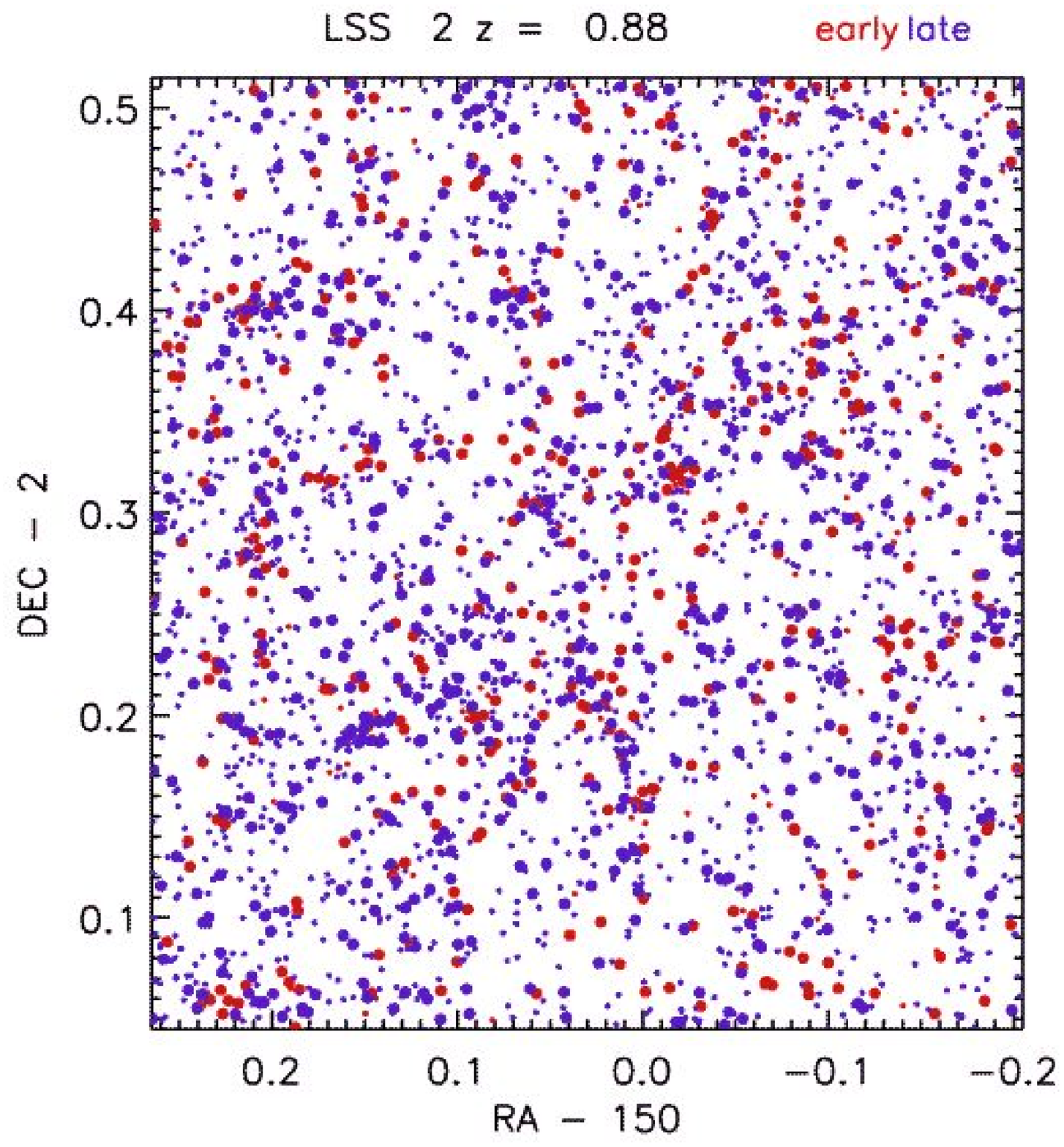}

\vskip 0cm
\plottwo{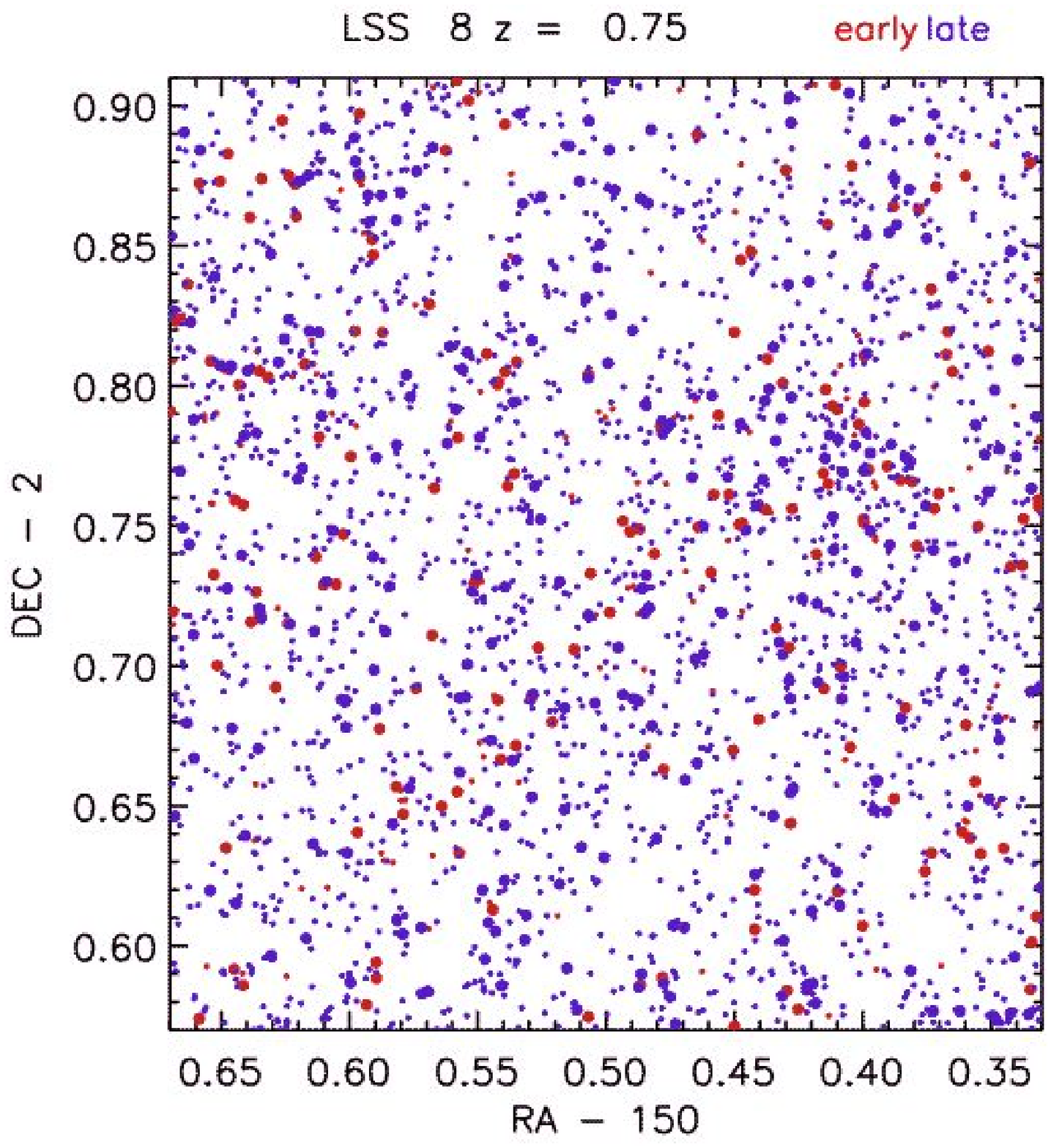} {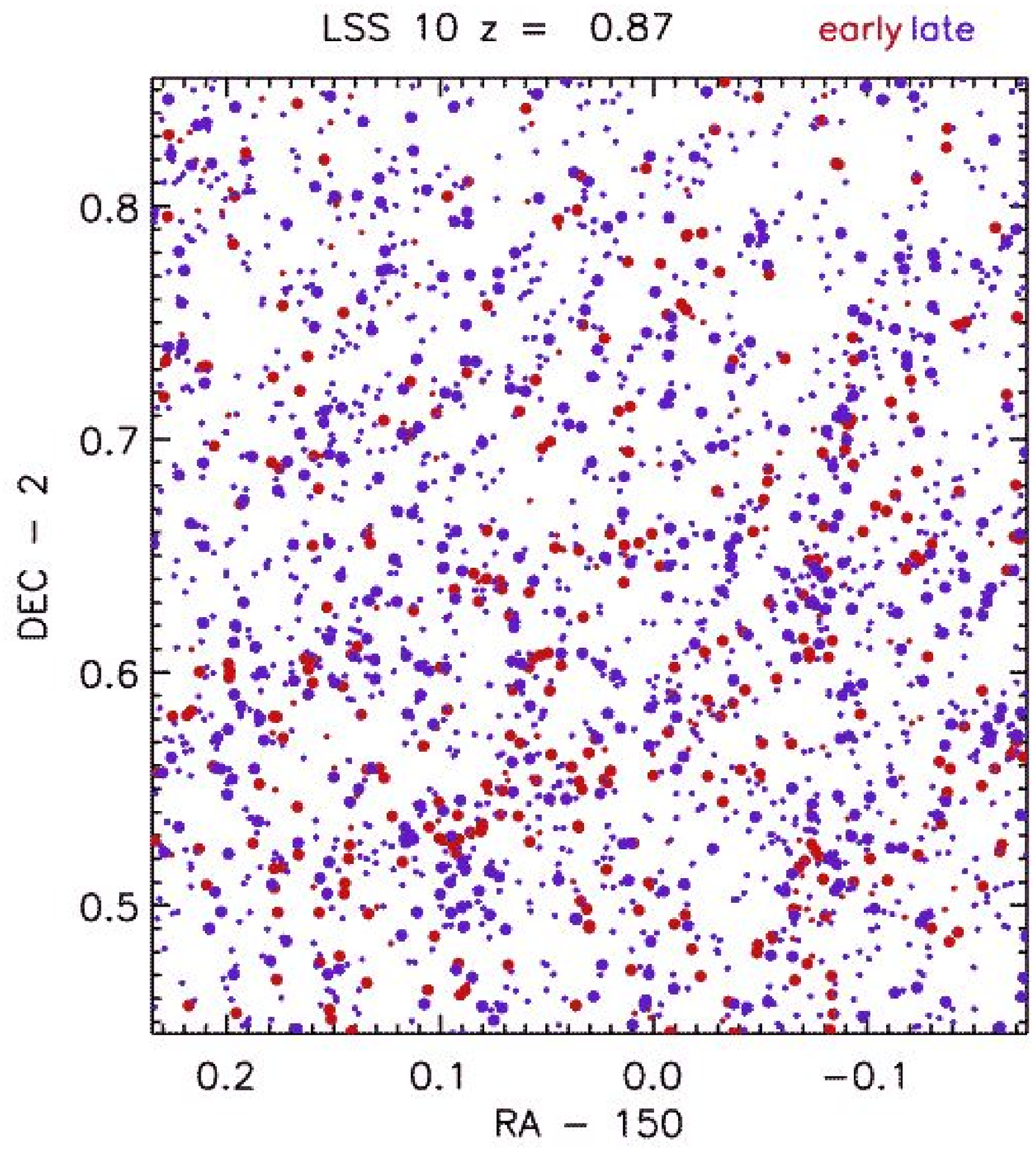}

\vskip 0cm
\plottwo{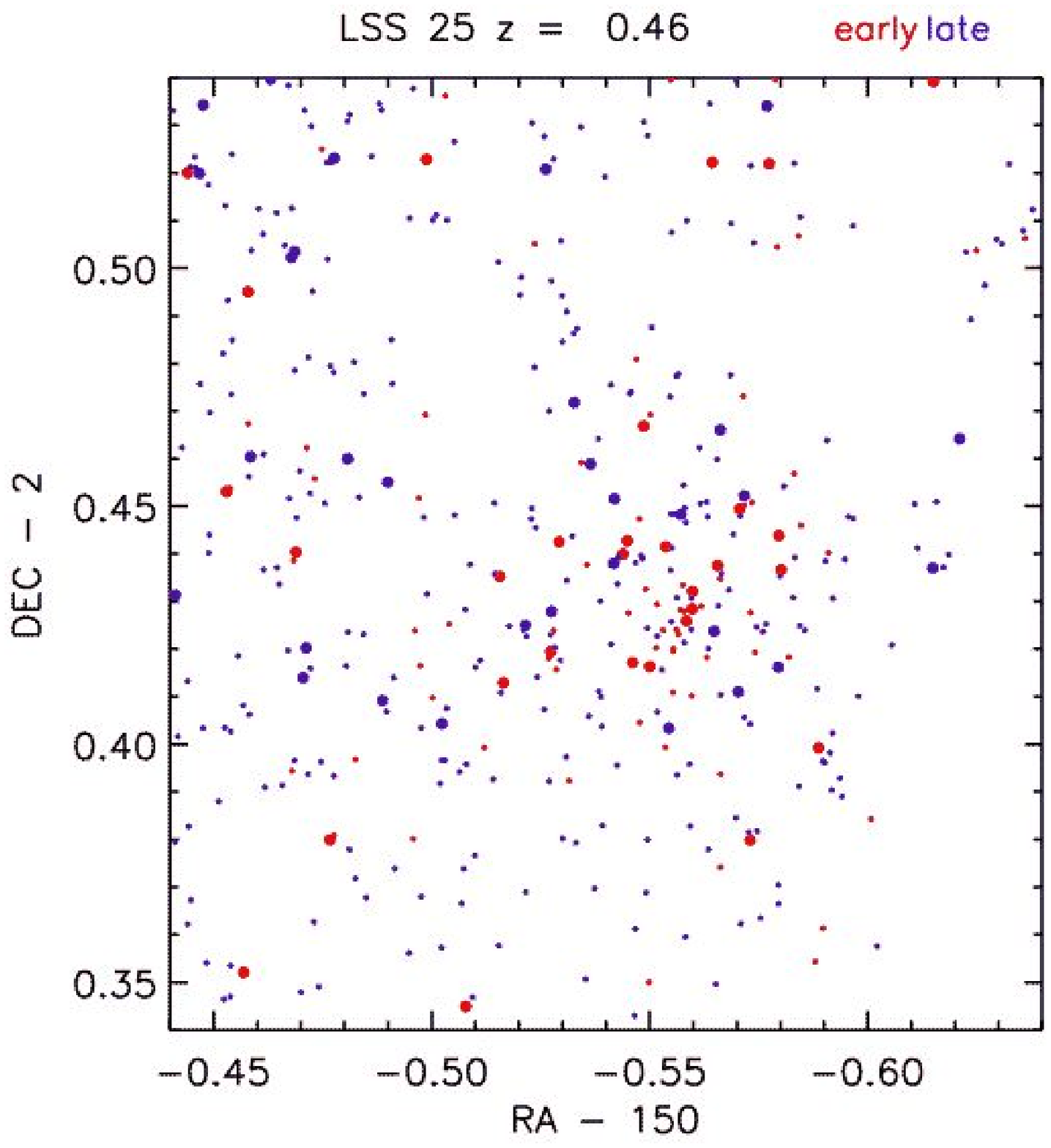} {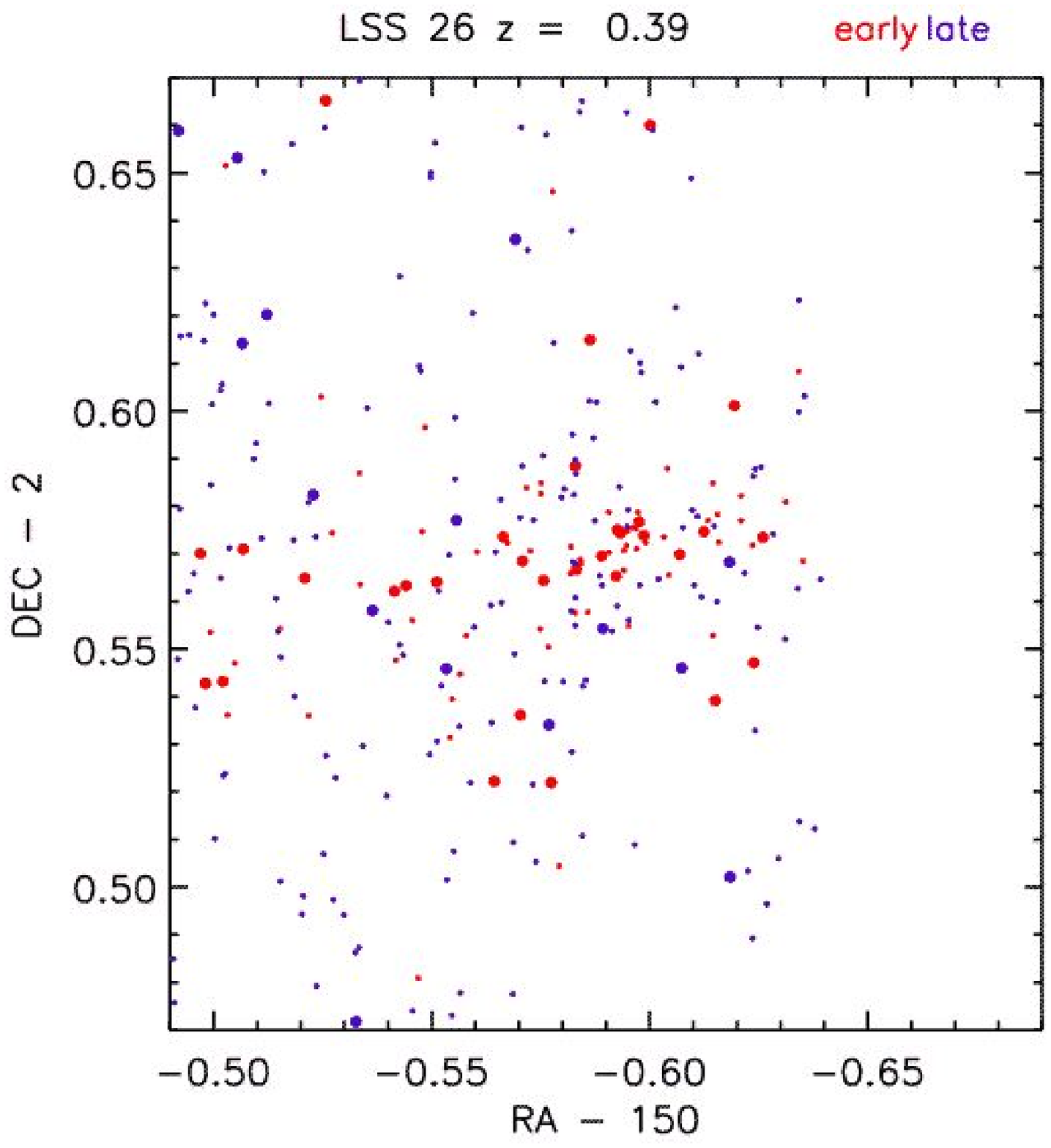}

\caption{Galaxy distributions over the areas of 6 representative LSS (ranging from the most massive 
to a quite low mass structure). The red points indicate early type (SED $< 2.5$)
and the blue point later types. The larger dots are galaxies with M$_V < -21$ mag.
 The galaxy points plotted in the figures have 
their redshifts within the $\Delta z$ range given in Table \ref{struct_param} for each structure. The blank 
areas are either at the edge of the field or where stellar masking occurs. } 
\label{samp}
\end{figure}

 \begin{figure}[ht]
\epsscale{1.} 
\plottwo{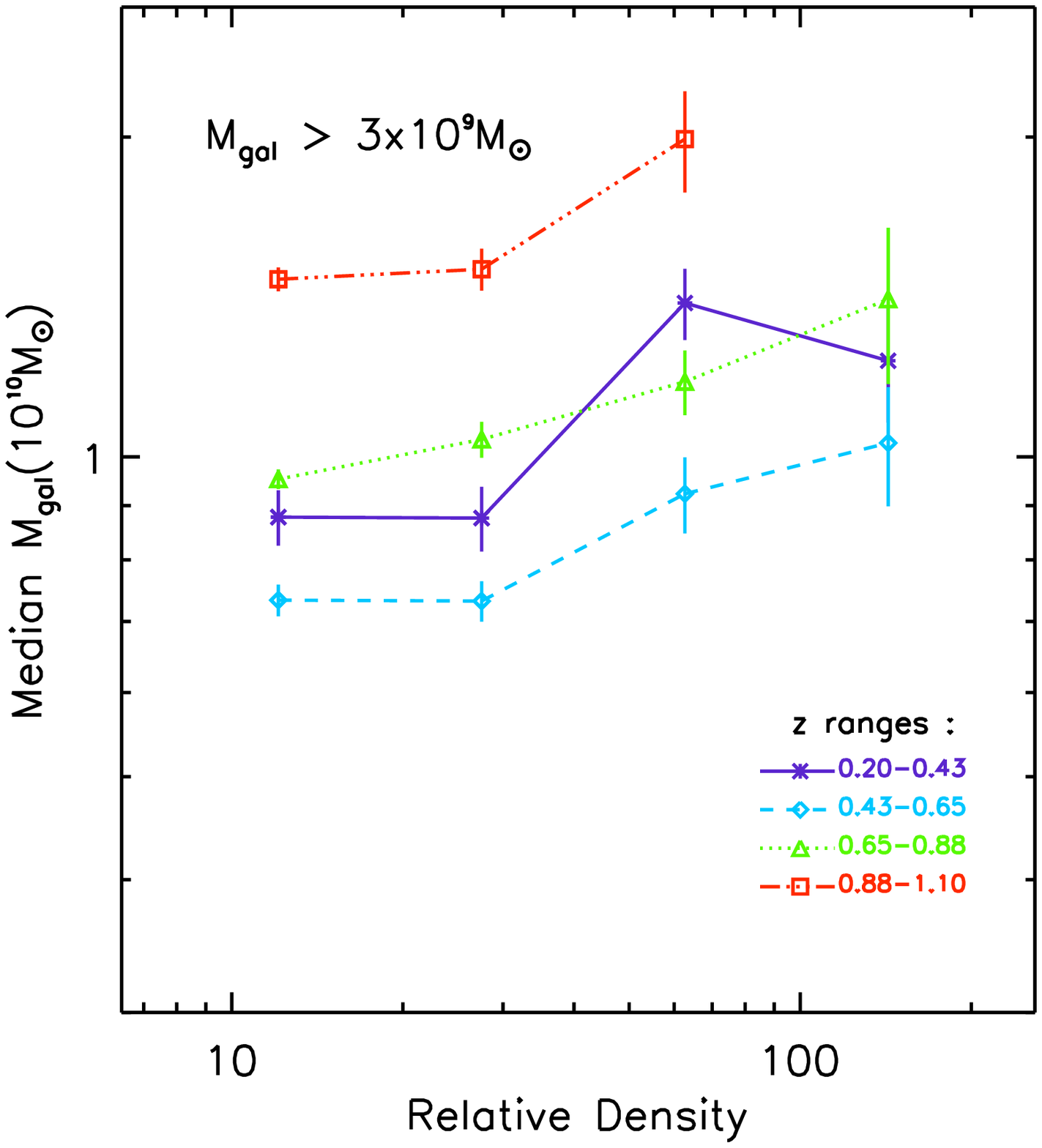}{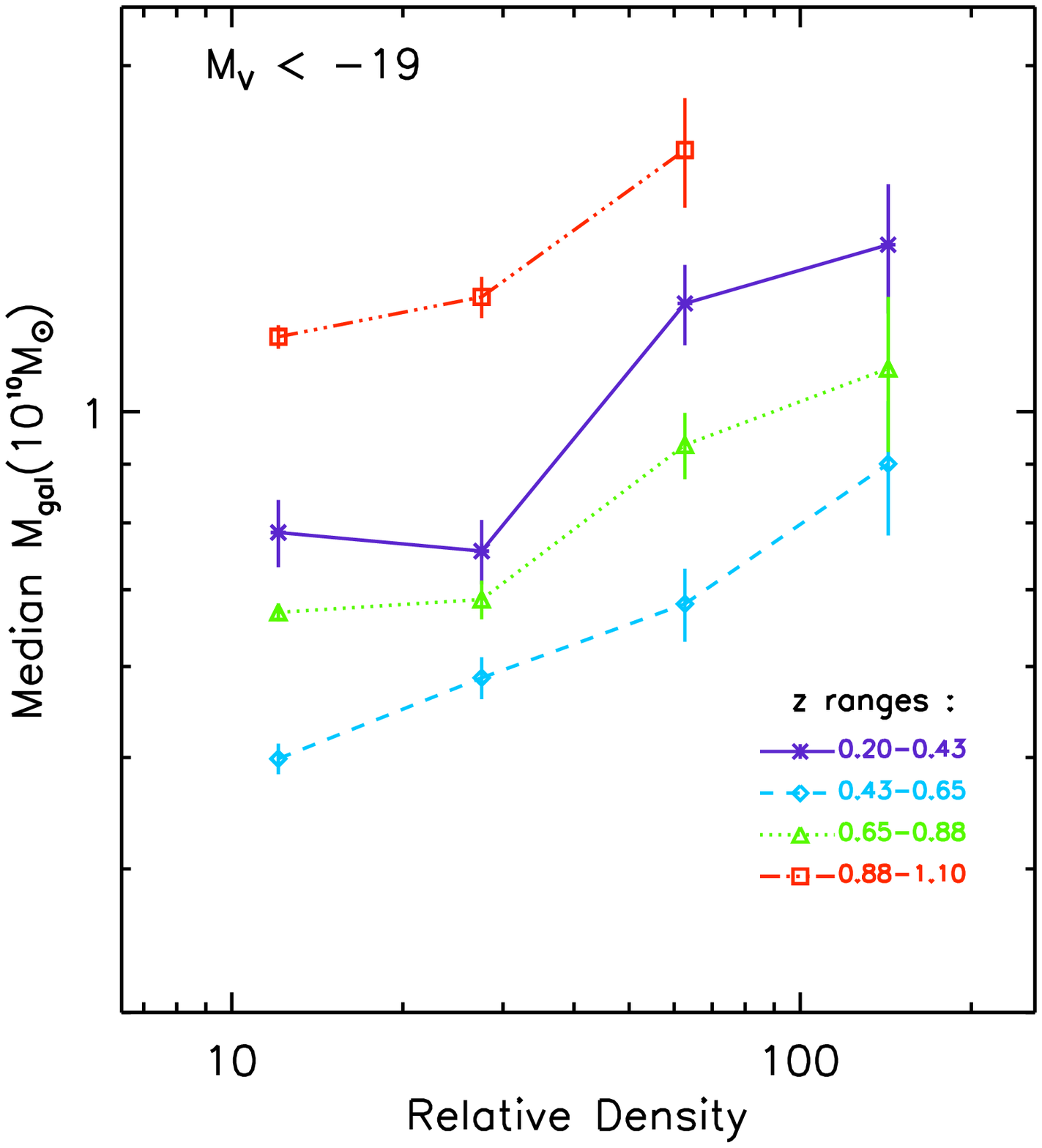}

\caption{The median stellar mass (M$_*$) is shown for galaxies 
binned in redshift (the curves) as a function of environmental density. Two samples are
shown : M$_V < -19$ mag (left panel) and  M$_* > 3\times10^9$\msun (right panel). (The densities have 
been normalized separately at each redshift for these plots (Equation \ref{eqrho}); however, above 
z $\sim 0.25$ the normalizations are similar and a relative 
density of 10 corresponds to a surface densities $\sim 100$ Mpc$^{-2}$ -- see text.)} 
\label{mass_dens}
\end{figure}

 \begin{figure}[ht]
\epsscale{1.} 
\plottwo{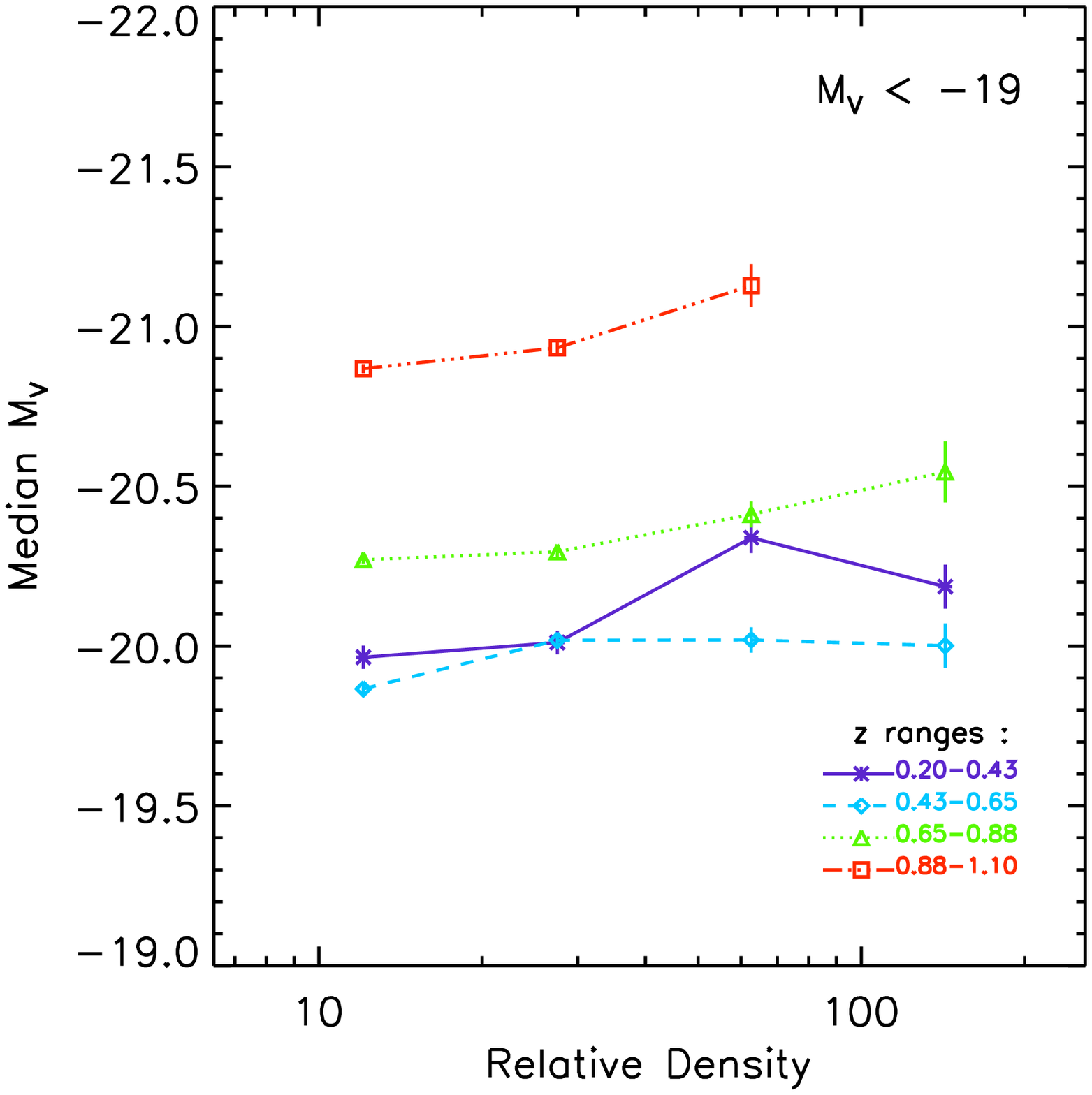}{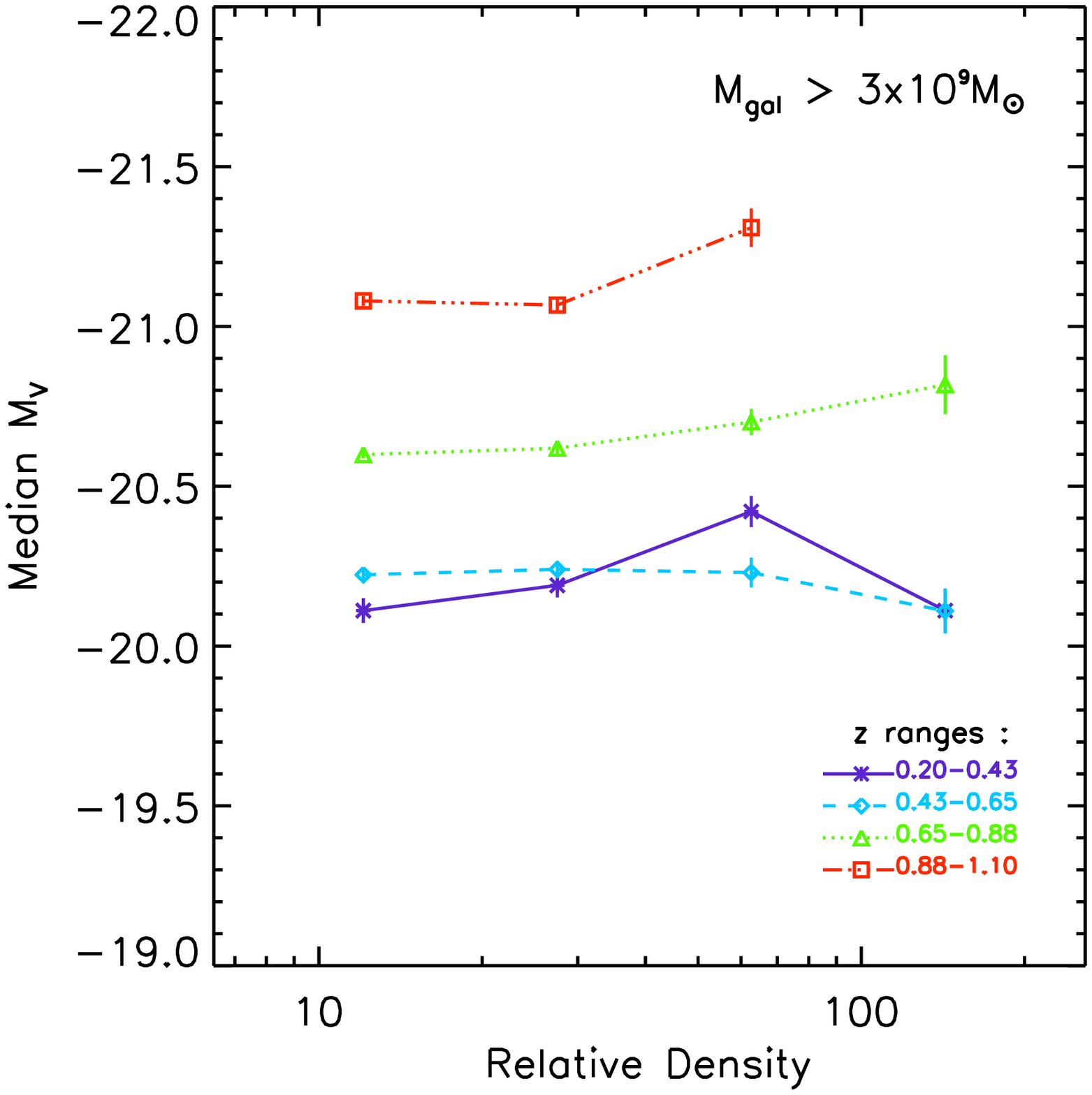}

\caption{The median absolute magnitude (M$_V$) is shown for galaxies 
binned in redshift (the curves) as a function of environmental density. Two samples are
shown : M$_V < -19$ mag (left panel) and  M$_* > 3\times10^9$\msun (right panel).} 
\label{mv_dens}
\end{figure}

 \begin{figure}[ht]
\epsscale{1.} 
\plottwo{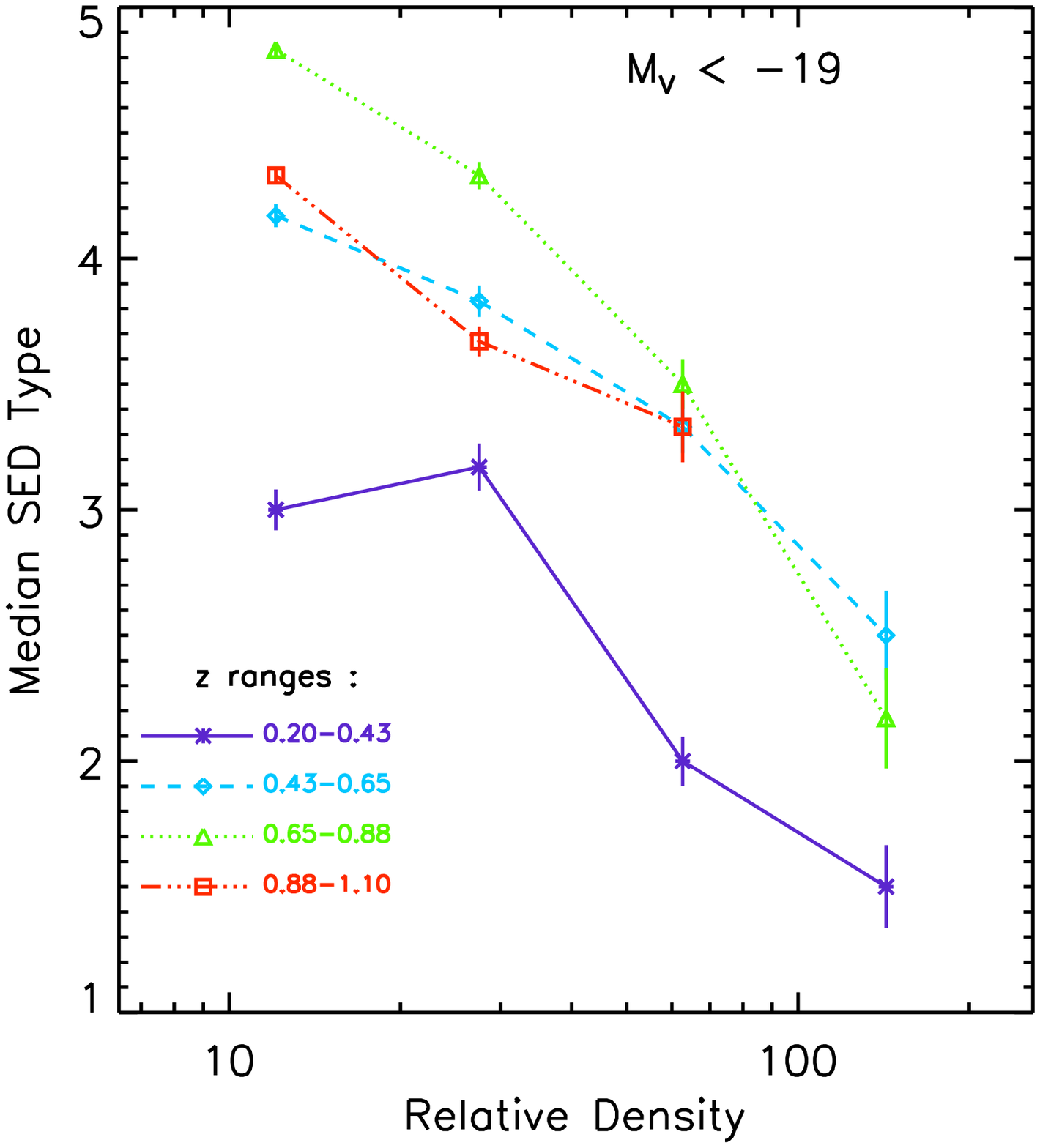}{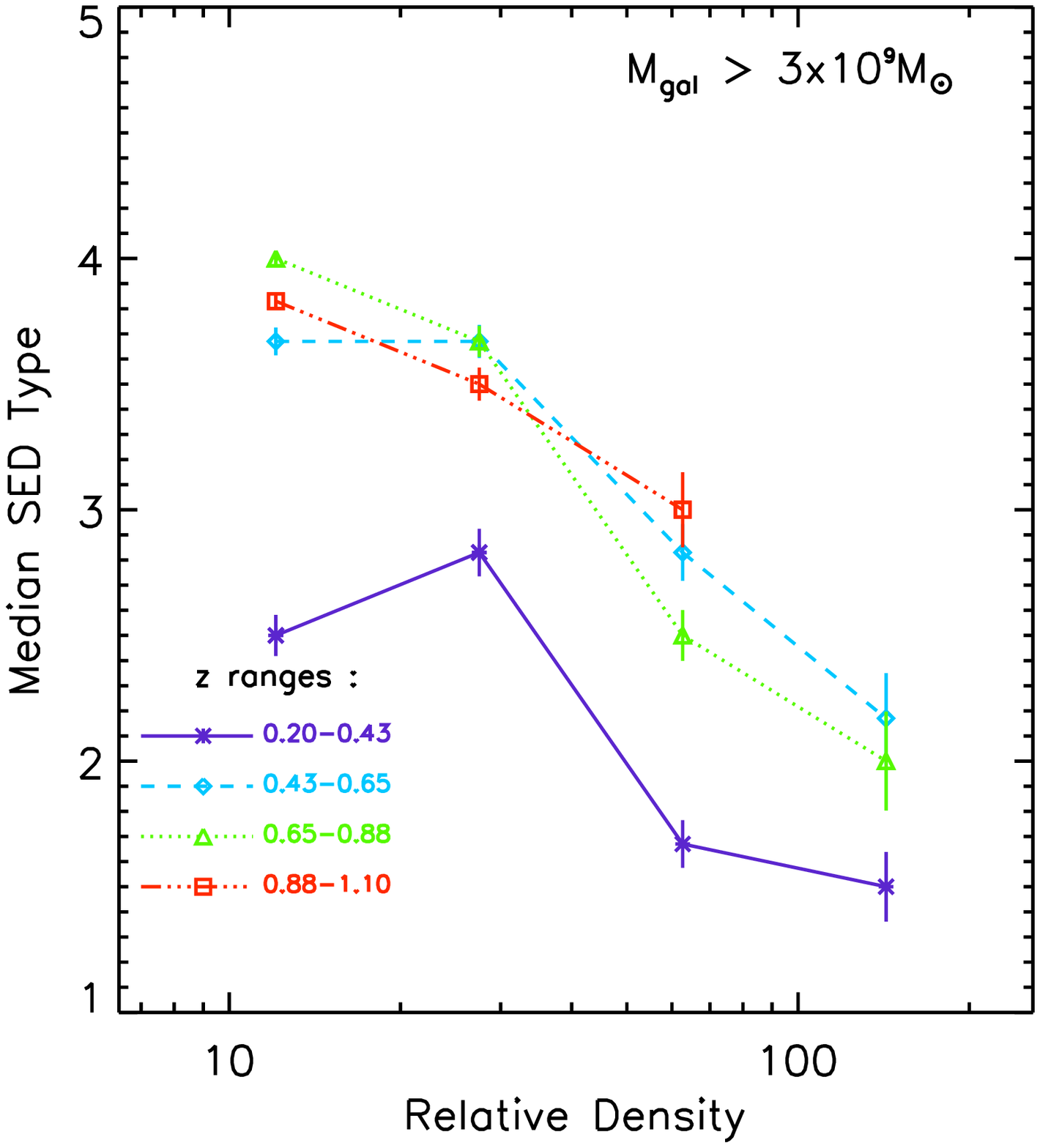}
\caption{The median spectral type (SED) is shown for galaxies 
binned in redshift (the curves) as a function of environmental density. Two samples are
shown : M$_V < -19$ mag (left panel) and  M$_* > 3\times10^9$\msun (right panel).} 
\label{sed_dens}
\end{figure}

 \begin{figure}[ht]
\epsscale{1.} 
\plottwo{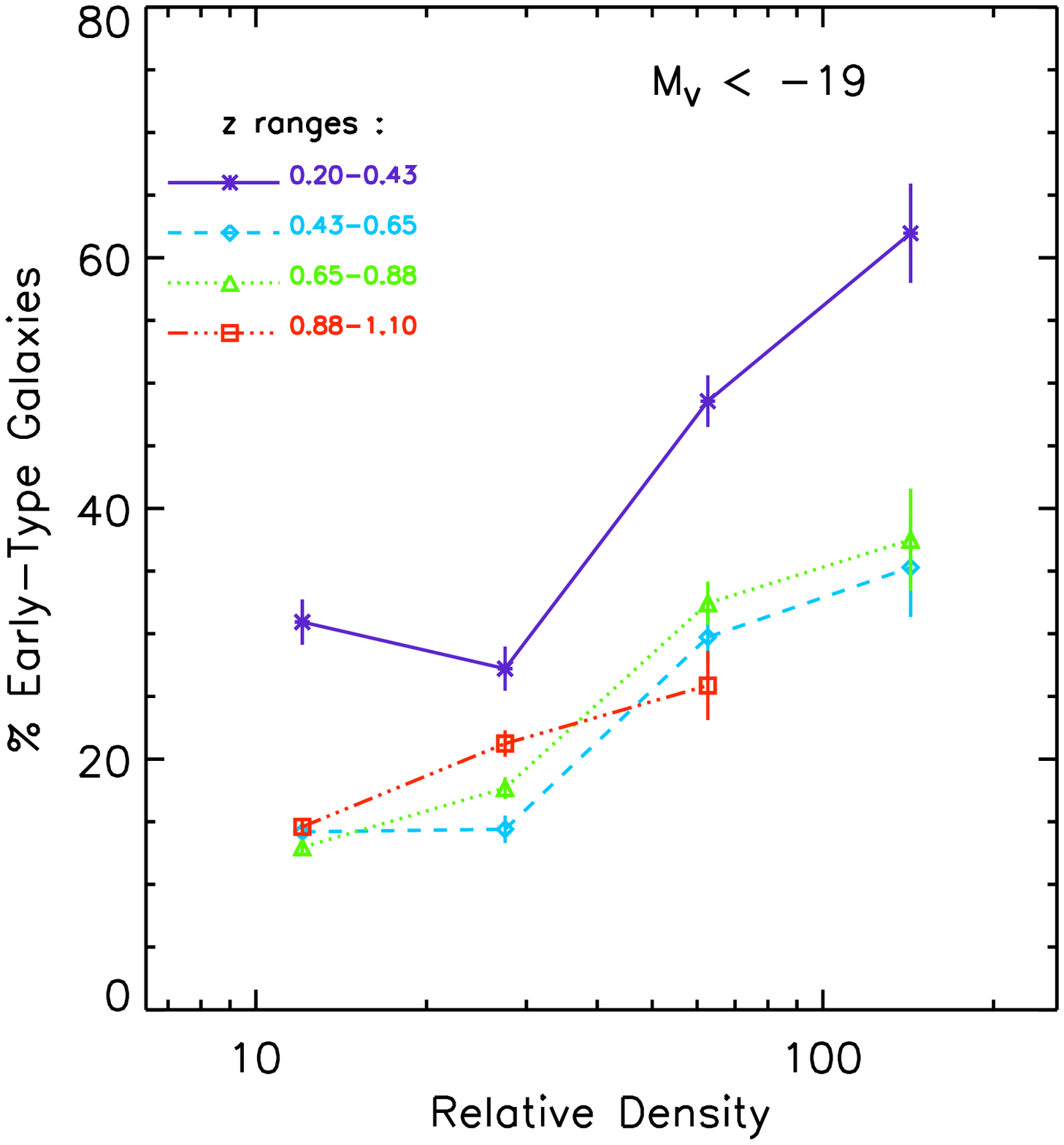}{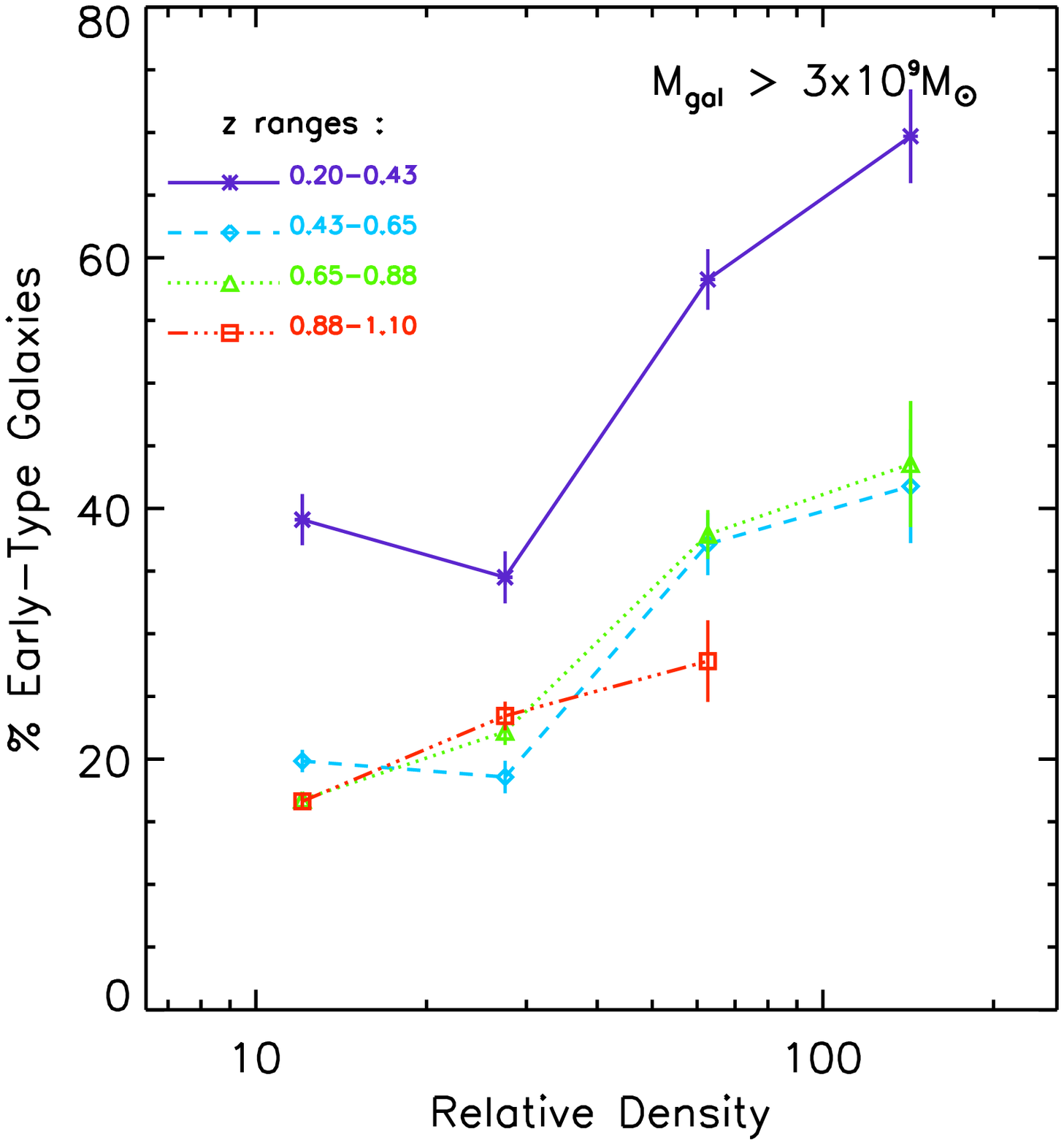}
\caption{ The median early galaxy fraction (or red galaxy fraction with type $< 1.9$) is shown for galaxies 
binned in redshift (the curves) as a function of environmental density. Two samples are
shown : M$_V < -19$ mag (left panel) and  M$_* > 3\times10^9$\msun (right panel).} 
\label{red_dens}
\end{figure}

 \begin{figure}[ht]
\epsscale{1.} 
\plottwo{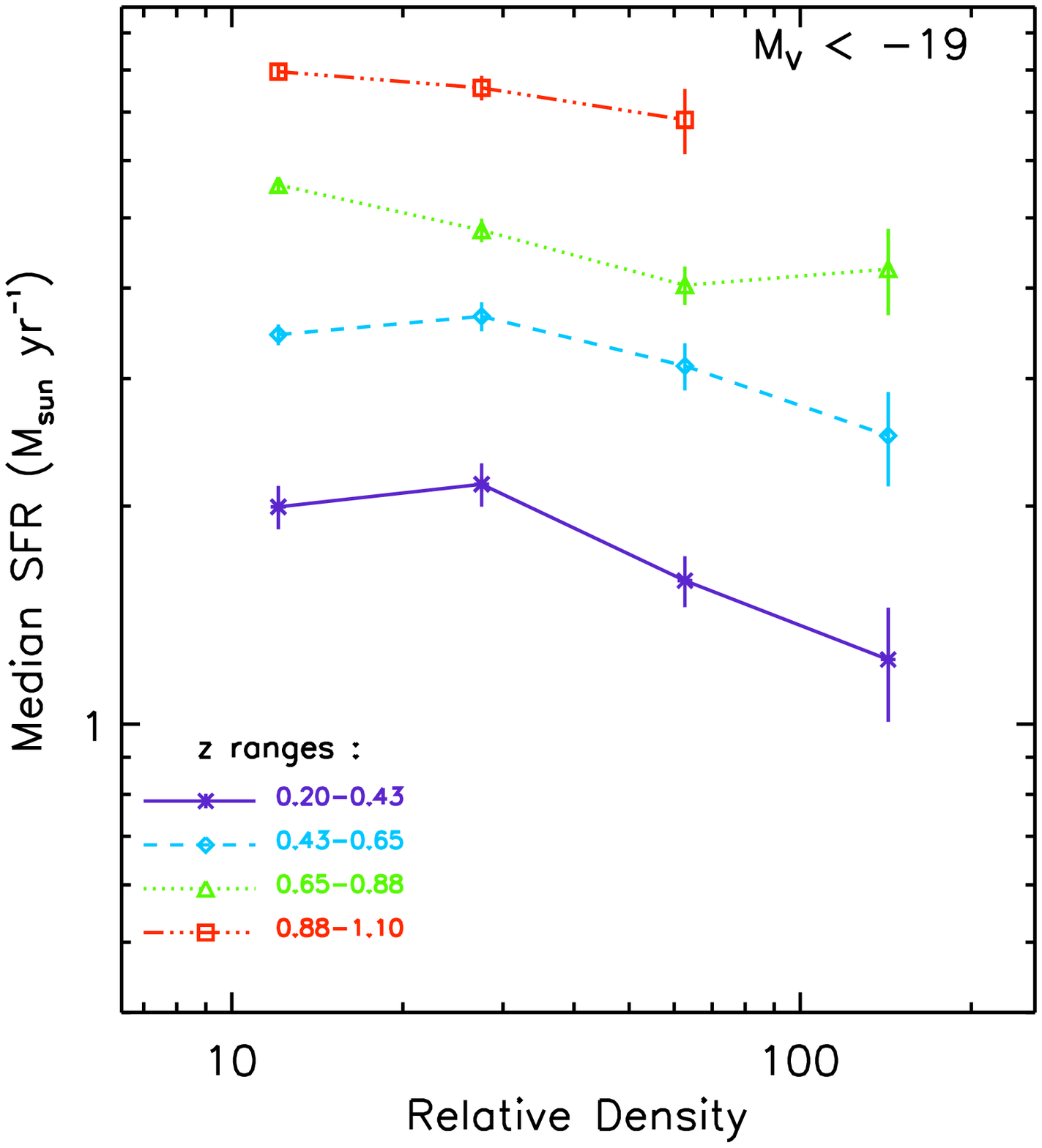}{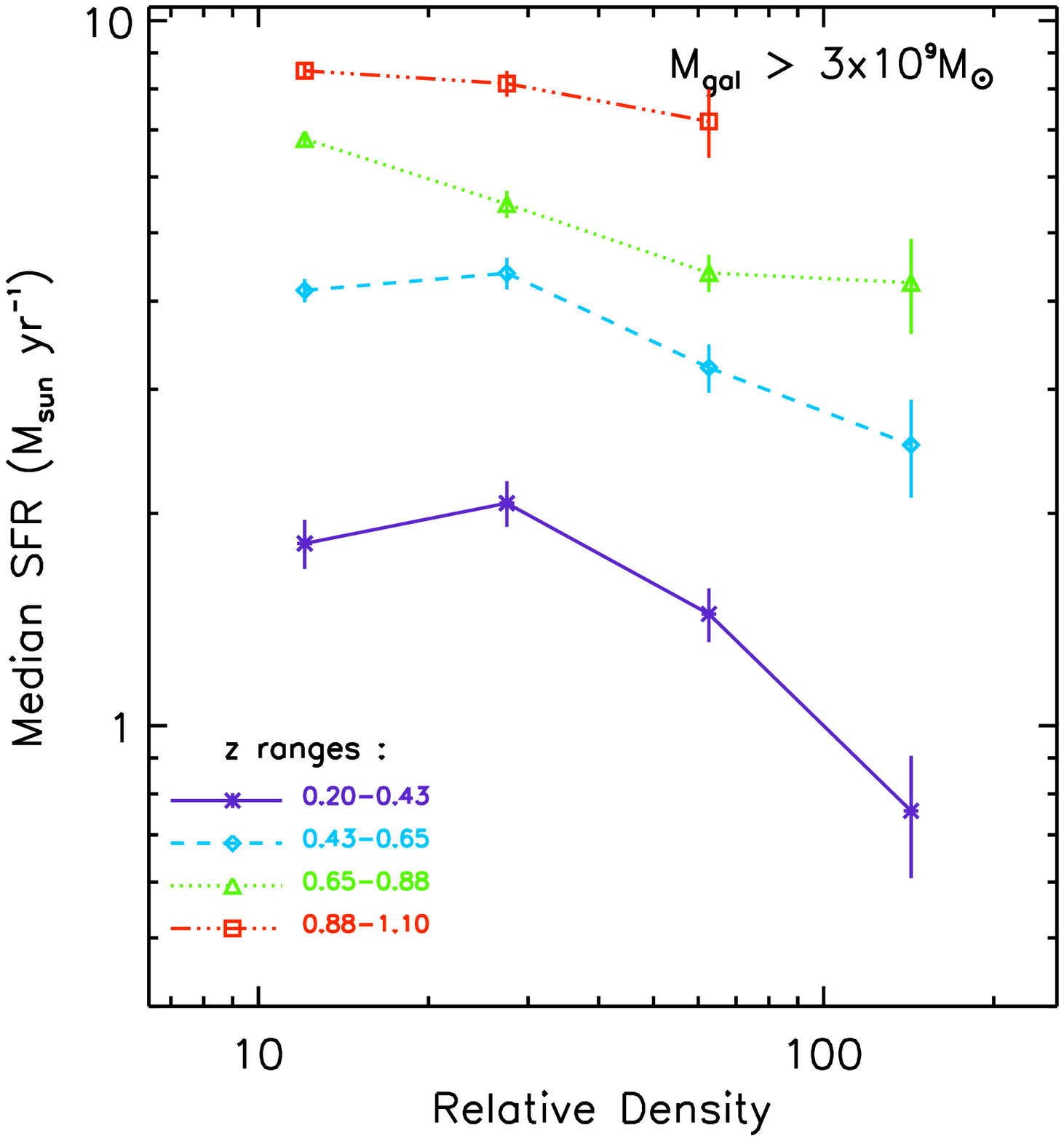}

\caption{The median star formation rate (SFR, \msun yr$^{-1}$) is shown for galaxies 
binned in redshift (the curves) as a function of environmental density. Two samples are
shown : M$_V < -19$ mag (left panel) and  M$_* > 3\times10^9$\msun (right panel).} 
\label{sfr_dens}
\end{figure}

 \begin{figure}[ht]
\epsscale{1.} 
\plottwo{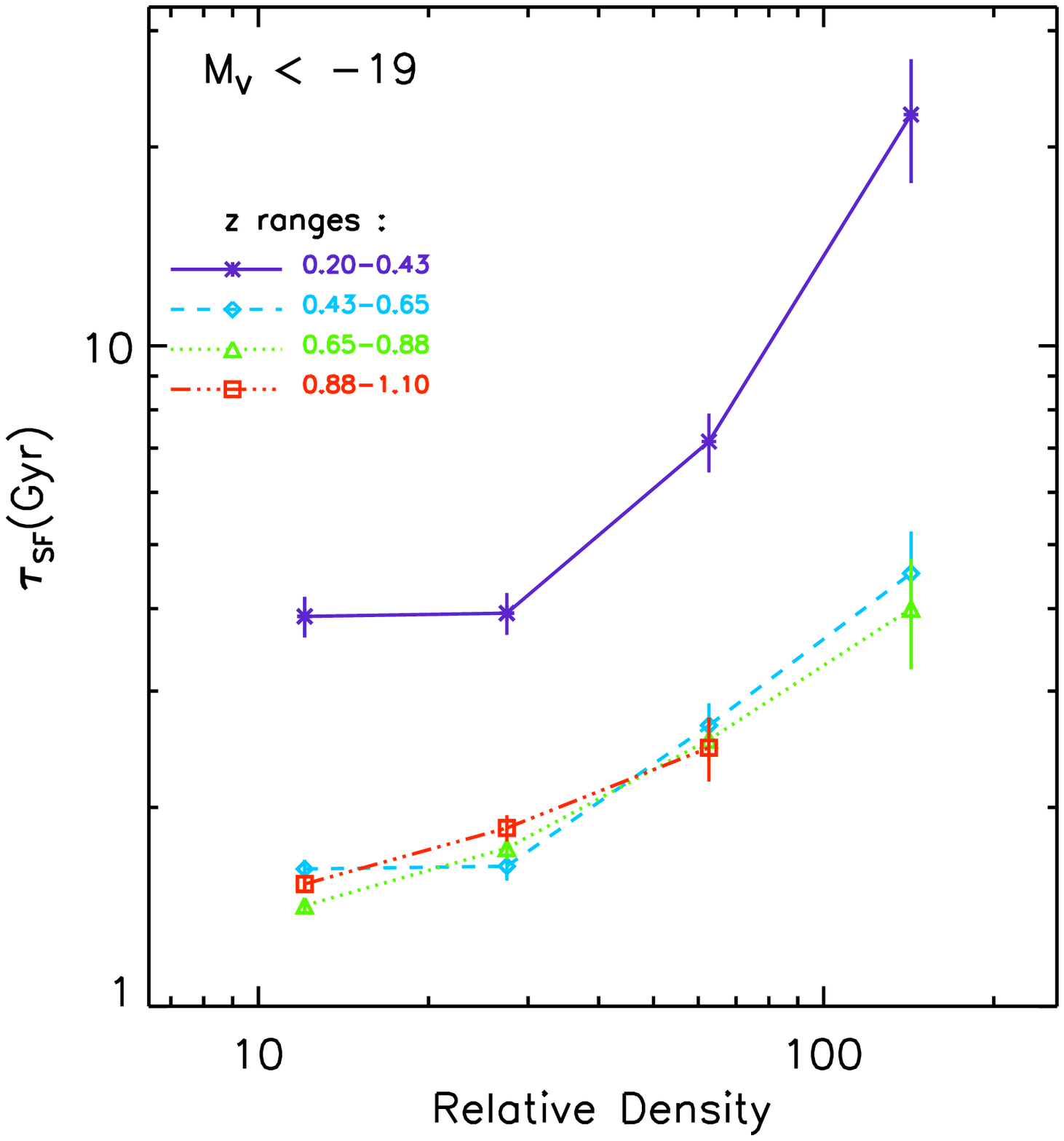}{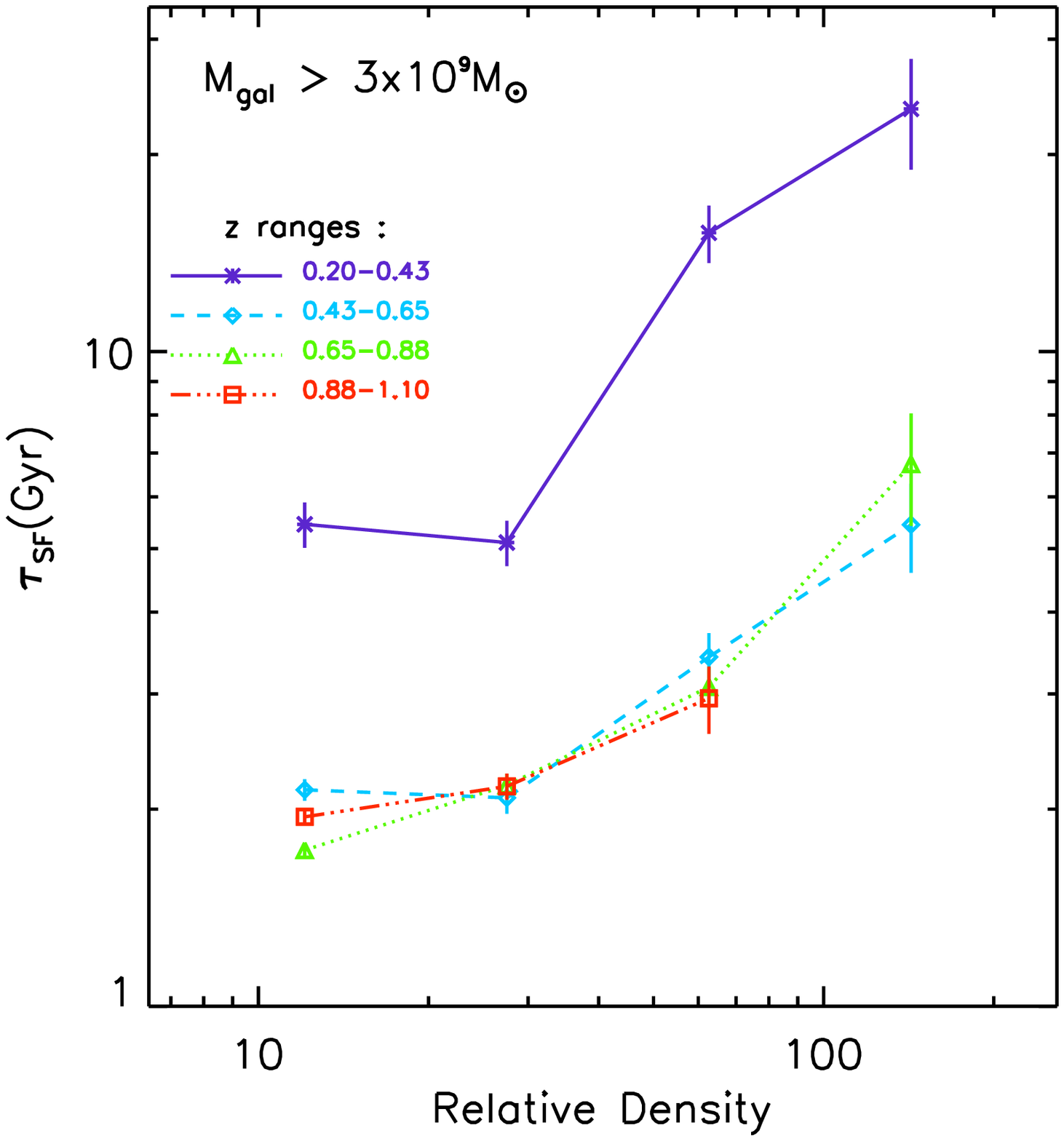}

\caption{The median star formation timescale ($\tau_{SF} = $M$_*$ / SFR) is shown for galaxies 
binned in redshift (the curves) as a function of environmental density. Two samples are
shown : M$_V < -19$ mag (left panel) and  M$_* > 3\times10^9$\msun (right panel).} 
\label{tau_dens}
\end{figure}

 \begin{figure}[ht]
\epsscale{1.} 
\plotone{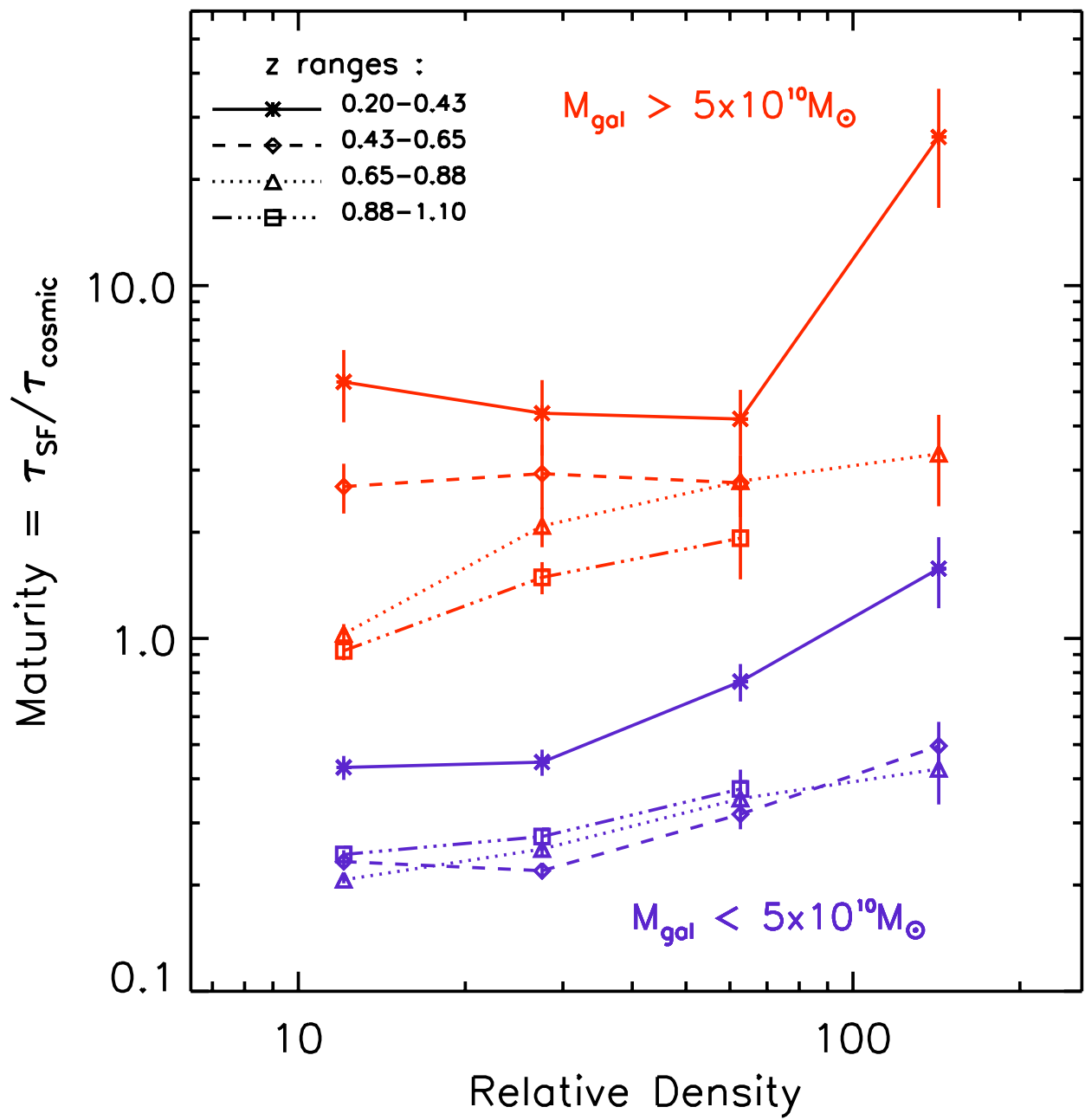}

\caption{The median Maturity (=$\tau_{SF}/\tau_{cosmic}$) is shown for galaxies 
binned in redshift (the curves) as a function of environmental density. M $< 1$ correspond to active,
star forming galaxies and M $>1$ corresponds to galaxies which formed most of their
stars at earlier epochs but see text for a more detailed discussion for interpretation of the Maturity. } 
\label{tau_dens_color}
\end{figure}

 \begin{figure}[ht]
\epsscale{0.9} 
\plottwo{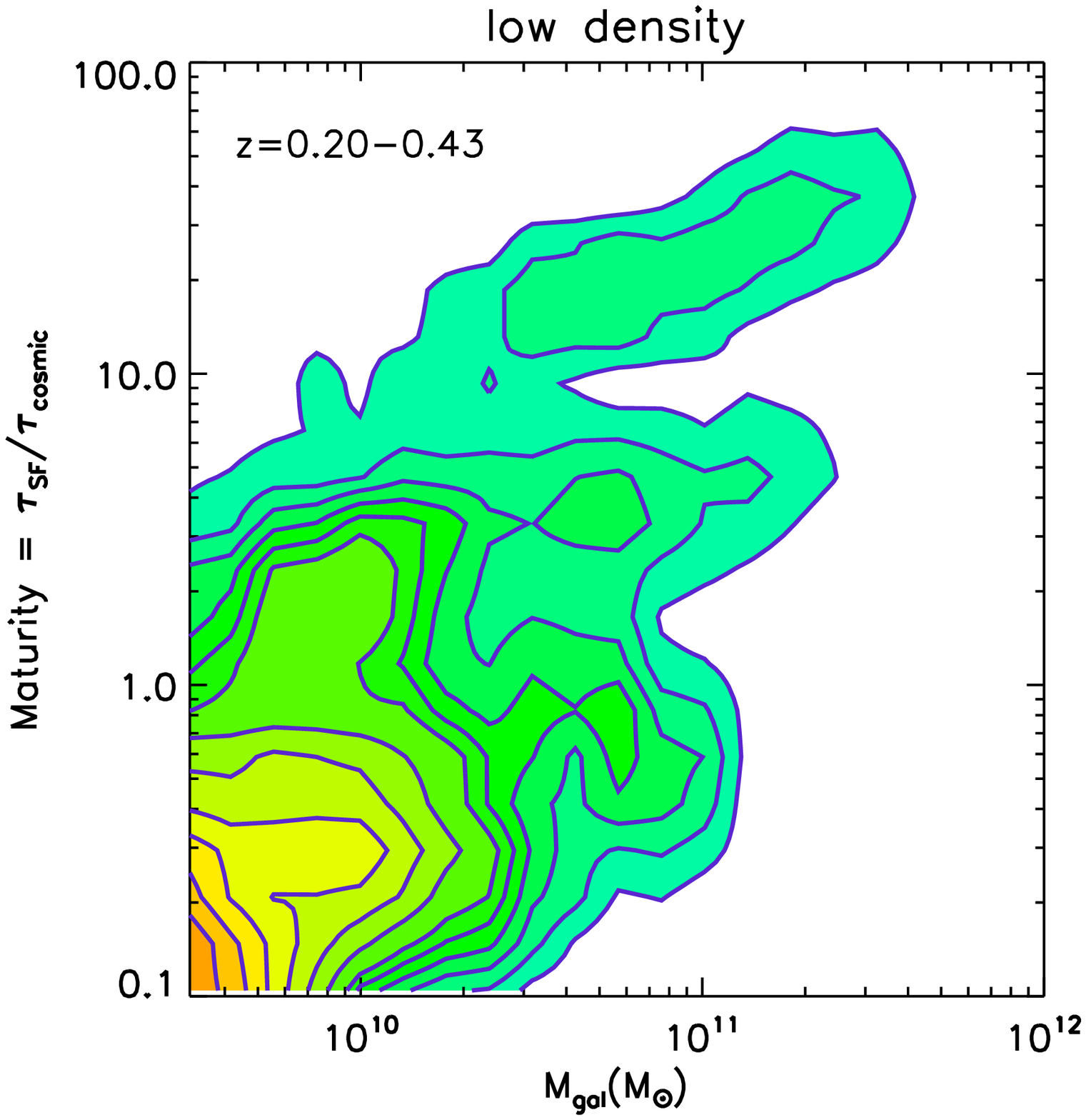}{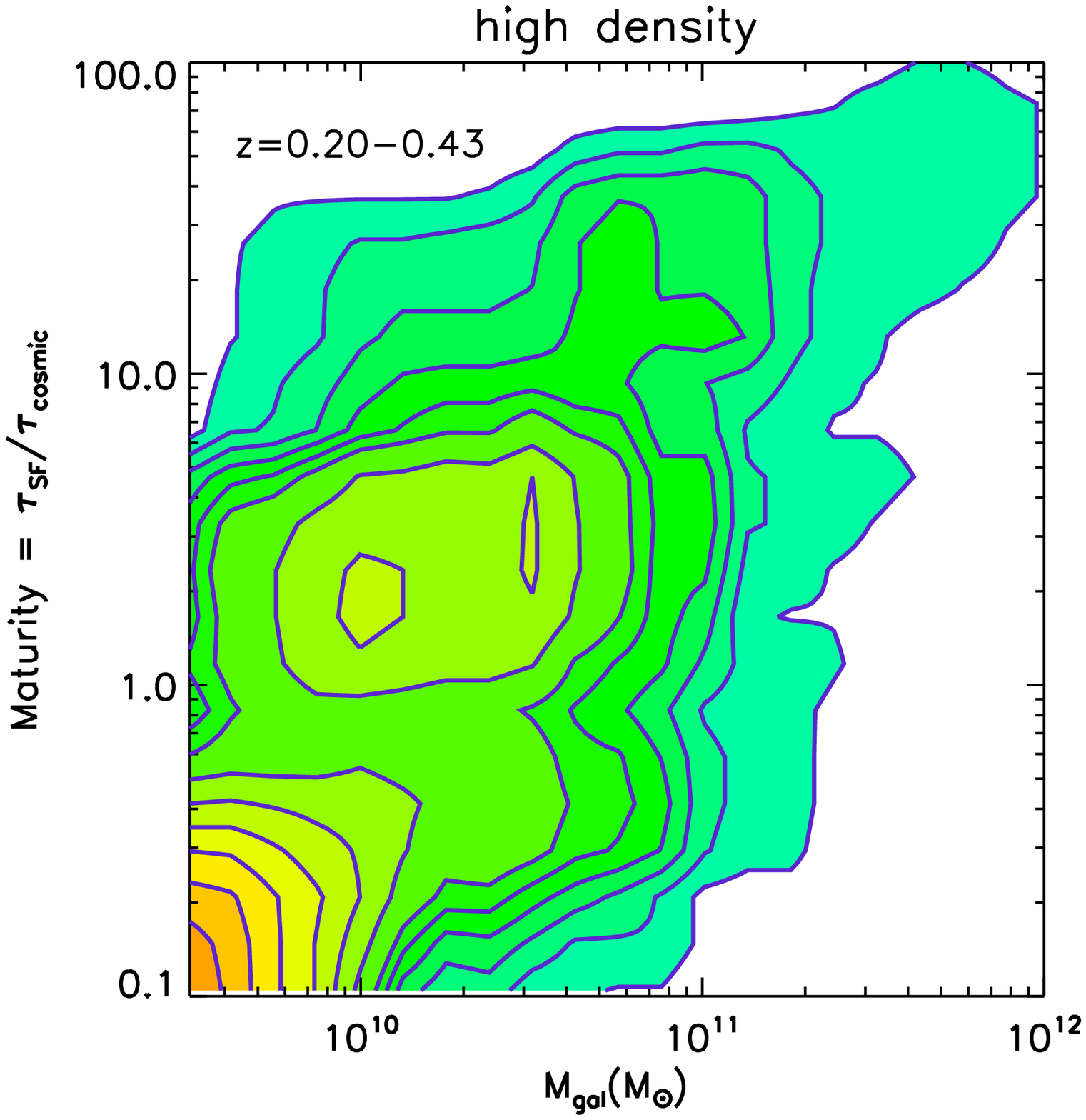}
\plottwo{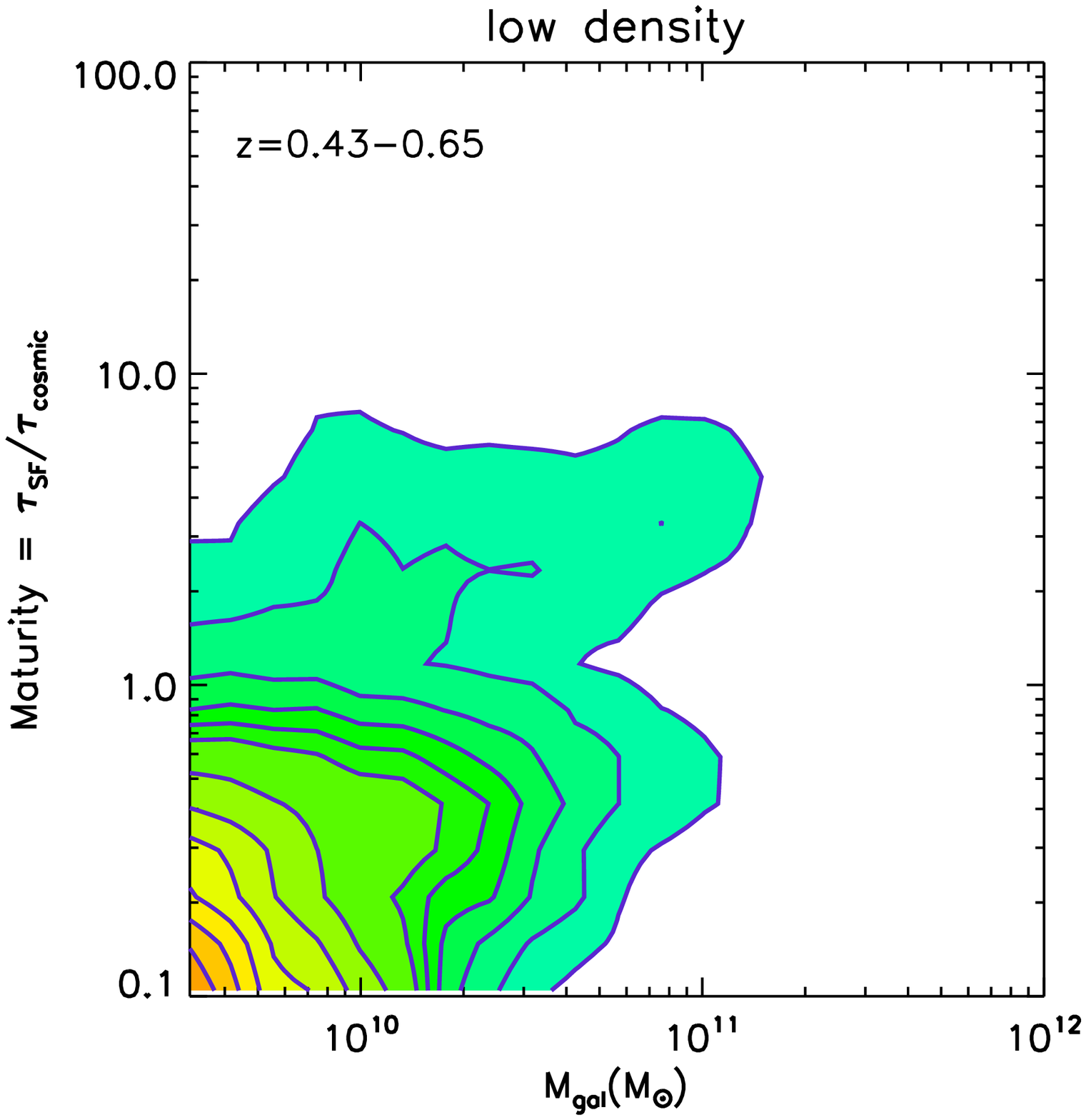}{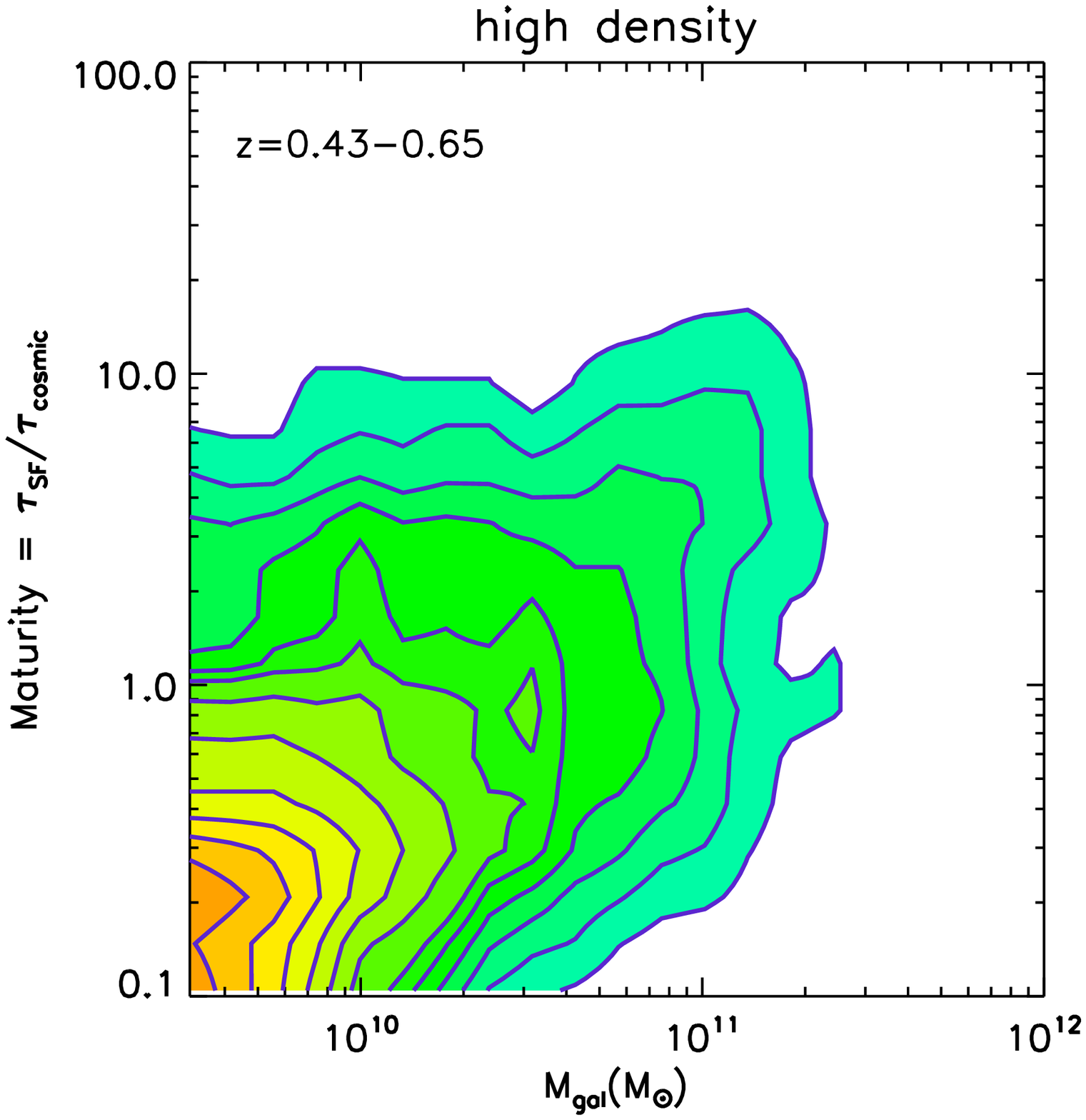}

\plottwo{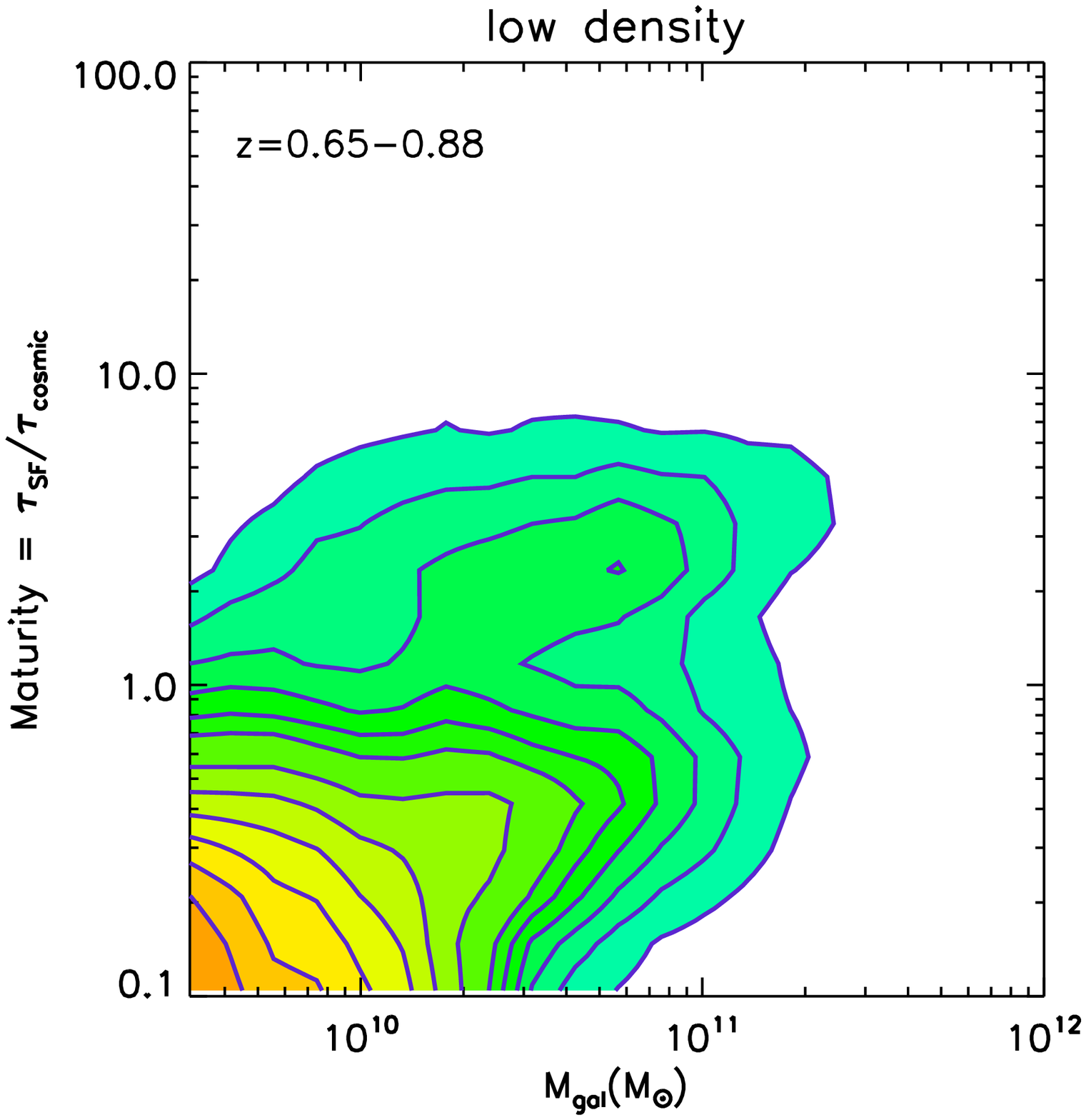}{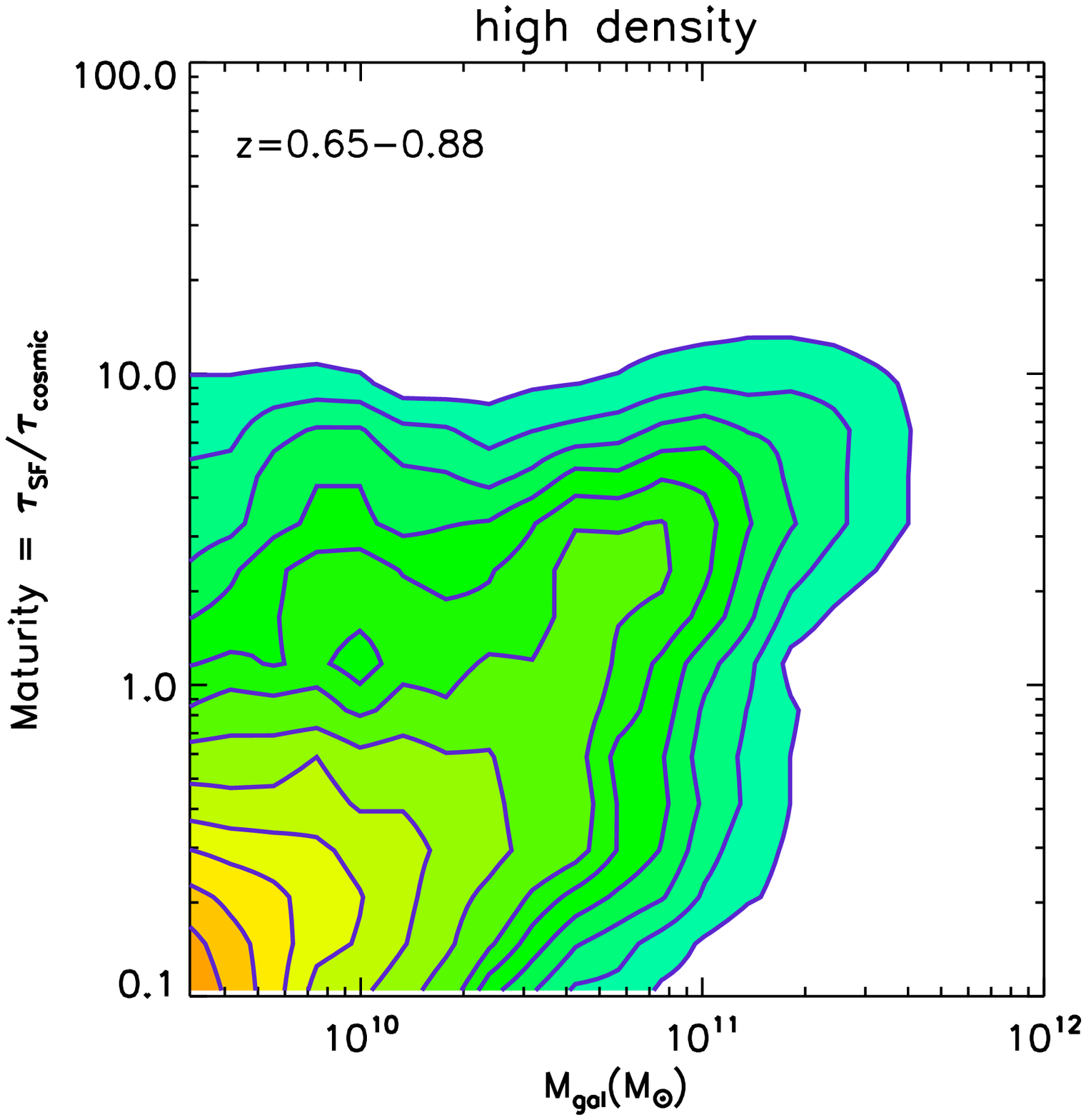}

\caption{The Maturity ($= \tau_{SF}/\tau_{cosmic}$) is shown as a function 
of galactic mass for high and low density environments in three redshift bins. The cut between high and low density was taken at 
relative density $\rho = 45$, i.e. between the middle two bins of the 4 density bins
used earlier. M $< 1$ correspond to active,
star forming galaxies and M $>1$ corresponds to galaxies which formed most of their
stars at earlier epochs but see text for a more detailed discussion for interpretation of the Maturity. Contours are normalized at
0.05, 0.1, 0.15, 0.2, 0.25, 0.3, 0.4, 0.5, 0.6, 0.7, 0.8, 0.9 times the peak.} 
\label{mtrans}
\end{figure}

\begin{figure}[ht]
\epsscale{0.6} 

\includegraphics[width=3.3in,angle=90]{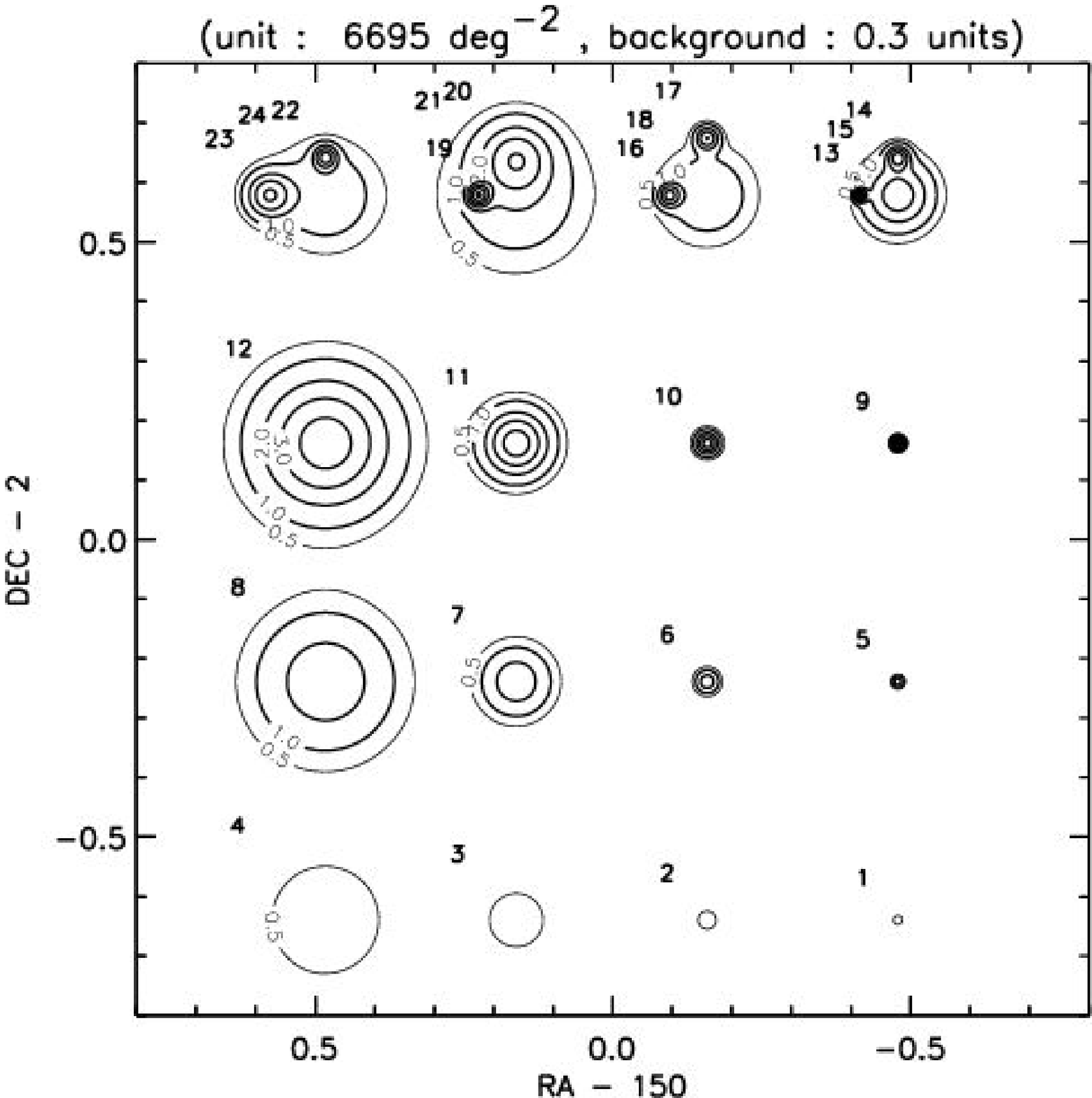}
\includegraphics[width=3.3in,angle=90]{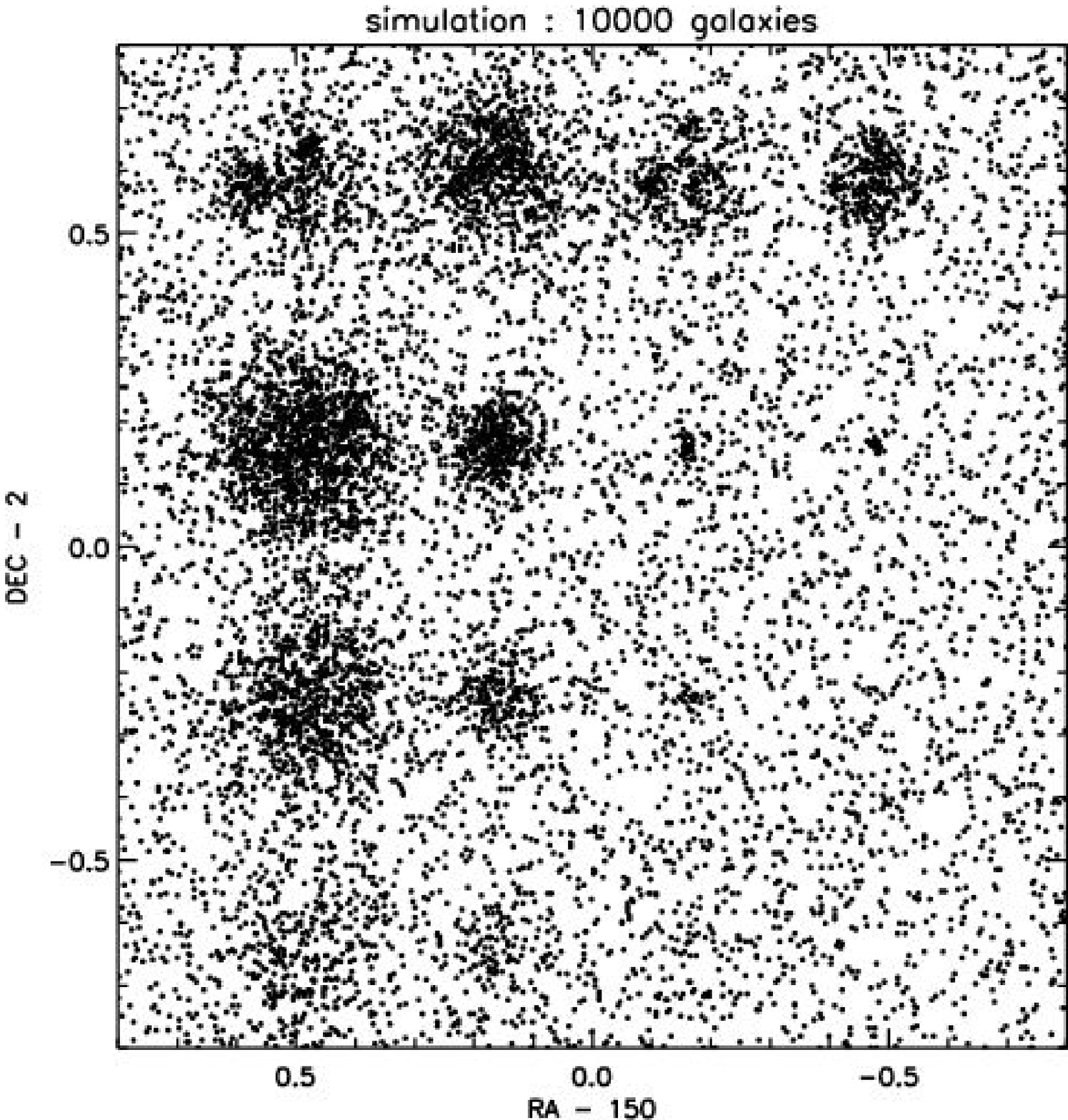}
\caption{a) Top : The projected model densities of the clusters are shown for the 
parameters given in Table \ref{sim_param}. The number to the upper left of 
each feature are the structure ID's given in Table \ref{sim_param}. The contour units 
are given in the top legend, along with the background $\Sigma$ in the same units.  b) Bottom : The positions of 10,000 galaxies (including 5260 randomly placed galaxies) are shown for 
the cluster distribution in the top figure.} 
\label{simulation}
\end{figure}

\begin{figure}[ht]
\epsscale{1.0} 
\includegraphics[angle=90]{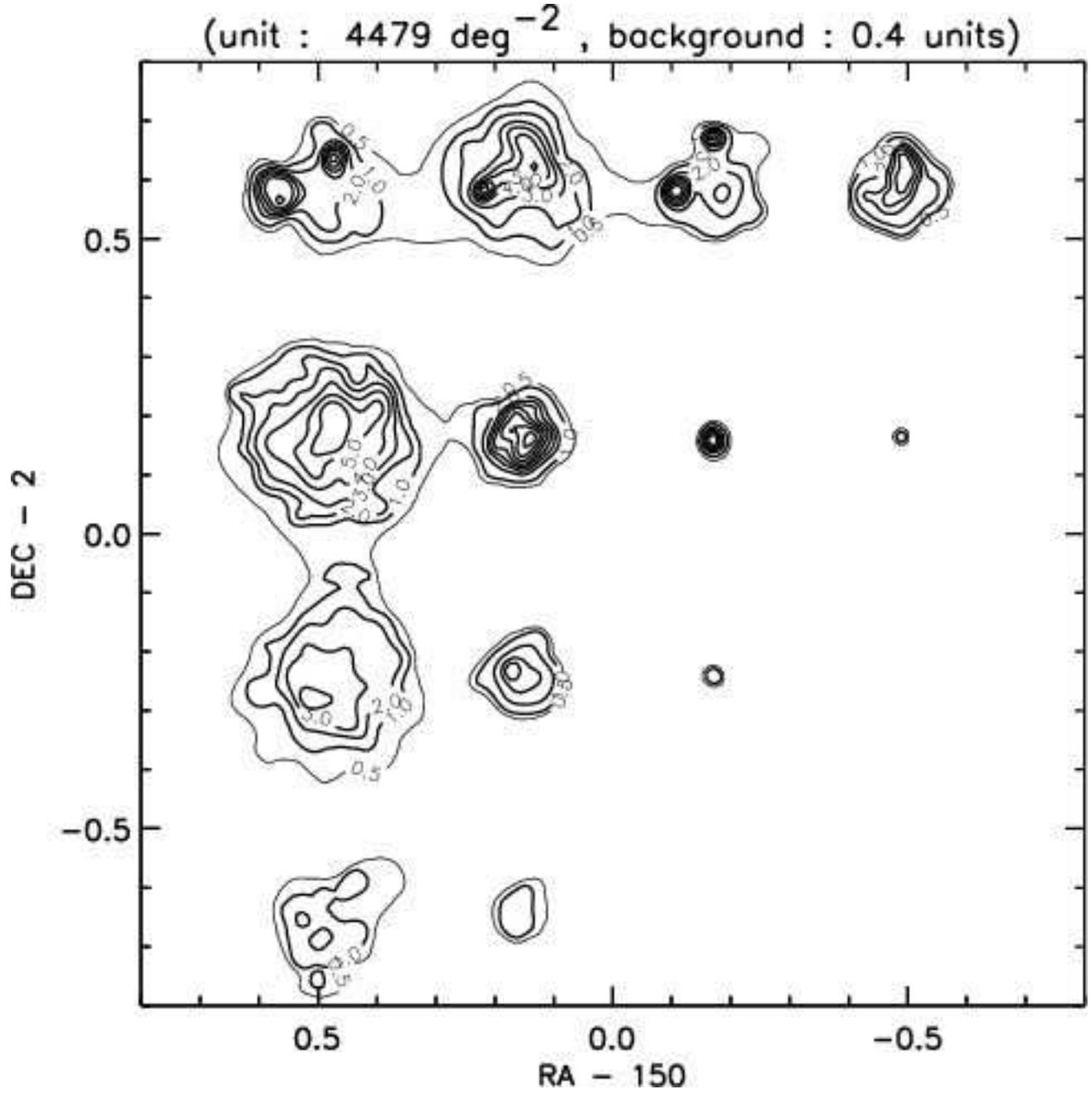}

\caption{The density distribution recovered using the adaptive smoothing algorithm.} 
\label{recover}
\end{figure}

 \clearpage

 \begin{figure}[ht]
\epsscale{1.} 
\plottwo{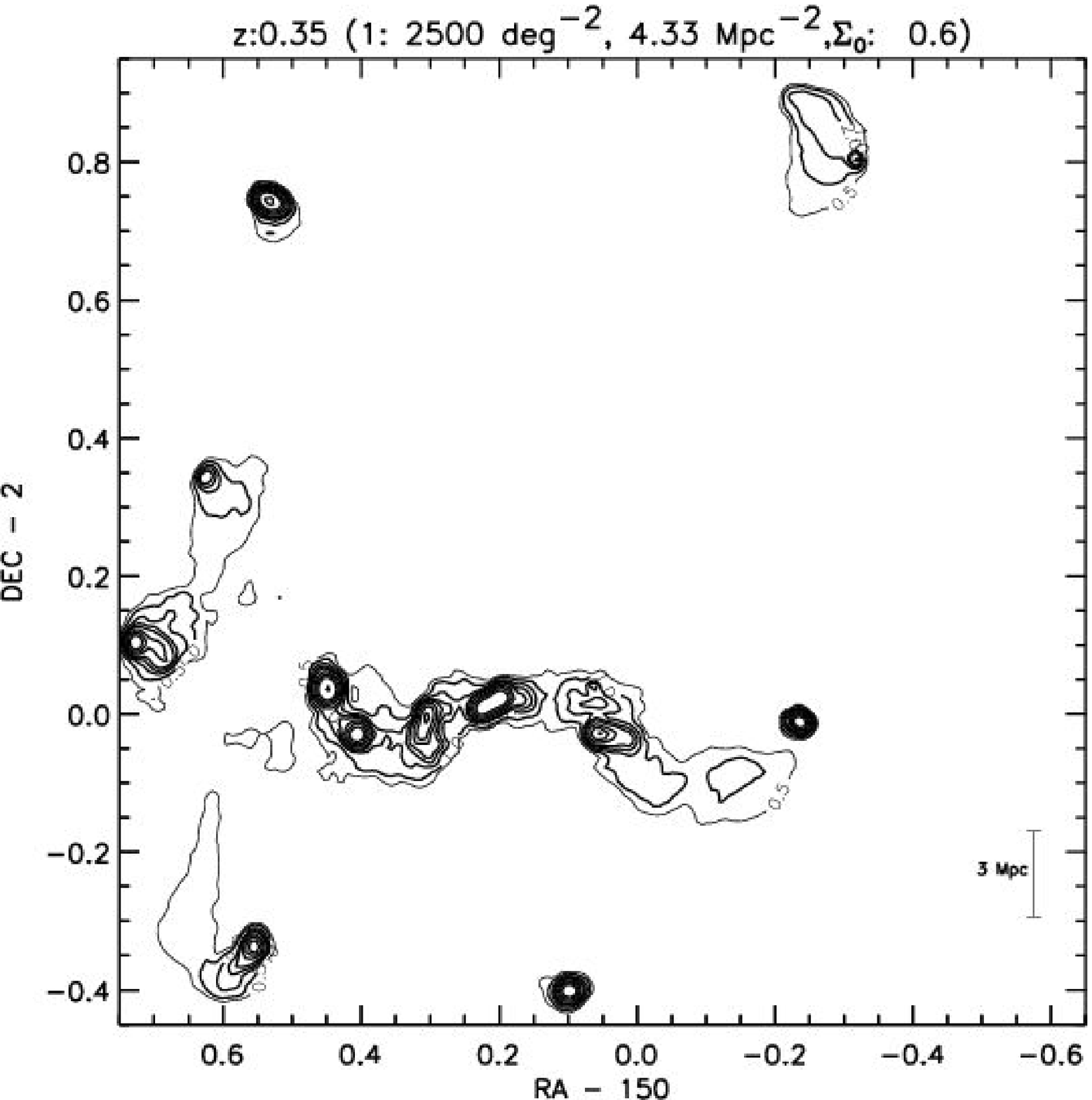}{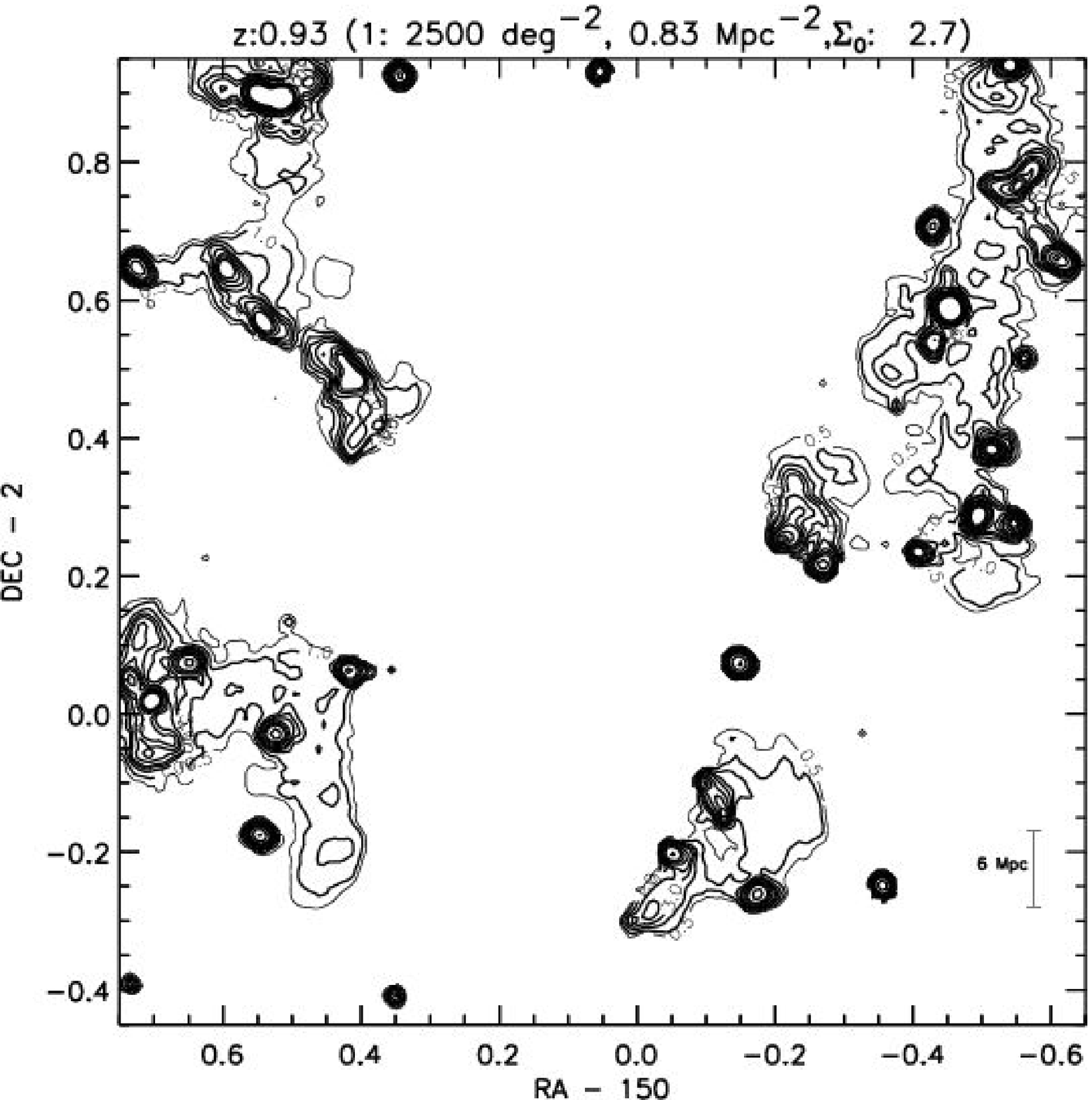}
\caption{Application of  the adaptive smoothing algorithm to the dark matter particle 
distribution in one of the Virgo consortium $\Lambda$CDM simulations \citep{ben01}. The algorithm 
detects virtually all significant structures seen visually in the simulation. More importantly,
the characterisitcs of the detected structures (their scales and changes as a function of 
redshift) correspond closely to those detected in the galaxy overdensities from the 
photometric redshift catalog.} 
\label{lcdm}
\end{figure}

 \begin{figure}[ht]
\plotone{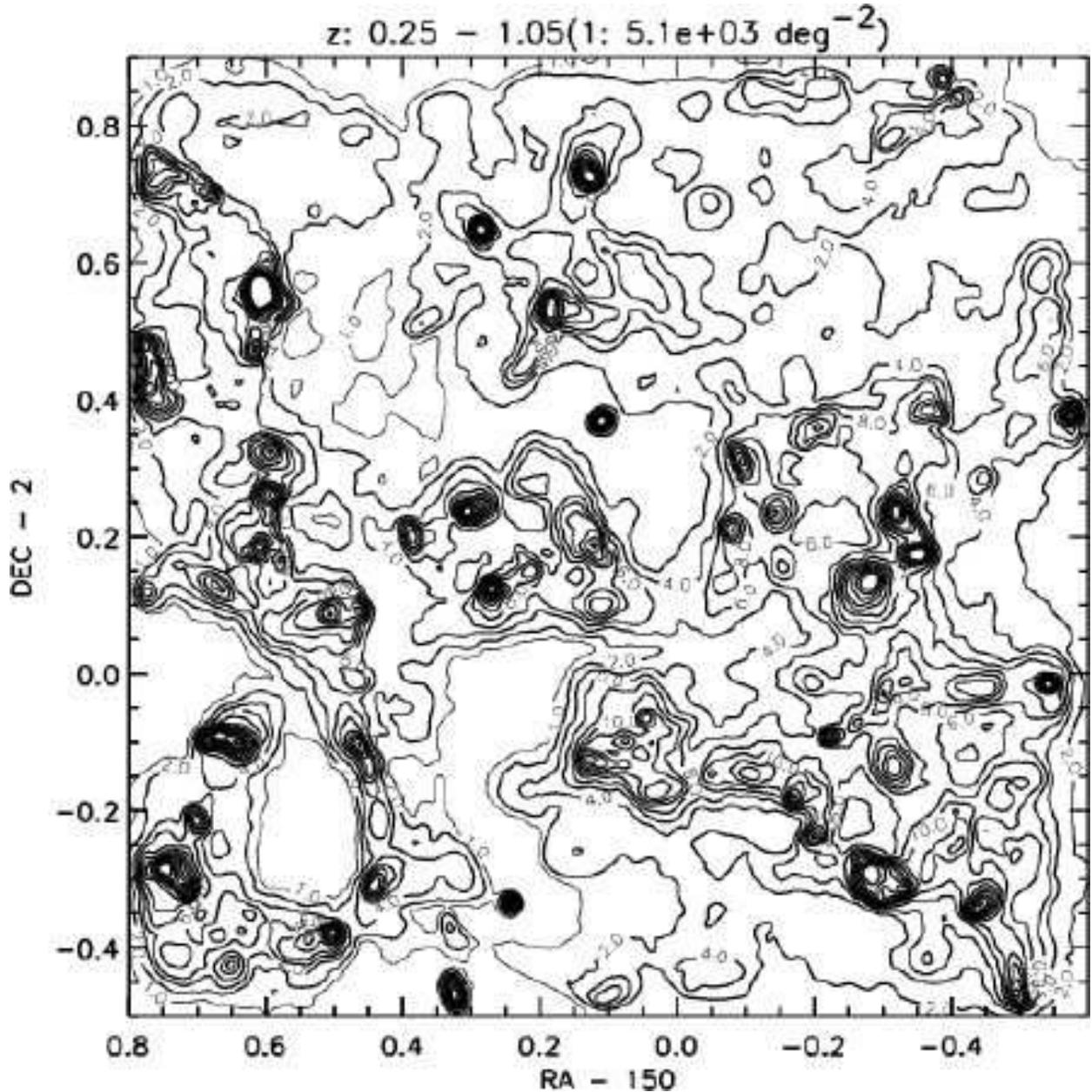}
\caption{Application of  the adaptive smoothing algorithm to the galaxies 
generated in the Millennium Simulation COSMOS light cone were processed 
identically to the observed galaxy distribution, including the imposition of 
redshift uncertainties like those in the observations and using the same 
smoothing parameters. The figure shows the projected distribution 
of overdensities integrated from z = 0.2 to 1.05 and with contours 
identical to those employed for the observations in Figure \ref{lss_2d}. 
} 
\label{mil}
\end{figure}

 
 \end{document}